\documentclass[twocolumn,tighten]{aastex62}
\usepackage{natbib}
\usepackage{epsfig}
\usepackage{amsmath}
\usepackage{xcolor}
\usepackage[encapsulated]{CJK}
\usepackage{ucs}
\usepackage[utf8x]{inputenc}

\newcommand{\sigsfrunits}{\mbox{M$_\odot$ yr$^{-1}$ kpc$^{-2}$}}

\newcommand{\mtolwise}{\mbox{$\Upsilon_\star^{3.4}$}}
\newcommand{\mtolunits}{\mbox{${\rm M}_\odot~{\rm L}_\odot^{-1}$}}

\graphicspath{{./}{graphics/astroph/}}

\newcommand{\zomgs}{$z$0MGS}
\newcommand{\CASS}{\affiliation{Center for Astrophysics and Space Sciences, Department of Physics, University of California, San Diego\\9500 Gilman Drive, La Jolla, CA 92093, USA}}
\newcommand{\OSU}{\affiliation{Department of Astronomy, The Ohio State University\\4055 McPherson Laboratory, 140 West 18th Ave, Columbus, OH 43210, USA}}
\newcommand{\CCAPP}{\affiliation{Center for Cosmology and Astroparticle Physics, The Ohio State University, Columbus, OH 43210, USA}}
\newcommand{\Dunlap}{\affiliation{Department of Astronomy \& Astrophysics and Dunlap Institute, University of Toronto, 50 Saint George Street, Toronto, ON, M5S 3H4, Canada; visitor, Department of Physics \& Astronomy, University of Waterloo, 200 University Avenue West, Waterloo, ON, N2L 3G1, Canada; visitor, Perimeter Institute for Theoretical Physics, 31 Caroline Street North, Waterloo, ON, N2L 2Y5, Canada}}
\newcommand{\Indiana}{\affiliation{Department of Astronomy, Indiana University, Bloomington, IN 47404, USA}}

\shorttitle{IR and UV Maps of Local Galaxies}
\shortauthors{Leroy, Sandstrom et al.}

\begin{document}

\title{A $z=0$ Multi-wavelength Galaxy Synthesis I: \\ 
A WISE and GALEX Atlas of Local Galaxies}

\correspondingauthor{Adam K. Leroy}
\email{leroy.42@osu.edu}

\author[0000-0002-2545-1700]{Adam K. Leroy}
\OSU
\author[0000-0002-4378-8534]{Karin M. Sandstrom}
\CASS
\author{Dustin Lang}
\Dunlap
\author{Alexia Lewis}
\OSU
\CCAPP
\author{Samir Salim}
\Indiana
\author{Erica A. Behrens}
\OSU
\author{J\'{e}r\'{e}my Chastenet}
\CASS
\author[0000-0003-2551-7148]{I-Da Chiang \begin{CJK*}{UTF8}{bkai}(江宜達)\end{CJK*}}
\CASS
\author{Molly J. Gallagher}
\OSU
\author{Sarah Kessler}
\OSU
\author[0000-0003-4161-2639]{Dyas Utomo}
\OSU

\begin{abstract}
We present an atlas of ultraviolet and infrared images of $\sim 15,750$ local ($d \lesssim 50$~Mpc) galaxies, as observed by NASA's WISE and GALEX missions. These maps have matched resolution (FWHM $7.5''$ and $15''$), matched astrometry, and a common procedure for background removal. We demonstrate that they agree well with resolved intensity measurements and integrated photometry from previous surveys. This atlas represents the first part of a program (the $z=0$ Multi-wavelength Galaxy Synthesis) to create a large, uniform database of resolved measurements of gas and dust in nearby galaxies. The images and associated catalogs will be publicly available at the NASA/IPAC Infrared Science Archive. This atlas allows us estimate local and integrated star formation rates (SFRs) and stellar masses (M$_\star$) across the local galaxy population in a uniform way. In the appendix, we use the population synthesis fits of \citet{SALIM16,SALIM18} to calibrate integrated M$_\star$ and SFR estimators based on GALEX and WISE. Because they leverage an SDSS-base training set of $> 100,000$ galaxies, these calibrations have high precision and allow us to rigorously compare local galaxies to Sloan Digital Sky Survey results. We provide these SFR and M$_\star$ estimates for all galaxies in our sample and show that our results yield a ``main sequence'' of star forming galaxies comparable to previous work. We also show the distribution of intensities from resolved galaxies in NUV-to-WISE1 vs. WISE1-to-WISE3 space, which captures much of the key physics accessed by these bands.
\end{abstract}


\section{Introduction}
\label{sec:intro}

Surveys that simultaneously resolve stellar structure, recent star formation, and ISM content across galaxies represent powerful tools to test models of chemical evolution, galaxy assembly, molecular cloud and star formation, and dust production. Building on the {\em Spitzer} Infrared Nearby Galaxy Survey \citep[SINGS,][]{KENNICUTT03}, the last fifteen years have seen deep multiwavelength studies of dozens of galaxies. Pairing infrared observations of dust and star formation with radio maps of the ISM has taught us about how gas forms stars \citep[see review in][]{KENNICUTT12} and dust evolution \citep[e.g.][]{DRAINE07B,REMYRUYER14}, among many other topics. 

A major limitation of most resolved multiwavelength studies to date (with a few key exceptions as discussed below) has been their limited sample size. Deep multiwavelength studies that heavily resolve galaxy disks have tended to target dozens of galaxies. Meanwhile multiwavelength surveys that capture the integrated light from galaxies have moved into the regimes of large samples and robust statistics \citep[e.g.,][]{SAINTONGE11,SALIM16}. A next step will be to expand resolved studies to the regime of large samples, which is currently accessed primarily by integrated light studies.

This paper presents an atlas of $\sim 15,750$ local galaxies in ultraviolet, near-IR, and mid-IR light. To first order, the mid-IR and ultraviolet light trace recent massive star formation. The near-IR light traces stellar structure. This is the first part of a larger project, the $z=0$ Multi-wavelength Galaxy Synthesis (\zomgs ), that aims to synthesize a large sample of resolved maps of stellar structure, massive star formation, dust, and gas content in galaxies. Here ``resolved'' means placing on the order of ten resolution elements across the galaxy or achieving resolution of order one kiloparsec. \zomgs\ aims to expand the set of local galaxies tractable for panchromatic studies from dozens to hundreds, and in some dimensions, to thousands. The core components of \zomgs\ are:

\begin{enumerate}
\item \textit{This paper:} An atlas of ultraviolet, near-, and mid-infrared (IR) maps of local galaxies, and a description of the survey infrastructure. These maps are useful to trace recent massive star formation and the distribution of stellar mass.
\item New and archival VLA \textsc{Hi} 21-cm mapping of several hundred local galaxies, including all northern galaxies with resolved {\em Herschel} mapping (D. Utomo et al. in preparation).
\item A multi-resolution atlas of resolved IR spectral energy distributions (J. Chastenet et al. in preparation) in several hundred galaxies based on archival multi-band IR mapping from {\em Herschel}.
\item Resolved CO maps of a subset of the 21-cm targets.
\end{enumerate}

This paper describes infrastructure for the project and the atlas of ultraviolet, near-IR, and mid-IR maps. To build this atlas, we use imaging from the Galaxy Evolution Explorer \citep[GALEX,][]{MARTIN05} and the Wide-field Infrared Survey Explorer \citep[WISE,][]{WRIGHT10}. GALEX observed a large fraction of the sky in near- ($\lambda \approx$ 231 nm, hereafter NUV) and far- ($\lambda \approx$ 154 nm, hereafter FUV) ultraviolet light. WISE mapped the entire sky in near- and mid-IR emission at $\lambda \approx 3.4,~4.6,~12,~{\rm and}~22\mu$m (hereafter WISE1, WISE2, WISE3, and WISE4). The broad sky coverage of both surveys means that they resolved thousands of galaxies.

After selecting a sample of local galaxies from the Lyon Extragalactic Database \citep[LEDA,][]{PATUREL03,PATUREL03B,MAKAROV14}, we construct spatial cutouts for each WISE and GALEX band. We beam-match the images, place them on the same astrometric grid, and subtract local background emission. We use these to carry out integrated-light photometry, allowing each of these local galaxies to be placed in global relations. We also sample the disks of our targets beam-by-beam, constructing a database of resolved surface brightness measurements at all 6 bands.

This paper describes the construction of the sample, cutout images, and linked databases. All of these data products are publicly available online at \texttt{irsa.ipac.caltech.edu/data/WISE/z0MGS }. The DOI associated with these data is \texttt{ DOI: 10.26131/IRSA6}.

\subsection{Complementary Literature Efforts}

\zomgs\ resembles or builds on several other efforts over the last decade.

\begin{enumerate}
\item The \textit{Spitzer} Survey of Stellar Structure in galaxies \citep[S4G;][]{SHETH10} obtained \textit{Spitzer} IRAC $3.6\mu$m and $4.5\mu$m imaging for $\sim 2,350$ galaxies in a volume-limited ($d < 40$~Mpc), LEDA-selected sample. Our sample selection heavily resembles that used by S4G, though we do not impose a Galactic latitude cut. We will frequently reference their results. At overlapping wavelengths, the S4G IRAC imaging has higher quality than WISE, but over a smaller field of view. \citet{BOUQUIN18} combined S4G with GALEX imaging, extending the survey to also cover some of the star-formation tracing bands that we analyze here.

\item The Local Volume Legacy survey \citep[LVL;][]{LEE09,DALE09} obtained {\em Spitzer} and GALEX imaging of all galaxies within 11~Mpc. LVL has more complete coverage of the local volume than \zomgs, including dwarf galaxies, but targets $\sim 100$ times smaller volume. The LVL {\em Spitzer} observations are much more sensitive than those achieved by WISE, which struggles to detect many of the faint, small galaxies within 11~Mpc.

\item The \textit{Herschel} Reference Survey \citep[HRS,][]{BOSELLI10} built infrared maps of a volume-limited sample of $\sim 320$ galaxies in the range $d \approx 15{-}25$~Mpc. The HRS team has carried out a wide range of multiwavelength analysis and follow-up observations. We will make frequent reference to their results.

\item The GALEX-SDSS-WISE Legacy Catalog \citep[GSWLC,]{SALIM16,SALIM18} carried out spectral energy distribution fitting on the integrated optical and UV light for $\sim 700,000$ low-redshift galaxies and combined these with WISE photometry. This represents a key point of comparison for linking \zomgs\ to integrated light studies.

\end{enumerate}

In addition to those projects, the Star Formation Reference Survey \citep[SFRS]{ASHBY11} used {\em Spitzer} and GALEX to pursue similar goals. \citet{JARRETT13} also construct WISE maps of extended sources and consider the translation of WISE intensities to star formation rate estimates \citep[see an application in][]{ELSON18}. And Dustpedia \citep{CLARK18} is also constructing panchromatic spectral energy distributions for hundreds of galaxies. A large number of other surveys have made use of WISE and GALEX intensities or fluxes and considered how they should be used for physical parameter estimation.

\section{Sample}
\label{sec:sample}

\subsection{Super-sample}

Our final \zomgs\ sample consists of $\sim 15,750$ local galaxies with luminosity greater than the LMC and likely to lie within $\sim 50$~Mpc. Before defining this sample, we constructed a larger database of local galaxies that we use to carry out the sample selection. This database is built around the Lyon/Meudon Extragalactic Database \citep[LEDA,][]{PATUREL03,PATUREL03B,MAKAROV14}. We use LEDA's ``PGC'' number as the unique identifier for each galaxy because it is available for all galaxies in the super-sample. 

Our supersample consists of every object identified as a galaxy (\texttt{objtype='G'}) with a heliocentric recessional velocity less than 30,000~km~s$^{-1}$ (\texttt{v < 30000}). We adopt LEDA's list of aliases, galaxy position, and homogenized estimates for the shape, orientation, rotation velocity, and apparent magnitudes, including IR (from IRAS) and \textsc{Hi} fluxes.

\subsection{Adopted Distance and Uncertainty}

\begin{figure*}[t!]
\centering
\plottwo{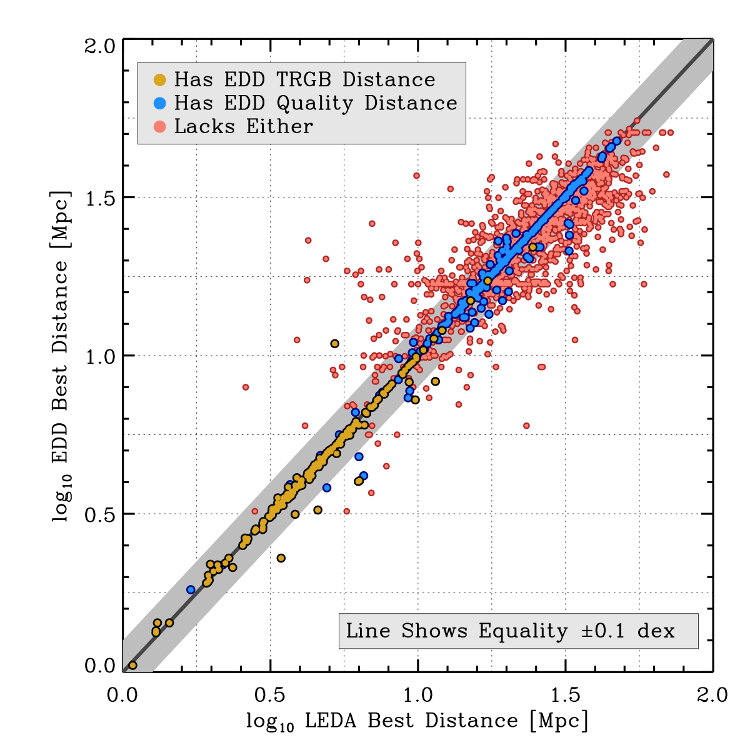}{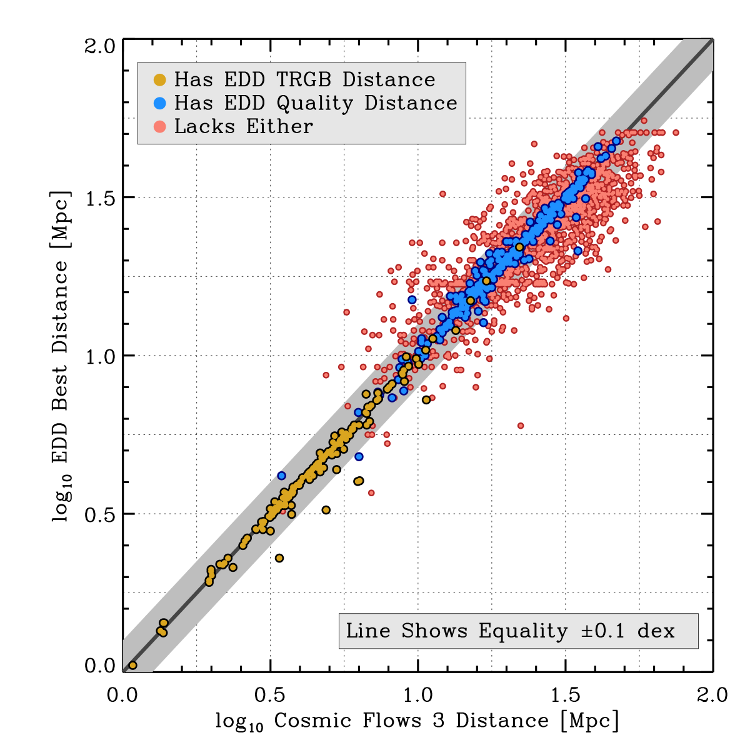}
\plottwo{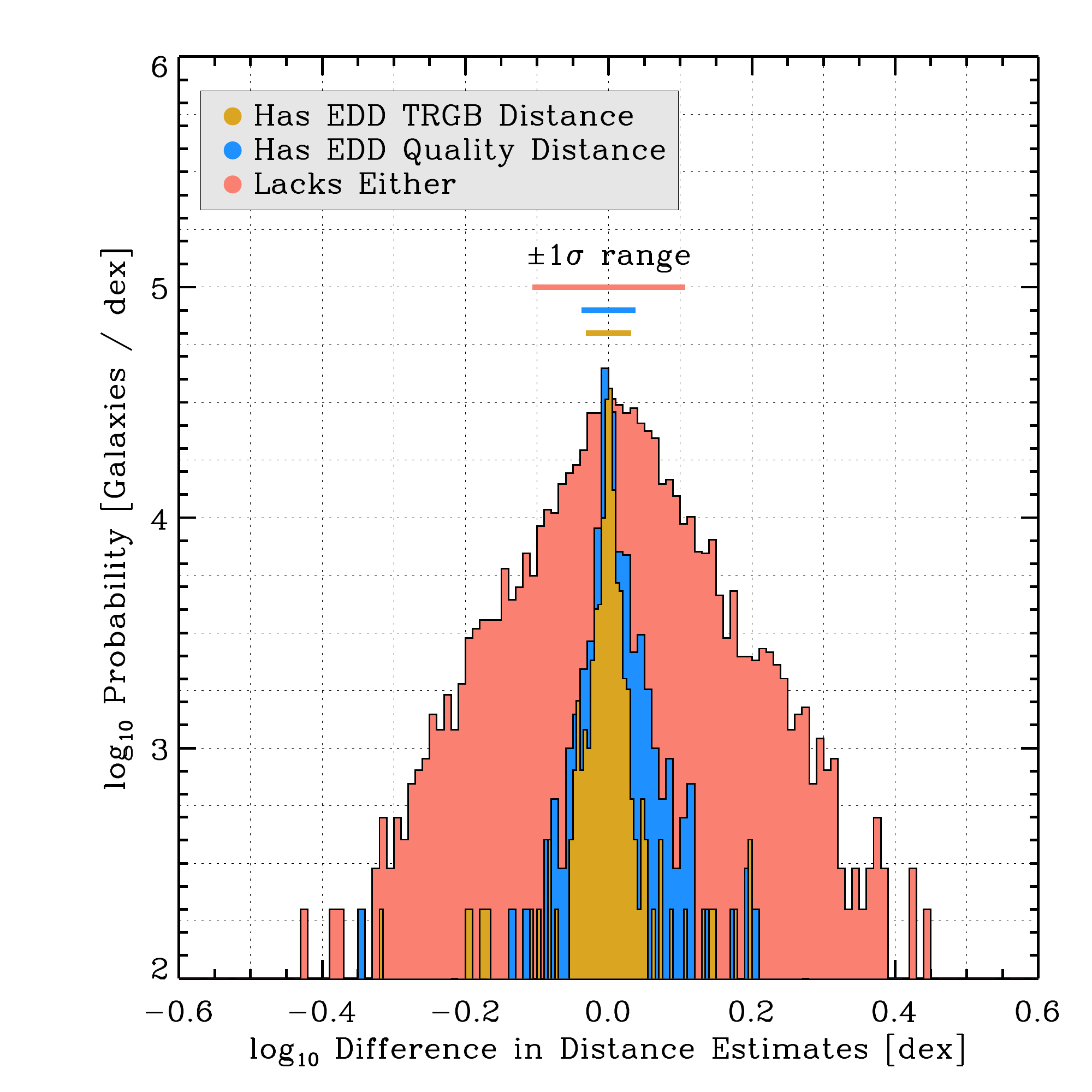}{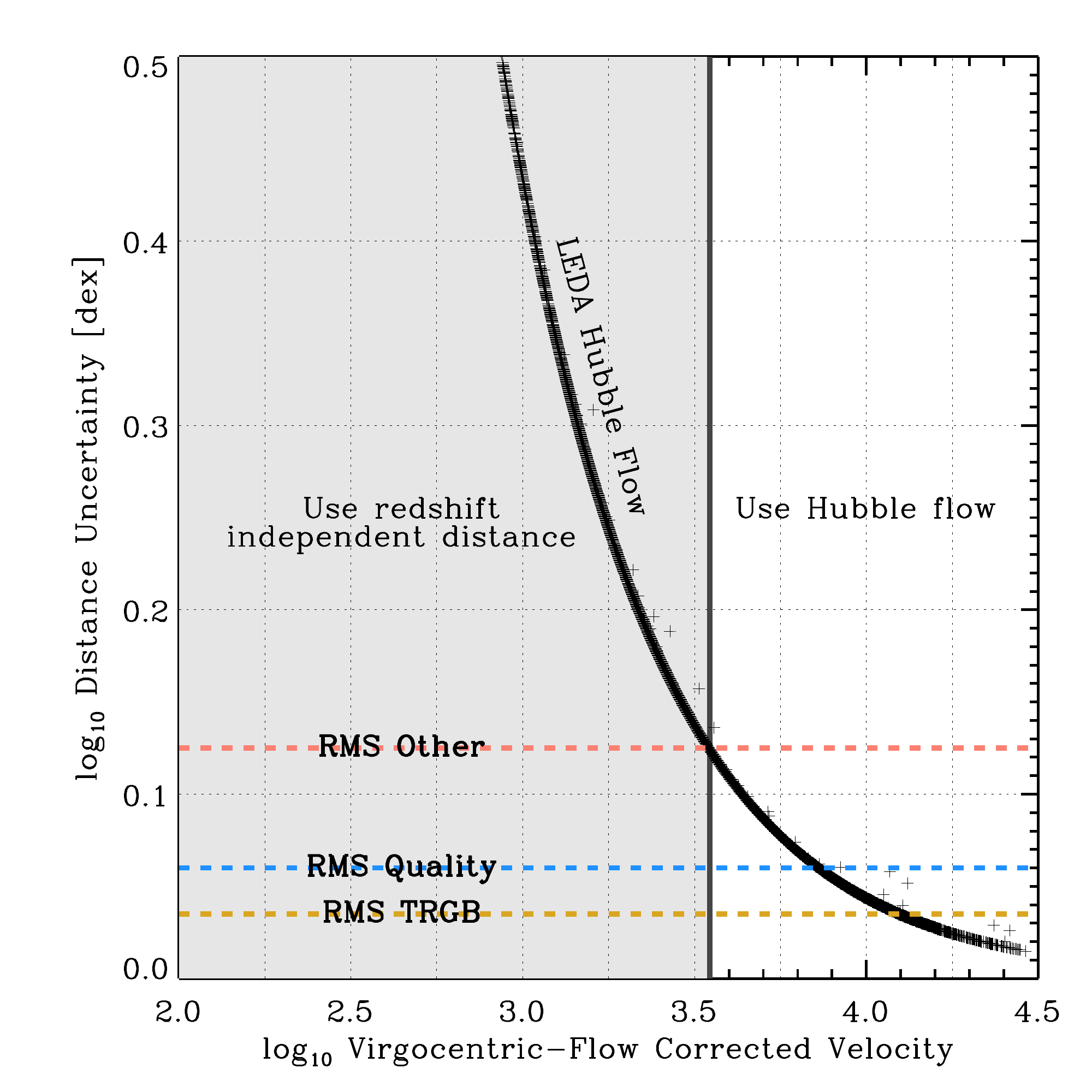}
\caption{{\bf Approach to Distances.} {\em Top row}: Comparison of distance estimates for targets of the Extragalactic Distance Database \citep[EDD,][]{TULLY09}. The $y$ axis in both plots shows the EDD best distance estimate. The $x$ axis shows the best distance estimate from LEDA ({\em left}, \citealt{MAKAROV14}) and CosmicFlows ({\em right}, \citealt{COURTOIS12}, \citealt{TULLY16}). We color code galaxies by the quality of their distance estimate in EDD: gold=TRGB, blue=``quality,'' red=other. The striping visible in the plot reflects the use of (shared) group distances by EDD. {\em Bottom left:} Histograms showing the distribution of differences in distance estimates from different source for the same galaxy. We consider all galaxies that have distance estimates in the Extragalactic Distance Database and either CosmicFlows or LEDA. The bars show the $\pm1\sigma$ width of the distribution. Based on this comparison, we adopt a minimum uncertainty of $0.125$~dex~$\approx 33\%$ for galaxies without a quality distance estimate, $0.06$~dex~$\approx 15\%$ for those with a quality distance determination, and $0.03$~dex~$\approx 7\%$ for those with a TRGB distance. {\em Bottom right:} Our adopted approach to distances and uncertainties, illustrated. Points show the LEDA Virgo-centric flow corrected velocity ($x$-axis) and LEDA's nominal uncertainty in the distance ($y$-axis), and so illustrate the uncertainty in distance using the Hubble flow according to LEDA. Red, blue, and gold lines show our adopted uncertainties for redshift-independent distance estimates. We use redshift-independent distance indicators in the gray region, and Hubble flow distances for $v > 3,500$~km~s$^{-1}$. Because we focus on nearby galaxies, those with $v > 3,500$~km~s$^{-1}$ represent only a small part of the sample.}
\label{fig:distances}
\end{figure*}

Distance plays a key role in our analysis and sample selection. Distance estimates for nearby galaxies remain heterogeneous because of the inability to derive precise distance estimates from only the Hubble flow. High quality distances to individual nearby galaxies, e.g., using the tip of the red giant branch (TRGB), require effort and still constitute an important current area of research \citep[e.g.,][]{MCQUINN17,KARACHENTSEV18}. Unfortunately, the high cost of these measurements means that distances tend to be measured for a few galaxies at a time, and so remain scattered across the literature.

Fortunately, there are several large ongoing efforts to synthesize distance determinations to nearby galaxies. LEDA offers one set of distances, combining redshift-independent distance estimates with Hubble flow distances, corrected for local motions towards the Virgo cluster \citep[``Virgocentric flow corrected'', see][]{MAKAROV14}. The NASA Extragalactic Database (NED) also compiles distance estimates to galaxies \citep{STEER17}. The Extragalactic Distance Database \citep[EDD,][]{TULLY09} and the CosmicFlows project \citep{COURTOIS12,TULLY16} are also compiling distances and estimating the location and motions to many thousands of local galaxies.

These efforts harness a large amount of expertise, and they condense a large, complex literature into accessible databases. Therefore, we adopt distances from these compilations. We specifically focus on EDD, CosmicFlows, and LEDA. We omit NED because it requires the user to apply their own homogenization. We prefer to trust the homogenization already carried out by other, more expert teams. For nearby galaxies with ``low-quality'' redshift-independent distances, we prefer those from EDD and CosmicFlows to those from LEDA because they incorporate knowledge of galaxy groups.

We attempt to impose realistic uncertainties on our adopted distances by comparing different distance estimates for the same galaxy. Figure \ref{fig:distances} illustrates our approach. In the top row, we compare distances for the same galaxy from the EDD, LEDA, and CosmicFlows. In the bottom left panel, we plot histograms of the difference between different distance estimates for the same galaxy.

In all three panels, we color code galaxies by the type of distance estimate in the EDD. Gold and blue show TRGB and other ``quality'' distance indicators. For these targets, LEDA, CosmicFlows, and EDD all agree well, with only modest ($\approx 0.03$~dex) scatter among sources. This largely reflects that the three compilations all heavily weight the same primary source. The other distance estimates, in red, mostly come from some version of Tully-Fisher, group assignment, or a local velocity field model. For these, the different sources show more scatter, $\approx 0.1$~dex.

LEDA also provides a homogenized uncertainty in the redshift-independent distances. LEDA's estimate for the typical uncertainty in a TRGB distance (gold) is $< 0.02$~dex, while ``quality'' distances (blue) have mean uncertainty $\sim 0.06$~dex, and low quality ones $\sim 0.125$~dex. These are similar but slightly larger than the scatter that we find. Again, we expect that this reflects the use of the same primary sources by the different databases.

Finally, the curved line of points in the bottom right panel show the uncertainty in distance from only using LEDA's version of the Hubble flow. As expected, the fractional uncertainty drops with increasing recessional velocity because peculiar motions represent a smaller and smaller fraction of the overall velocity. The uncertainty in a Hubble flow distance becomes $\sim 0.1$~dex, comparable to the uncertainty in a low quality redshift-independent distance, at $v_{\rm vir} \sim 3,500$~km~s$^{-1}$. This is also about the outer extent of the EDD.

Based on these comparisons, we adopt the following approach to distances:

\begin{enumerate}
\item {\bf Has a quality distance:} We adopt TRGB and ``quality'' distances from the Extragalactic Distance Database (EDD) whenever available. Based on the comparison above, we impose an uncertainty of $\pm 0.03$~dex or $\sim 7\%$ to these cases with TRGB and $0.06$, or $\sim 15\%$ to other ``quality'' indicators. In a few cases, we are aware of TRGB distances that have been published since the EDD was updated. We input these by hand when we are aware of them.

\item {\bf Far enough, so that the Hubble flow yields a good estimate:} When a quality EDD distance is not available for galaxies with Virgocentric-flow-corrected recessional velocities $> 3,500$~km~s$^{-1}$, we prefer the ``best'' distances from the Lyon Extragalactic Database (LEDA). At this distance, these are primarily based on the Hubble flow. For these cases, we adopt the LEDA-suggested uncertainty. This is $\approx 0.1$~dex at $v_{\rm vir} \sim 3,500$~km~s$^{-1}$ and becomes smaller with increasing distance. Because we focus on nearby galaxies, galaxies with recessional velocity $> 3,500$~km~s$^{-1}$ represent only a small part of our working sample. We mostly adopt redshift-independent distances.

\item {\bf Nearby, but no quality distance:} For galaxies with recessional velocity $< 3,500$~km~s$^{-1}$ but where the EDD distance is neither TRGB nor a ``quality'' distance measurement, we adopt the EDD best value when available. If this is not available, we take the Cosmic Flows group value. We adopt the LEDA best distance estimate if neither are available. These distances include a mixture of Tully-Fisher, group, and ``numerical action model'' estimates \citep[see][]{TULLY09}. We assign these distances a minimum uncertainty of $\approx 0.125$~dex, $\sim 33\%$. For these cases, particularly those that are members of important subsamples, we will add ``quality'' distances by hand when we become aware of them.
\end{enumerate}

\subsection{\zomgs\ GALEX and WISE Sample}

\begin{figure}[t!]
\centering
\includegraphics[width=0.45\textwidth]{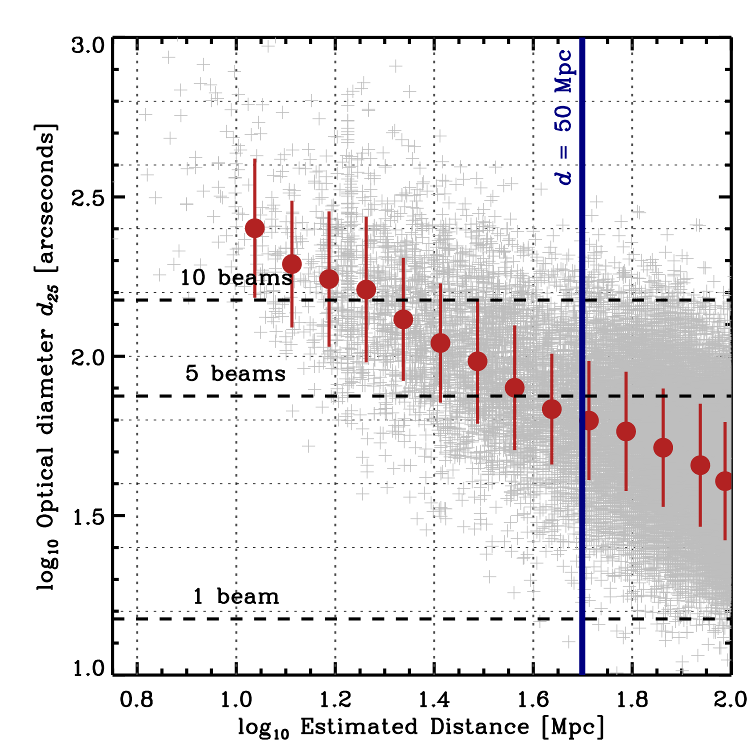}
\caption{{\bf Motivation for Sample Selection --- Resolution.} Optical angular size, expressed as $d_{25}$, of all galaxies with $M_B < -18$~mag as a function of estimated distance. Horizontal dashed lines indicate angular sizes corresponding to $1$, $5$, and $10$ beams at our FWHM $15\arcsec$ resolution. Points with associated error bars show the median optical size and rms scatter in bins of fixed distance. Our adopted distance cut $d < 50$~Mpc corresponds to placing $\gtrsim 5$ beams across each galaxy, on average. At the sharper $7.5\arcsec$ resolution of the $\lambda \leq 12\mu$m data (i.e., WISE3 and shorter) we place $\gtrsim 10$ beams across each target.}
\label{fig:resolution_cut}
\end{figure}

\begin{figure}[t!]
\centering
\includegraphics[width=0.45\textwidth]{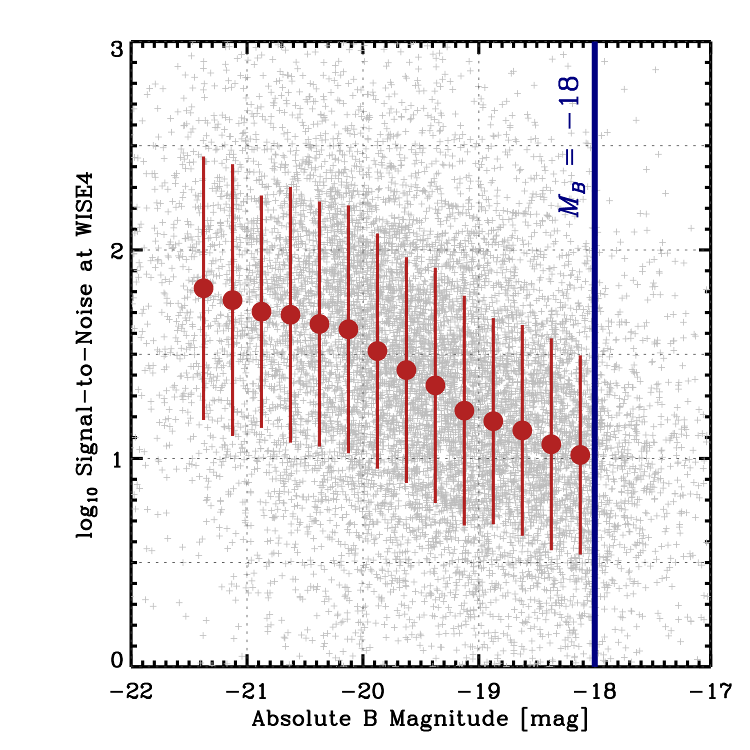}
\caption{{\bf Motivation for Sample Selection Criteria --- Sensitivity.} Signal-to-noise in integrated WISE4 flux a function of absolute $B$ magnitude. We plot galaxies in our final sample that have $d > 25$~Mpc and adopt the signal-to-noise from the photometry presented in this paper. Red points and lines show the median and rms scatter in the signal-to-noise in bins of fixed $M_B$. The typical signal-to-noise drops with decreasing $B$-band luminosity. Near our cutoff of $M_B \sim -18$~mag the typical integrated signal-to-noise for WISE4, our least sensitive band, drops to $\sim 10$.}
\label{fig:sensitivity_cut}
\end{figure}

\begin{figure*}[t!]
\centering
\includegraphics[width=0.45\textwidth]{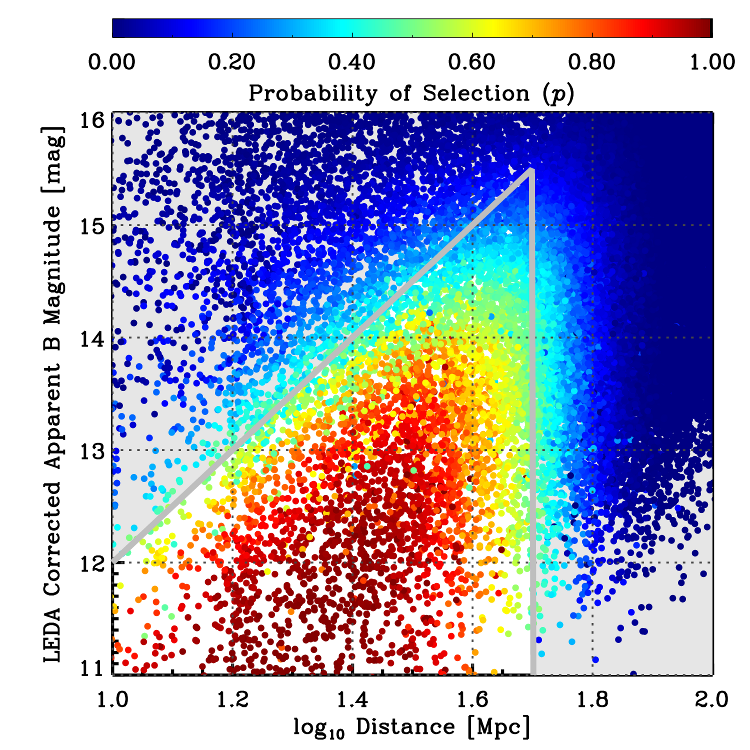}
\includegraphics[width=0.45\textwidth]{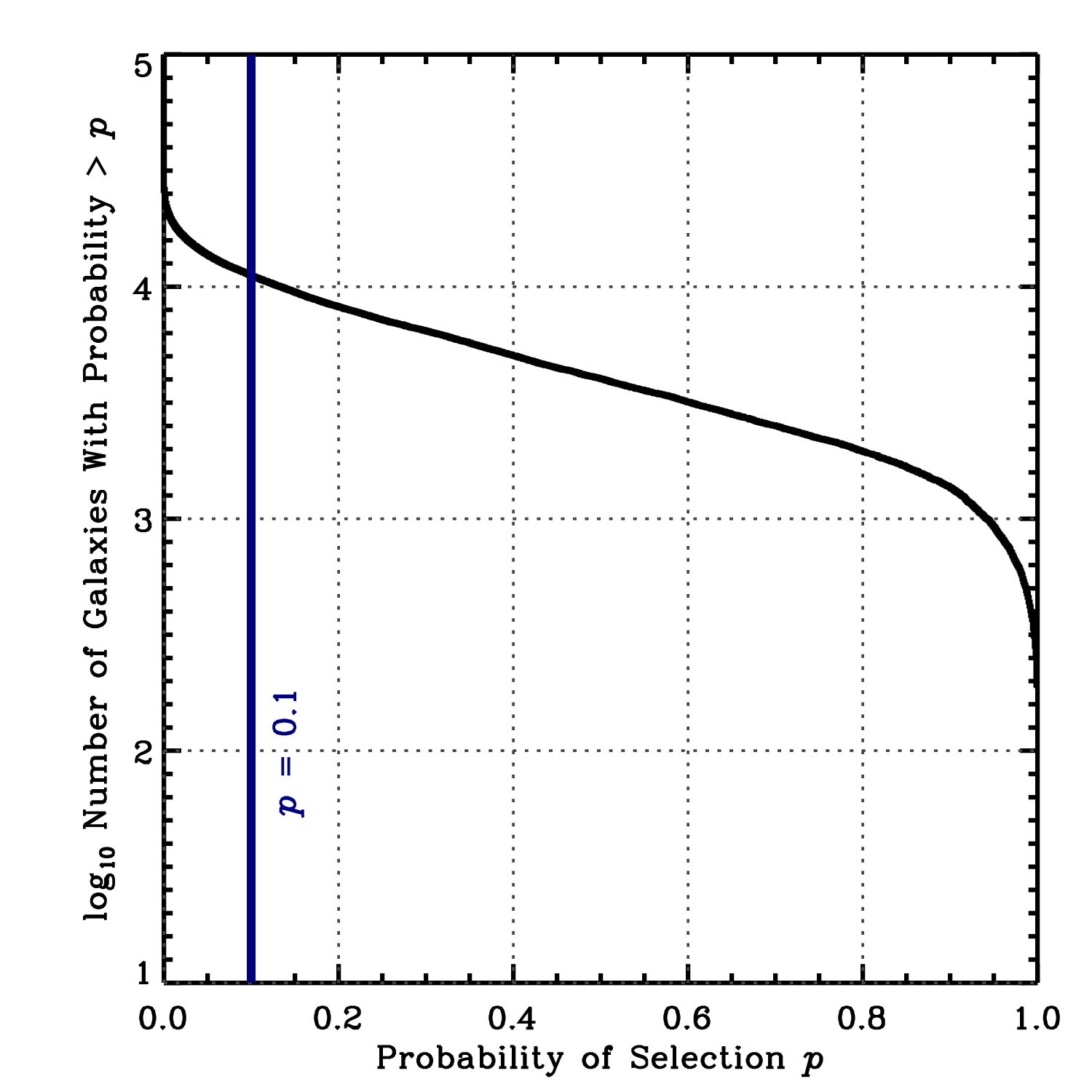}
\caption{{\bf Selection of Targets.} ({\em left}) A superset of possible targets from LEDA in the $B$ magnitude-distance plane. A gray wedge shows our selection criteria. We color code points by the probability that they lie within the gray wedge given uncertainties in their distance and apparent magnitude. The {\em right} panel shows the number of galaxies that have probability $>p$ of meeting our selection as a function of $p$. We build maps for galaxies with $p > 0.1$.}
\label{fig:selection}
\end{figure*}

We aim to build resolved UV, near-IR, and mid-IR maps for all reasonably luminous, resolved local star-forming galaxies. We define our sample based on proximity and luminosity, following the logic illustrated in Figures \ref{fig:resolution_cut} and \ref{fig:sensitivity_cut}.

{\bf Proximity ($d < 50$~Mpc):} The point spread function (PSF) of WISE at $22\mu$m limits our atlas to a common Gaussian PSF with FWHM $\sim 15\arcsec$ (see Section~\ref{sec:conv} for details on the convolution to common PSF). Even if we consider only bands with $\lambda < 12\mu$m, our atlas has a common resolution of $\sim 7.5\arcsec$. This is also about the resolution of a typical VLA {\sc Hi} map \citep[e.g.,][]{WALTER08} and near the diffraction limit of {\em Herschel} near the peak of the IR SED. 

At $\sim 50$ Mpc, $15\arcsec$ translates to $\sim 3.6$~kpc and $7.5\arcsec \sim 1.8$~kpc. This, $\sim 1{-}3$ kpc, is also the scale length for the stellar disk in a massive star-forming galaxy. Most star formation occurs within the inner few stellar scale lengths \citep[within $\sim 0.4~d_{25}$, e.g.,][]{YOUNG95,YOUNG96,SCHRUBA11}. We consider matching a beam to a scale length a reasonable lower limit to ``resolving'' a galaxy and adopt $50$~Mpc as our distance cut.

Figure \ref{fig:resolution_cut} illustrates this choice. We plot the $d_{25}$, the diameter at B band surface brightness of $25$~mag~\arcsec$^2$, as a function of distance for LEDA galaxies with $M_B < -18$~mag. Horizontal lines illustrate angular sizes corresponding to $1$, $5$, or $10$ of our $15\arcsec$ (FWHM) resolution elements. We can place $\sim 5$ resolution elements across a typical galaxy out to $\sim 50$~Mpc.

{\bf Luminosity ($M_B < -18$~mag):} We select galaxies with $M_B < -18$~mag, motivated primarily by sensitivity and completeness. While GALEX and WISE cover a large fraction of the sky, they do so at limited sensitivity. In practice, this prevents us from pushing far into the dwarf galaxy regime. 

This limitation is strongest for WISE. Dwarf galaxies tend to have less dust and fainter infrared emission than more massive galaxies \citep[see][]{MELISSE94,LISENFELD98,DALE09,REMYRUYER14}. As a result, the 12$\mu$m and 22$\mu$m emission from low mass galaxies can be faint relative to the depth of WISE. Figure \ref{fig:sensitivity_cut} shows the signal to noise at WISE4 of galaxies in our final sample (from photometry carried out in this paper) as a function of $M_B$. We plot only galaxies outside $25$~Mpc, i.e., the more distant targets in our sample. The plot shows that given our resolution cut, we can detect galaxies with $M_B < -18$~mag at signal to noise $\gtrsim 10$ on average. Extending to fainter magnitudes would lead to frequent nondetections and lower quality maps.

$M_B = -18$~mag is about the magnitude of the Large Magellanic Cloud (LMC), and one magnitude fainter than M33. Thus, our selection can be thought of as extending down to dwarf spirals, with stellar masses $\gtrsim 10^9$~M$_\odot$ and metallicities $\sim 1/3{-}1/2~Z_\odot$. This is a much higher mass cut than, for example, LVL \citep{DALE09}, which employed much more sensitive {\em Spitzer} mid-IR observations.

{\bf Selection:} We select all galaxies that we estimate to have $> 10\%$ chance of meeting our $d < 50$~Mpc and $M_B < -18$~mag selection criteria. To estimate the probability of a galaxy meeting our criteria, we generate 1,000 random combinations of its distance and apparent magnitude, distributed according to their uncertainties. Based on LEDA, we take $0.5$~mag as a typical uncertainty in the apparent $B$ magnitude. We describe the uncertainties in the distance above. We combine the apparent $B$ magnitude and distant to calculate $M_B$. Then we check whether the galaxy would be selected for that realization. Based on 1,000 realizations, we derive the probability, $p$, that a galaxy meets our selection criteria given the uncertainties on its distance and luminosity.

We visualize this process in Figure \ref{fig:selection}. The left panel shows our selection criteria in a plane of $M_B$ and distance. We color points by the probability, $p$, that they meet our selection criteria over 1,000 draws. The right panel shows the number of galaxies that exceed each given value of $p$. We define our sample by a cut of $p > 0.1$, which yields $\sim 11,000$ galaxies. These are all galaxies that we deem to have a $>10\%$ chance of lying within 50~Mpc and having $M_B < -18$~mag. We supplement these by $\sim 5,000$ extra maps (see below) that do not meet this criteria (but did match earlier sample definitions) to create our final sample of $\sim 15,750$.

We manually excluded the LMC from our atlas because its large angular size renders it as a special case for image construction.

\begin{figure*}[t!]
\centering
\includegraphics[width=0.45\textwidth]{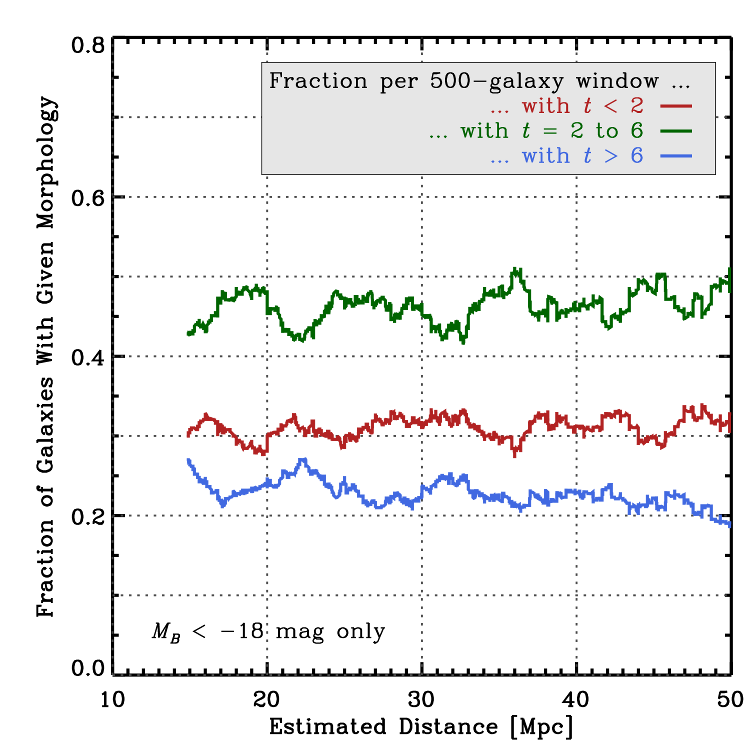}
\includegraphics[width=0.45\textwidth]{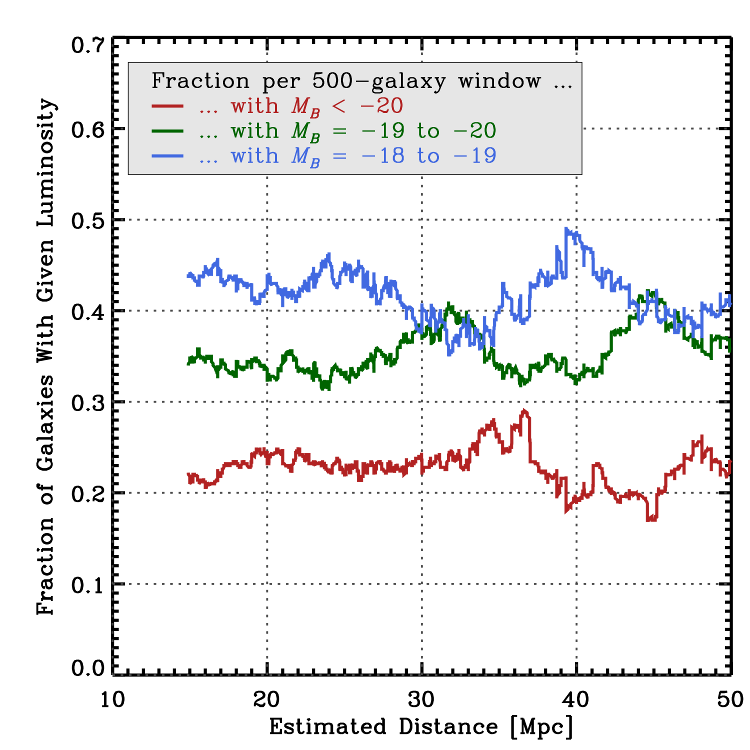}
\caption{{\bf Completeness.} Distributions of ({\em left}) morphology, parameterized by $t$ type, and ({\em right}) $M_B$ as function of distance. We consider galaxies with $M_B < -18$~mag, $d < 50$~Mpc. We sort these targets by distance. Then, in a running 500 galaxy window we plot ({\em left}) the fraction of galaxies with $t < 2$ (early types), $2 < t < 6$ (intermediate type spirals), and $t > 6$ (late type spirals and irregulars). The right panel shows a similar calculation but for bins of absolute $B$ magnitude. Both plots show mild incompleteness at high $d$, with late type, low luminosity galaxies slightly less common near $d \sim 50$~Mpc than at $d \sim 20$~Mpc.}
\label{fig:completeness}
\end{figure*}

{\bf Completeness and Supplemental Targets: } Given our conservative $M_B$ and $d$ cuts, we expect our atlas to be reasonably complete. Figure \ref{fig:completeness} shows fraction of galaxies in bins of $B$-band luminosity and morphology as a function of distance. We see some sign of decreasing completeness in both plots. Late-type galaxies and low luminosity galaxies decrease in prevalence with increasing distance. But the effect appears relatively mild. The fraction of late type galaxies drops from $\sim 25\%$ in the nearest bins to $\sim 20\%$ near $d \sim 50$~Mpc.

Note that Figure \ref{fig:completeness} considers only galaxies with best estimate $M_B < -18$~mag and $d < 50$~Mpc. We do not plot galaxies with fainter $M_B$ or larger $d$. That is, our probabilistic selection includes galaxies for which the best estimate is $d > 50$~Mpc or $M_B > -18$~mag, but the uncertainties allow a 10\% chance that the galaxy might meet our selection criteria. If we included these galaxies in Figure \ref{fig:completeness}, the plots would show many more late type, low luminosity galaxies at low $d$ and many more high luminosity galaxies at high $d$ (see the points outside the gray lines in Figure \ref{fig:selection}). In total, there are $\sim 11,000$ galaxies that meet our selection criteria of $p > 0.1$ for $M_B < -18$~mag and $d < 50$~Mpc.

We iterated this process of defining the sample and estimating distances. As a result, we constructed maps for $\sim 5,000$ additional galaxies that do not meet our selection criteria. These maps have scientific utility, and we include them in the atlas. Practically, these just represent ``extra'' local galaxies that are either too distant or not luminous enough to be selected given our present estimates of their properties.

\textit{These extra galaxies, completeness concerns, probabilistic selection, and use of $B$ band for selection mean that our atlas should not be taken as a complete sample. It does, however, \textbf{contain} reasonably complete subsamples. For projects focused on completeness, we recommend defining subsamples within the atlas.} We adopt the $B$-band cut because $B$ magnitudes remain the most available for nearby galaxies across the whole sky \citep[though thanks to 2MASS, $J$, $H$, and $K_S$ are almost as common][]{SKRUTSKIE06}.

\begin{figure*}[t!]
\centering
\includegraphics[width=0.45\textwidth]{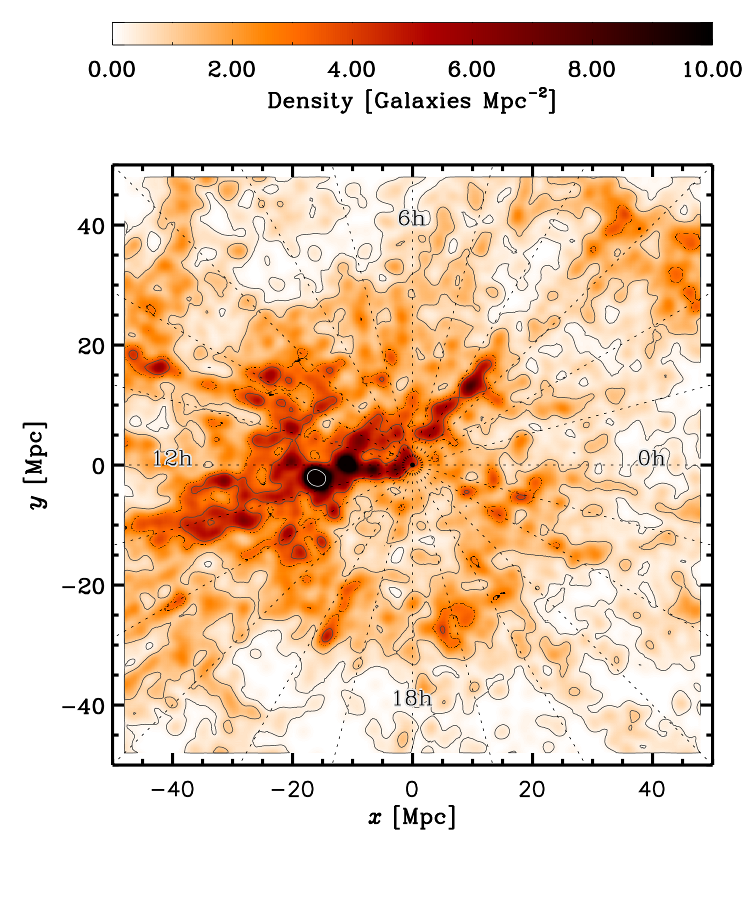}
\includegraphics[width=0.45\textwidth]{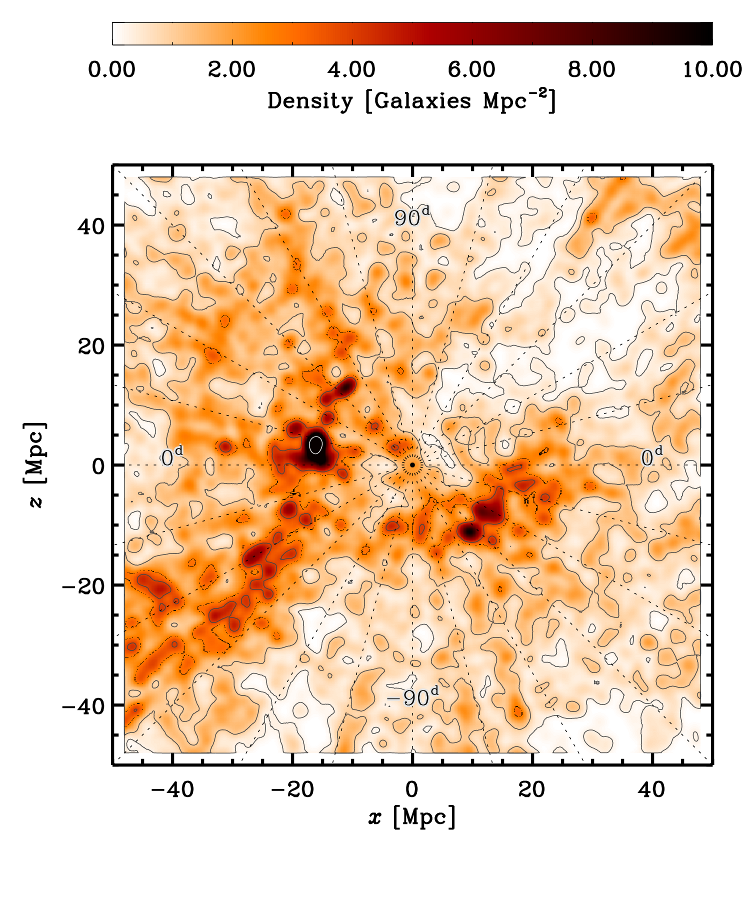}
\caption{{\bf Locations of Selected Galaxies.} Density of selected targets, in units of galaxies per Mpc$^2$ after convolution with a FWHM 2~Mpc gridding kernel. We show projections into the equatorial (declination 0) ({\em left}) plane and onto a slice along the $0-12^{\rm h}$ plane ({\em right)}.}
\label{fig:location}
\end{figure*}

{\bf Selected Targets:} Figure \ref{fig:location} visualizes the distribution of our targets in two projections. In both plots the Milky Way lies at $(0,0)$. The left panel shows all targets projected onto the equatorial plane. The right panel shows all targets projected onto a plane aligned with the declination ($z$) axis and the $0^{\rm h}$-$12^{\rm h}$ axis ($x$). As expected, the Virgo Cluster and associated structures appear prominent in both panels. ``Finger of god'' effects also manifest in both images, highlighting the significant distance uncertainties discussed above.

\section{Some Reference Conversions to Physical Quantities}

Our atlas has two main goals. First, we aim to create resolved maps of the recent star formation rate surface density, $\Sigma_{\rm SFR}$, and stellar mass surface density, $\Sigma_\star$. Second, we attempt to place the nearby galaxy population in the context of the large surveys, e.g., the Sloan Digital Sky Survey (SDSS). 

In this section we present simple conversions that translate from intensities in the WISE and GALEX bands to $\Sigma_{\rm SFR}$ and $\Sigma_{\star}$. We also quote some literature conversions between luminosity and integrated quantities, SFR and M$_\star$. Note that we provide improved versions of these integrated conversions later in the paper. We use the GSWLC \citep{SALIM16,SALIM18} to derive empirical relationships that relate SFR and $M_{\star}$ to integrated WISE and GALEX photometry. These allow us to compare \zomgs\ to the larger sample from the GSWLC. Our suggested approaches to the WISE1 mass to light ratio, \mtolwise , and SFR estimation are summarized in Tables \ref{tab:mtolfits} and \ref{tab:sfr_gswlc}, described in the Appendix, and applied in Section \ref{sec:sfms}.  

The new calibrations derived in this paper consider integrated photometry, based on integrated population synthesis modeling by \citet{SALIM16,SALIM18}. We defer development of a unified stellar mass and star formation rate estimation scheme that works self-consistently on integrated and resolved measurements to future work. We note, in general, that we expect large parts of galaxies to more or less act as whole galaxies. However global prescriptions can fail dramatically in small regions of highly resolved, i.e., very nearby, targets \citep[e.g.,][]{BOQUIEN16,TOMICIC19}. We also emphasize that these prescriptions represent \textit{mean} calibrations. Individual galaxies scatter significantly about them; see the appendix for quantification of this scatter.

We give intensity to surface density conversions appropriate for a face-on geometry. When applied to a disk with inclination $i$, the $\Sigma_{\rm SFR}$ and $\Sigma_\star$ estimated by these formulae should be adjusted by a factor of $\cos i$.

\subsection{WISE3 and WISE4 to Star Formation Rate}

\citet{KENNICUTT12} and \citet{JARRETT13} provide conversions between $22\mu$m or $24\mu$m luminosity, $\nu L_\nu$, and SFR. They adopt the stellar initial mass function (IMF) of \citet{KROUPA03} with bounds of $0.1$ and $100$~M$_\odot$, slope $-2.35$ between 1 and 100~$M_\odot$, and slope $-1.3$ between $0.1$ and $1$~M$_\odot$. The two works give practically identical conversions, with

\begin{equation}
\label{eq:ke12mips24}
{\rm SFR}~\left[ {\rm M_\odot~yr}^{-1} \right] \approx \frac{\nu L_\nu [{\rm erg~s}^{-1}]}{10^{42.7}}~.
\end{equation}

Following \citet{KENNICUTT12}, in the Appendix we refer to the conversion from $\nu L_\nu$ in erg~s$^{-1}$ to SFR in M$_\odot$~yr$^{-1}$ as $C$. Thus, $C$ has units of M$_\odot$~yr$^{-1}$ (erg~s$^{-1}$)$^{-1}$. For brevity, we take these units to be implicit through this section. For Equation \ref{eq:ke12mips24}, $\log_{10} C$ for WISE4 is the \citet{KENNICUTT12} value of $-42.7$. Using WISE4 only, i.e., not combining with any UV band, we suggest $\log_{10} C = -42.55$ for best consistency with the SDSS-based GSWLC estimates (see the Appendix for more detail).

For an intensity map in units of MJy~sr$^{-1}$, the \citet{KENNICUTT12} and \citet{JARRETT13} $\log_{10} C = -42.7$ implies


\begin{eqnarray}
\frac{\Sigma_{\rm SFR}}{1~\sigsfrunits } \approx \\
\nonumber 3.24 \times 10^{-3} \left( \frac{C}{10^{-42.7}} \right) \left(\frac{I_{22\mu m}}{1~{\rm MJy~sr^{-1}}} \right)
\end{eqnarray}

\noindent for WISE Band 4, taking $\nu$ to correspond to $\lambda = 22\mu$m. If we instead take $\nu$ to correspond to $\lambda = 24\mu$m, appropriate for MIPS on {\em Spitzer}, then


\begin{eqnarray}
\frac{\Sigma_{\rm SFR}}{1~\sigsfrunits } \approx \\
\nonumber 2.97 \times 10^{-3} \left( \frac{C}{10^{-42.7}} \right) \left(\frac{I_{24\mu m}}{1~{\rm MJy~sr^{-1}}} \right)~.
\end{eqnarray}

\noindent For reference, for $\log_{10} C=-42.7$ an intensity of $1$~MJy~sr$^{-1}$ at $\lambda \sim 22\mu$m corresponds to roughly the surface density of star formation in the Solar Neighborhood, $\sim 3 \times 10^{-3}$~M$_\odot$~yr$^{-1}$~kpc$^{-2}$. This value at $22\mu$m also typically corresponds to a line-integrated CO intensity of order $\sim 1$~K~km~s$^{-1}$.

\citet{JARRETT13} also provide a fiducial conversion from WISE Band 3 luminosity to SFR.

\begin{equation}
{\rm SFR}~\left[ {\rm M_\odot~yr}^{-1} \right] \approx \frac{\nu L_\nu [{\rm erg~s}^{-1}]}{10^{42.9}}
\end{equation}

\noindent This derives from the {\em Spitzer} 24$\mu$m results noted above, and so adopt the same IMF. In the Appendix we suggest $\log_{10} C = -42.67$ when using WISE3 alone. But we also caution that using WISE3 alone introduces significant systematic biases into the SFR estimate (see the Appendix for details).

The \citet{JARRETT13} value of $\log_{10} C=-42.9$ for WISE3 translates to


\begin{eqnarray}
\frac{\Sigma_{\rm SFR}}{1~\sigsfrunits } \approx \\
\nonumber 3.77 \times 10^{-3} \left( \frac{C}{10^{-42.9}} \right) \left(\frac{I_{12\mu m}}{1~{\rm MJy~sr^{-1}}} \right)
\end{eqnarray}

\noindent adopting $\lambda = 12\mu$m and converting this to $\nu$. Again, this can be adjusted to reflect a different conversion, $C$, from WISE3 luminosity to SFR.

\subsection{WISE1 to Stellar Mass}

The near-IR intensity, traced by IRAC's $3.6\mu$m band and WISE1 at $3.4\mu$m, is often taken to trace the total stellar mass of a population. As discussed by, e.g., \citet{MCGAUGH14}, \citet{MEIDT14}, \citet{QUEREJETA15}, and \citet{SIMONIAN17}, the appropriate mass to light ratio, \mtolwise , remains a topic of research. The variations in \mtolwise\ are generally believed to be modest, though not necessarily negligible \citep[see discussion below and in][]{SALIM16}.  In the Appendix, we use the GSWLC to calibrate prescriptions for integrated \mtolwise\ based on GALEX and WISE observables.  These allow us to place the \zomgs\ sample on the same SFR-M$_\star$ relationship with the larger galaxy sample probed by GSWLC. 

Here, we note some fiducial conversions between stellar mass surface density and WISE1 brightness. \citet{SCHOMBERT14} and \citet{MCGAUGH14} argue for $\Upsilon_\star^{3.6} = 0.47 \sim 0.5$~\mtolunits . They work in the IRAC ZMAG system, for which 0 magnitudes corresponds to 280.9~Jy and they adopt an absolute magnitude of the Sun of $3.24$~mag at $3.6\mu$m. The \citet{SCHOMBERT14} and \citet{MCGAUGH14} Solar magnitude translates to a flux density of

\begin{equation}
L_{\nu,\odot}^{3.6} \approx 1.7 \times 10^{18}~{\rm erg~s^{-1}~Hz^{-1}}
\end{equation}

\noindent so that

\begin{equation}
\nu L_{\nu,\odot}^{3.6} \approx 1.4 \times 10^{32}~{\rm erg~s}^{-1} \approx 0.0369~L_\odot~.
\end{equation}

\noindent In this case, one can convert from WISE1 or IRAC1 surface brightness to stellar mass surface density via

\begin{equation}
\label{eq:sigmastar_irac}
\frac{\Sigma_{\star}}{1~{\rm M_\odot~pc}^2 } \approx 3.5 \times 10^{2} \left( \frac{\Upsilon_\star^{3.6}}{0.5} \right) \left(\frac{I_{3.6\mu m}}{1~{\rm MJy~sr^{-1}}} \right)~.
\end{equation}

\noindent For comparison, \citet{LEROY08} adopted $\Sigma_\star = 280 I_{3.6}$, equivalent to $\Upsilon_\star^{3.6} \approx 0.4$~M$_\odot$/$L_\odot$.

IRAC1 has mean wavelength $\lambda \sim 3.6\mu$m, while WISE1 has $\lambda \sim 3.4\mu$m. The Sun has about the same Vega magnitude in both bands, but the difference matters in energy units. For WISE1, the luminosity of the Sun is
\begin{eqnarray}
\label{eq:w1_lsun}
L_{\nu,\odot}^{3.4} &\approx& 1.8 \times 10^{18}~{\rm erg~s^{-1}~Hz^{-1}} \\
\nonumber \nu L_{\nu,\odot}^{3.4} &\approx& 1.6 \times 10^{32}~{\rm erg~s}^{-1} \approx 0.042~L_\odot~.
\end{eqnarray}

so that 
\begin{equation}
\label{eq:sigmastar_wise}
\frac{\Sigma_{\star}}{1~{\rm M_\odot~pc}^2 } \approx 3.3 \times 10^{2} \left( \frac{\Upsilon_\star^{3.6}}{0.5} \right) \left(\frac{I_{3.4\mu m}}{1~{\rm MJy~sr^{-1}}} \right)~.
\end{equation}

\noindent We assume that $\Upsilon_\star^{3.6} = \Upsilon_\star^{3.4}$ and refer only to \mtolwise . Given the small change in wavelength, this seems reasonable. The main difference between IRAC1 and WISE1 will be moving slightly farther down the Rayleigh-Jeans part of the stellar spectrum. We expect that both filters include a contribution from the 3.3$\mu$m PAH band \citep[e.g., see][]{QUEREJETA15}.

There is good reason to expect \mtolwise\ to change systematically across the local galaxy population. In population synthesis models, the near-IR mass to light ratio increases with the age of a stellar population \citep[e.g.,][]{BELL01,BELL03,COURTEAU14,STANWAY18} and low mass star-forming galaxies show systematically younger stellar populations than high mass star-forming galaxies \citep[e.g., see][]{KANNAPPAN13}.

In the appendix, we quantify this effect and offer prescriptions to predict \mtolwise\ from GALEX and WISE data. We compare WISE photometry to the multi-band SED fitting by \citet{SALIM16} and \citet{SALIM18}. We find that \mtolwise\ changes systematically from $\sim 0.2$ to $\sim 0.5$~\mtolunits , with the sense that \mtolwise\ becomes low when SFR/M$_\star$ becomes high. Thus, star-forming galaxies show lower \mtolwise\ while quiescent galaxies show high \mtolwise . The sense of our results agrees with that found in earlier near-IR studies by, e.g., \citet{MEIDT14} and \citet{QUEREJETA15}. But those studies only had access to IRAC2-to-IRAC1 colors (similar to WISE2-to-WISE1) and so had a relatively weak handle on SFR/M$_\star$. As emphasized above, an SFR/M$_\star$-dependent \mtolwise\ also agrees with longstanding results from other population synthesis work.

\subsection{FUV and NUV to Star Formation Rate}

In a part of a galaxy with abundant recent star formation, the continuum at $\sim 154$~nm (FUV) and $231$~nm (NUV) will be dominated by the light from relatively young ($\lesssim 100$~Myr) stars. After correcting for the (often substantial) effects of extinction, GALEX FUV emission has been widely used as a tracer of the recent SFR. The exact prescription varies, with sensitivity to the assumed stellar populations and a star formation history. \citet{KENNICUTT12} recommend

\begin{equation}
{\rm SFR}~\left[ {\rm M_\odot~yr}^{-1} \right] \approx \frac{\nu L^{\rm FUV}_\nu [{\rm erg~s}^{-1}]}{10^{43.35}}~,
\end{equation}

\noindent to estimate SFR averaged over $\sim 100$~Myr timescales.  This reflects the same \citet{KROUPA03} IMF described above. Following the notation above and in \citet{KENNICUTT12}, this is $\log_{10} C = -43.35$ for FUV. \citet{SALIM07} argue for a slightly different conversion, with $\log_{10} C \approx -43.45$ instead. In the Appendix, we use the GSWLC to check these values and suggest $\log_{10} C = -43.42$ for FUV, close to both values and virtually identical to \citet{SALIM07}.

For the NUV band, \citet{KENNICUTT12} recommend

\begin{equation}
{\rm SFR}~\left[ {\rm M_\odot~yr}^{-1} \right] \approx \frac{\nu L^{\rm NUV}_\nu [{\rm erg~s}^{-1}]}{10^{43.17}}~,
\end{equation}

\noindent or $\log_{10} C = -43.17$. \citet{SALIM07} suggest $\log_{10} C = -43.28$ and we recommend $\log_{10} C = -43.24$ in the Appendix. We highlight the \citet{SALIM07} values because the xGASS and xCOLDGASS surveys adopt this system \citep{JANOWIECKI17,SAINTONGE17} and these surveys represent important points of comparison for our sample. The recent star formation history matters to $C$ and will vary within and among galaxies. More sophisticated treatments adopt a population synthesis approach and combine ultraviolet with optical and even infrared data \citep[e.g.,][]{NOLL09,SALIM16}. Still, these single-value conversions give us a useful reference point.

Adopting the above equations, and taking $\lambda = 231$~nm for NUV and $154$~nm for FUV, the \citet{KENNICUTT12} calibrations translate to


\begin{eqnarray}
\frac{\Sigma_{\rm SFR}}{1~{\rm M_\odot~yr}^{-1}~{\rm kpc}^{-2} } \approx \\
\nonumber 1.04 \times 10^{-1} \left( \frac{C}{10^{-43.35}} \right) \left(\frac{I_{FUV}}{1~{\rm MJy~sr^{-1}}} \right)
\end{eqnarray}

\noindent and, because the UV SED of an extinguished young population is quite flat the conversion from $I_{\rm NUV}$ to $\Sigma_{\rm SFR}$ is almost identical to that for the FUV,


\begin{eqnarray}
\frac{\Sigma_{\rm SFR}}{1~{\rm M_\odot~yr}^{-1}~{\rm kpc}^{-2} } \approx \\ 
\nonumber 1.05 \times 10^{-1} \left( \frac{C}{10^{-43.17}} \right) \left(\frac{I_{NUV}}{1~{\rm MJy~sr^{-1}}} \right)~.
\end{eqnarray}

The appendix also discusses the use of linear combinations of WISE and GALEX data in ``hybrid'' tracers. We presents recommendations and recipes that place integrated measurements on a system in good agreement with \citet{SALIM16} and \citet{SALIM18}.

\section{WISE Maps}
\label{sec:wise}

\subsection{unWISE Cutouts}

We used the unWISE reprocessing \citep{LANG14} of the WISE all sky survey \citep{WRIGHT10}. As described by \citet{LANG14}, this reprocessing yields images with the original WISE PSF, sharper than those used to carry out the matched filter photometry that was the core mission of WISE.

In order to avoid ``bowling'' (i.e., oversubtraction of the local background due to contamination by real galactic emission) in the bands crucial to star formation rate estimation, we rebuilt the unWISE cutouts at 12$\mu$m and 22$\mu$m without the spatial median filter described by \citet{LANG14}. Instead, we applied our own background subtraction after image construction. In our atlas, bowling remains visible around M31, but only weakly present around M33 and our other targets are much smaller than M31.

We built our unWISE images with the same $2.75''$ pixel scale as \citet{LANG14}. By default each initial image is $1800''$ across, with the size reduced during later processing. For larger galaxies, we built an image with linear dimensions six times the LEDA-recorded optical radius, $r_{25}$. We used the ``masked'' data products described in \citet{LANG14}, in which outlying measurements are rejected during the image construction.

{\em Conversion to Intensity Units:} For each cutout we converted the original units to MJy~sr$^{-1}$ following the WISE documentation. We converted from Vega to AB magnitudes using offsets of $\Delta m = 2.683$, $3.319$, $5.242$, and $6.604$~mag for WISE bands 1, 2, 3, and 4. Then, we converted from AB magnitudes to Jy taking a flux density of $3,631$~Jy at an apparent AB magnitude of $0$~mag and using the unWISE zero point of $22.5$~mag. The conversion is

\begin{eqnarray}
I_\nu \left[ {\rm MJy~sr}^{-1} \right] = x I_\nu \left[ {\rm mag~pixel}^{-2} \right]~{\rm where}~ \\
\nonumber \log_{10} x = -\frac{\Delta m + 22.5}{2.5} + 7.31 + 2 \log_{10} \left( \frac{2.75\arcsec}{\delta} \right)
\end{eqnarray}

\noindent where $\delta$ represents the pixel scale in arcseconds.

For reference, a surface brightness of 25 AB~magnitudes~arcsecond$^{-2}$ corresponds to a surface brightness of 0.0154 MJy~sr$^{-1}$.

\subsection{Convolution}\label{sec:conv}

After image construction and unit conversion, we convolved the image with a custom kernel from the native resolution of WISE to a Gaussian PSF. We created images with full width at half maximum (FWHM) of $15\arcsec$ for all bands and FWHM $7.5\arcsec$ for WISE1 through WISE3.

We generated the kernels following \citet{ANIANO11}. They assume a single axisymmetric PSF for all WISE images. The $15\arcsec$ PSF represents the smallest common PSF that can accommodate both GALEX bands and all four WISE bands. The 22$\mu$m WISE4 band represents the limiting factor and we use a somewhat ``aggressive'' kernel \citep[see][]{ANIANO11} to reach this resolution for WISE4.  ``Aggressive'' kernels have significant negative values which can introduce artifacts in the case of bright point sources that approach the non-linearity thresholds of the instrument and can amplify noise in the images. The $7.5\arcsec$ resolution WISE1 and WISE2 images suffer from less confusion than the $15\arcsec$ images because foreground stars are better resolved. However, the sensitivity in the WISE3 band is notably worse at this resolution (see below). The convolution of WISE3 to $7.5\arcsec$ relies on another ``aggressive'' kernel, which is likely the source of the increased noise at this resolution. We suggest using the WISE3 images at $15\arcsec$ resolution unless the maximum angular resolution is essential to the scientific goals of the project.

\subsection{Region of Interest Extraction and Down-Sampling}

After convolution, we extracted a smaller region around the galaxy, usually $\pm 4~r_{25}$ and never less than $\pm 120''$. This smaller region was still always larger enough to encompass extended emission from the galaxy and give plenty of room for background subtraction. At this stage, we also rebinned the $15''$ resolution images to have pixel scale $5.5''$. Both steps reduce the data volume while still critically sampling the $7.5''$ and $15''$ beam and covering the galaxy.

\subsection{Mask Construction}

\begin{deluxetable}{lccc}
\tabletypesize{\scriptsize}
\tablecaption{Parameters for Star Mask Creation \label{tab:starmasking}}
\tablewidth{0pt}
\tablehead{
\colhead{Atlas Band} & 
\colhead{Magnitude} &
\colhead{$S$ at $7.5''$} &
\colhead{$S$ at $15''$} \\
}
\startdata
WISE1 & 2MASS $K_S$ & $5.22$ & $4.72$ \\
WISE1 & GAIA $G$ & $5.92$ & $5.32$ \\
WISE2 & 2MASS $K_S$ & $4.96$ & $4.44$ \\
WISE2 & GAIA $G$ & $5.65$ & $5.06$ \\
WISE3 & 2MASS $K_S$ & $4.23$ & $3.68$ \\
WISE3 & GAIA $G$ & $4.89$ & $4.28$ \\
WISE4 & 2MASS $K_S$ & \nodata & $3.19$ \\
WISE4 & GAIA $G$ & \nodata & $4.28$ \\
NUV\tablenotemark{a} & 2MASS $K_S$ & $2.31$ & $1.84$ \\
NUV\tablenotemark{a} & GAIA $G$ & $4.12$ & $3.53$ \\
FUV\tablenotemark{a} & 2MASS $K_S$ & $0.13$ & $-0.22$ \\
FUV\tablenotemark{a} & GAIA $G$ & $1.43$ & $1.21$ \\
\enddata
\tablenotetext{a}{Intensity prediction more uncertain for these bands due to lack of stellar color information.}
\tablecomments{Coefficient, $S$, in Equation \ref{eq:stars} used to translate magnitudes of known foreground stars to estimated intensities in the image atlas. These predictions are then used to generate masks of bright foreground stars.}
\end{deluxetable}

We constructed a series of masks that we use for analysis. We also distribute these as part of the atlas. These record:

\begin{enumerate}
\item {\bf Galactocentric Radius.} We created an image of deprojected galactocentric radius. We use this galactocentric radius image to define and subtract a background and carry out photometry. For this purpose, if the inclination of a galaxy exceeded $60^\circ$ then we set it to $i=60^\circ$. If either the position angle or inclination is not known, we set both to zero.

\item {\bf Other Galaxies in the Field.} Based on our LEDA-derived database of galaxies out to $z \sim 0.3$, we identified all galaxies other than the target that overlap the image. We create a mask of the expected footprint of each galaxy. For this purpose, we take each background galaxy to have semimajor axis $1.25~r_{25}$. We adopt a minimum galactocentric radius of $7.5''$ when the radius is unknown or less than this value.

\item {\bf Stars From 2MASS and GAIA.} Both 2MASS \citet{SKRUTSKIE06} and GAIA \citep{GAIA,GAIADR2} measured the position and brightness of stars across the whole sky. We noted the location of all 2MASS stars with apparent $K_S$ magnitudes brighter than $10$~mag that overlap our images. We also queried the GAIA DR2 for positions and $G$-band magnitudes of stars that lie inside each image footprint. We use these to construct star masks for each target. 

Especially for large, nearby galaxies, the fainter GAIA stars could be either foreground stars or bright stars and clusters in the target galaxy. To isolate Milky Ways stars, we considered only GAIA sources with a S/N$>3.5$ detection of parallax or proper motion in either R.A. or declination.

We measured the peak intensity of stars in our images but away from our targets. We used these to build a model a simple model to estimate the intensity of a star in our atlas given its GAIA $G$ magnitude or 2MASS $K_S$ magnitude. Plots illustrating this exercise for 2MASS appear below, in Section \ref{sec:starstacks}. Table \ref{tab:starmasking} gives the coefficient, $S$, in

\begin{eqnarray}
\label{eq:stars}
\log_{10} \left( \frac{I_\nu}{{\rm MJy~sr}^{-1}} \right) = S - \left( \frac{M}{2.5~{\rm mag}} \right)
\end{eqnarray}

\noindent where $M$ is the magnitude from 2MASS or GAIA, $S$ is our fit coefficient, and $I_\nu$ represents the peak intensity of the star in the atlas. $S$ depends on resolution and band. 

Using a single $S$ assumes that all foreground stars have similar colors. This approximation works reasonably well to predict the WISE intensities of unsaturated stars. For GALEX, more accurate prediction would require an estimate of stellar temperature; hot stars show much higher NUV-to-$G$ ratios than cool stars. GAIA only provide temperatures information for a subset stars, and we defer a more sophisticated multi-band star masking to future work. Because we mask regions of the image, rather than subtracting stars from the image, our masking is relatively insensitive to some scatter or inaccuracy in $S$.

We use the positions and magnitudes of GAIA and 2MASS stars to create an image of predicted intensity due to foreground stars. These have the intensity predicted by Equation \ref{eq:stars} and PSF of the image. We exclude stars within $3''$ of the galaxy center at this stage, because these often turned out to be nuclear features in the galaxy itself. We also use a bigger convolution kernel for cases where we expect saturation (see below). The size of these saturation correction kernels was chosen based on by-eye tests to match the extent of very bright stars.

After creating a predicted foreground star image, we derive a mask by imposing an intensity threshold on that image. The threshold is usually $\sim 10$ times the typical rms noise in that band. We chose the threshold to select regions where stars are likely to interfere with our measurements of the galaxy. Regions in the predicted foreground star intensity map with values above the threshold become the star mask for that band and resolution.

Because $S$ varies by band, stars that appear bright in WISE1 are often not blanked or are blanked with a much smaller footprint in the WISE3, WISE4, NUV, and FUV bands.

\item {\bf By-Hand Masks.} In a few cases, we also constructed masks by hand. We expect the number of such masks to grow as we use the atlas more.

\end{enumerate}

We used these masks during background subtraction and photometry and include them with the atlas. The ideal combination of masks will depend on the application. We use masks to derive flags, described below, that identify potential issues in the atlas images.

\subsection{Background Subtraction}

After convolution and creation of masks, we fit and subtract a background from the WISE images. During this fit, we excluded all points within galactocentric radius $2~r_{25}$ or $60''$, whichever is greater. For large galaxies, we inspected these apertures to make sure that the galaxy does not contaminate the background. Based on this inspection, we manually adjusted the background definition for about $10\%$ of the targets. We also excluded the footprints of other galaxies and foreground stars from the fit. In other words, for purposes of fitting the background, we applied all available masks.

We fit the background in several steps. First, we subtracted the median intensity calculated from the unmasked parts of the image. We iterated this process, masking $>2\sigma$ outliers about the median, recalculating the median and the noise, and adjusting the background. 

After this, for WISE 3 and 4 we fit and subtracted a plane to the unmasked data, again iteratively rejecting outliers about the fit. For WISE1 and WISE2 the backgrounds in these bands appear dominated by stars rather than cirrus or instrumental features, and in some cases the stars could fill a large part of the field of view. We fit a single value to the background, instead of a plane, for these bands.

After these fits, the distribution in the unmasked regions was usually still not a Gaussian centered on zero. Instead faint stars and other features skew the distribution towards positive values at a low ($\sim 1\sigma$) level. To account for this, after the initial background subtraction we made a histogram of intensities in the background (unmasked) region. We adjusted the background to sit at the mode of this histogram. 

Based on by-eye inspection of the residual image and histogram, this process yields a reasonable background centered at zero. 

\subsection{Statistical Properties of the WISE Atlas}

\begin{deluxetable}{lccc}
\tabletypesize{\scriptsize}
\tablecaption{Noise in the Atlas \label{tab:wisenoise}}
\tablewidth{0pt}
\tablehead{
\colhead{Band} & 
\colhead{Images} &
\colhead{Noise at $7.5''$} &
\colhead{Noise at $15''$} \\
\colhead{} & 
\colhead{} &
\colhead{(MJy~sr$^{-1}$)} &
\colhead{(MJy~sr$^{-1}$)}
}
\startdata
WISE1 & 15,748 & $3.2~(2.7{-}3.7) \times 10^{-3}$ & $2.5~(2.1{-}3.2)\times 10^{-3}$ \\
WISE2 & 15,748 & $5.8~(4.9{-}6.8)\times 10^{-3}$ & $3.0~(2.6{-}3.7)\times 10^{-3}$ \\
WISE3 & 15,721 & $7.5~(5.9{-}9.1)\times 10^{-2}$ & $1.8~(1.4{-}2.1)\times 10^{-2}$ \\
WISE4 & 15,749 & \nodata & $1.3~(1.0{-}1.5)\times 10^{-1}$ \\
\hline
\\
NUV\tablenotemark{a} & 11,688 & $2.6~(1.5{-}6.3)~\times 10^{-4}$ & $1.2~(0.8{-}2.5)~\times 10^{-4}$ \\
FUV\tablenotemark{a} & 10,754 & $3.1~(1.2{-}5.3)~\times 10^{-4}$ & $1.5~(0.6{-}2.4)~\times 10^{-4}$ \\
\enddata
\tablenotetext{a}{Before correction for Galactic extinction.}
\tablecomments{We report the median and $16{-}84$ percentile range (in parentheses) of the robustly estimated rms noise in our high Galactic latitude ($|b| > 40^\circ$) targets. The robust noise estimation uses the median absolute deviation to estimate the $1\sigma$ noise.}
\end{deluxetable}

\begin{figure*}
\centering
\plottwo{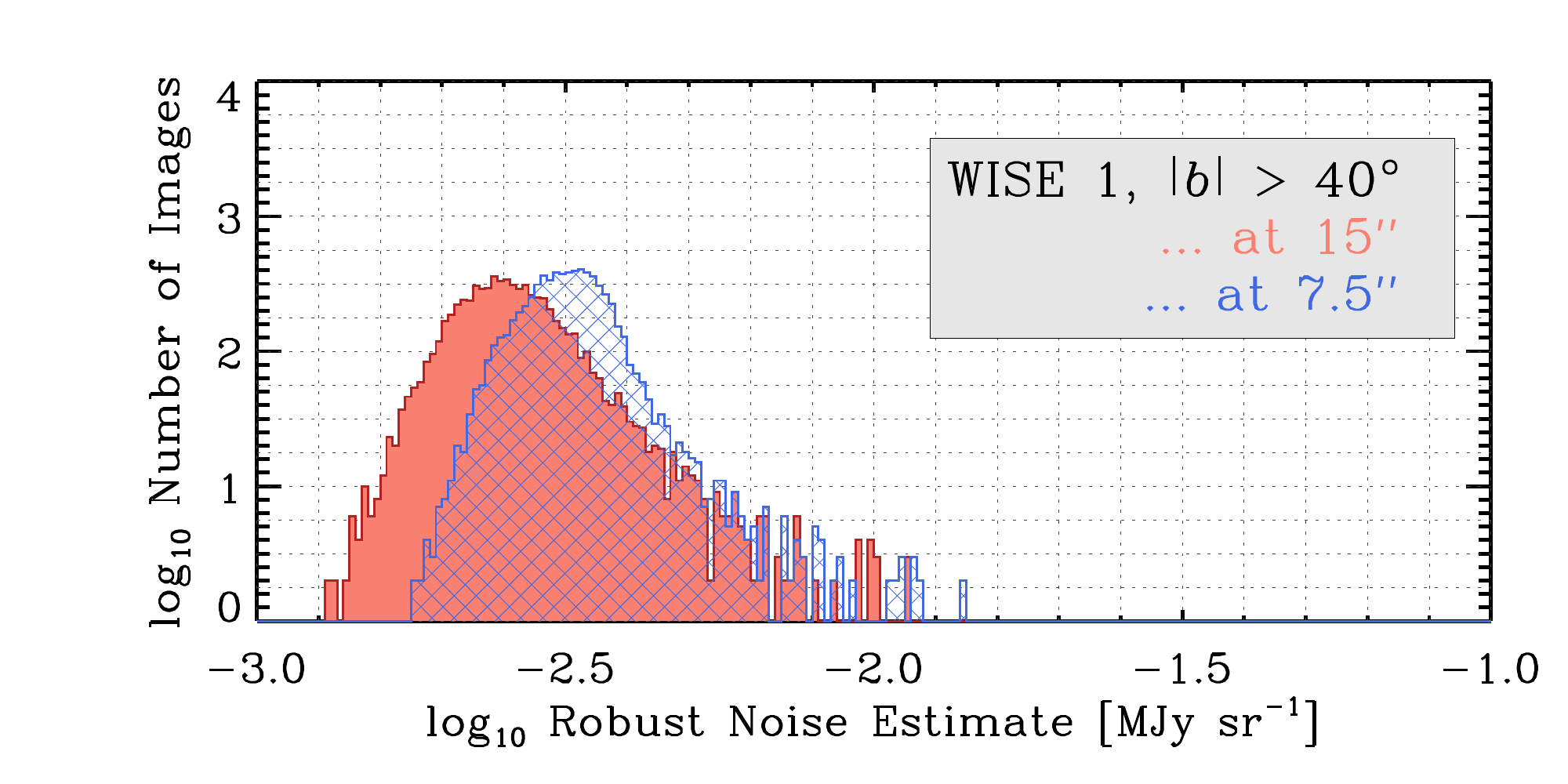}{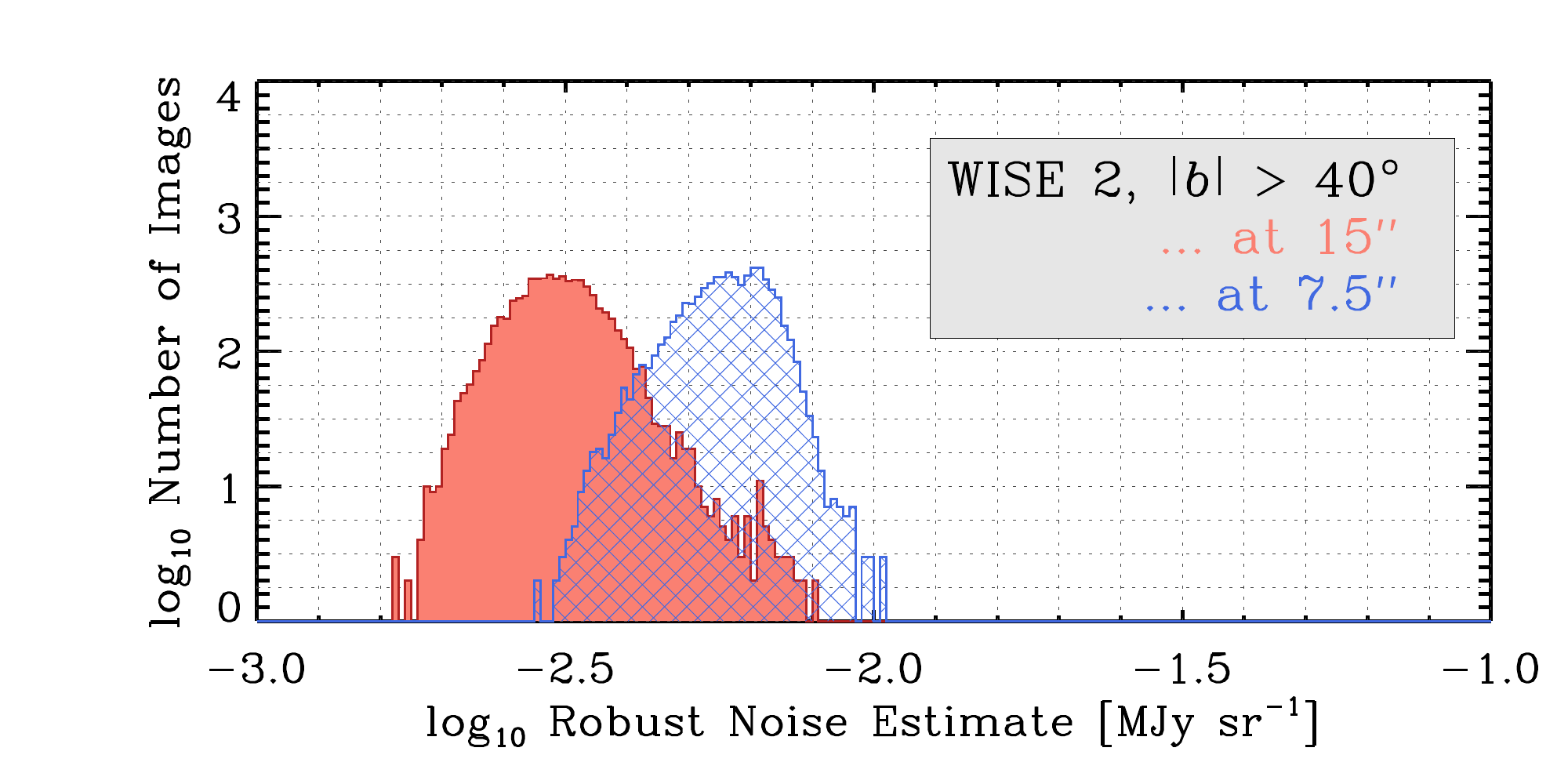}
\plottwo{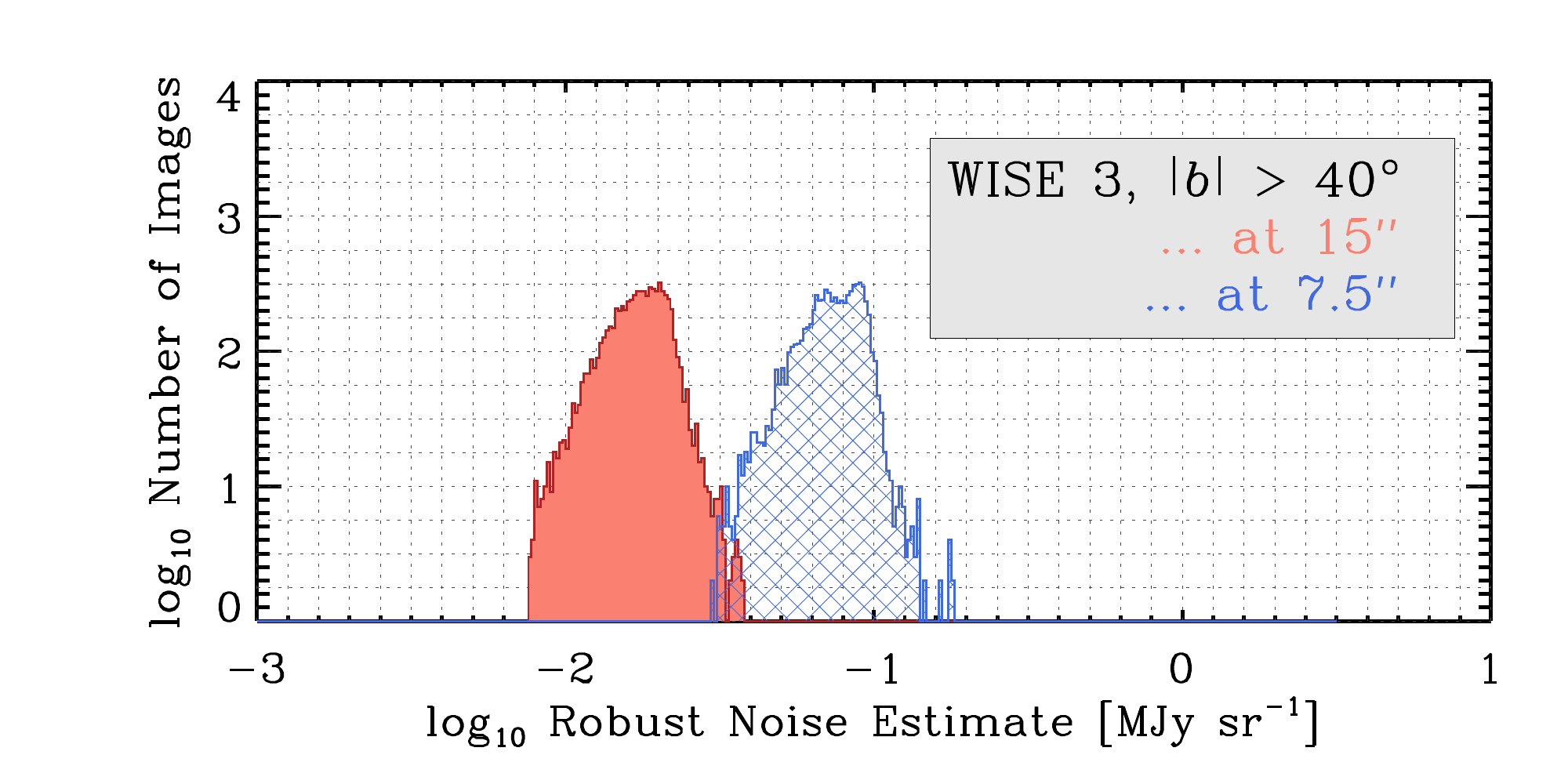}{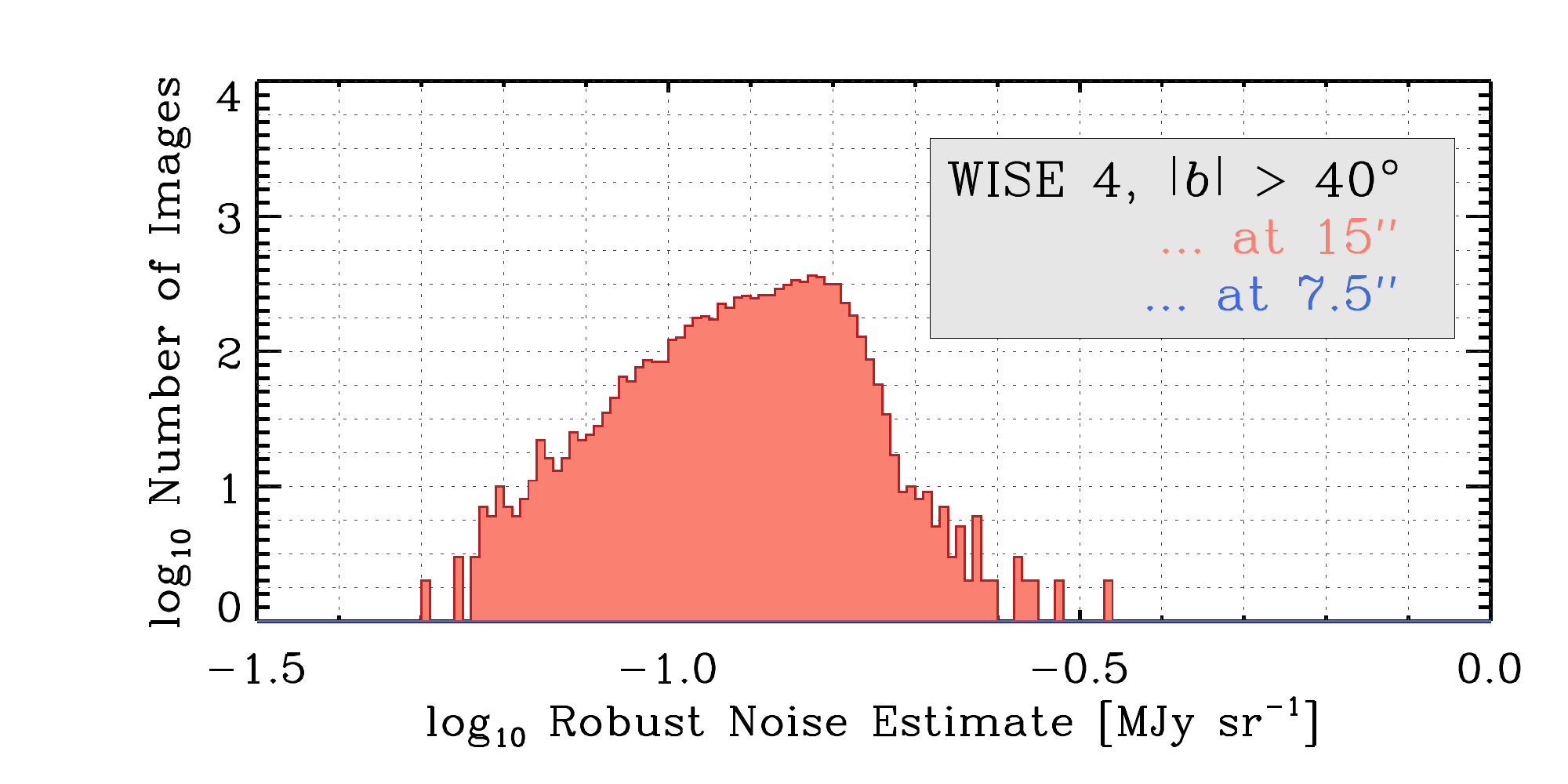}
\plottwo{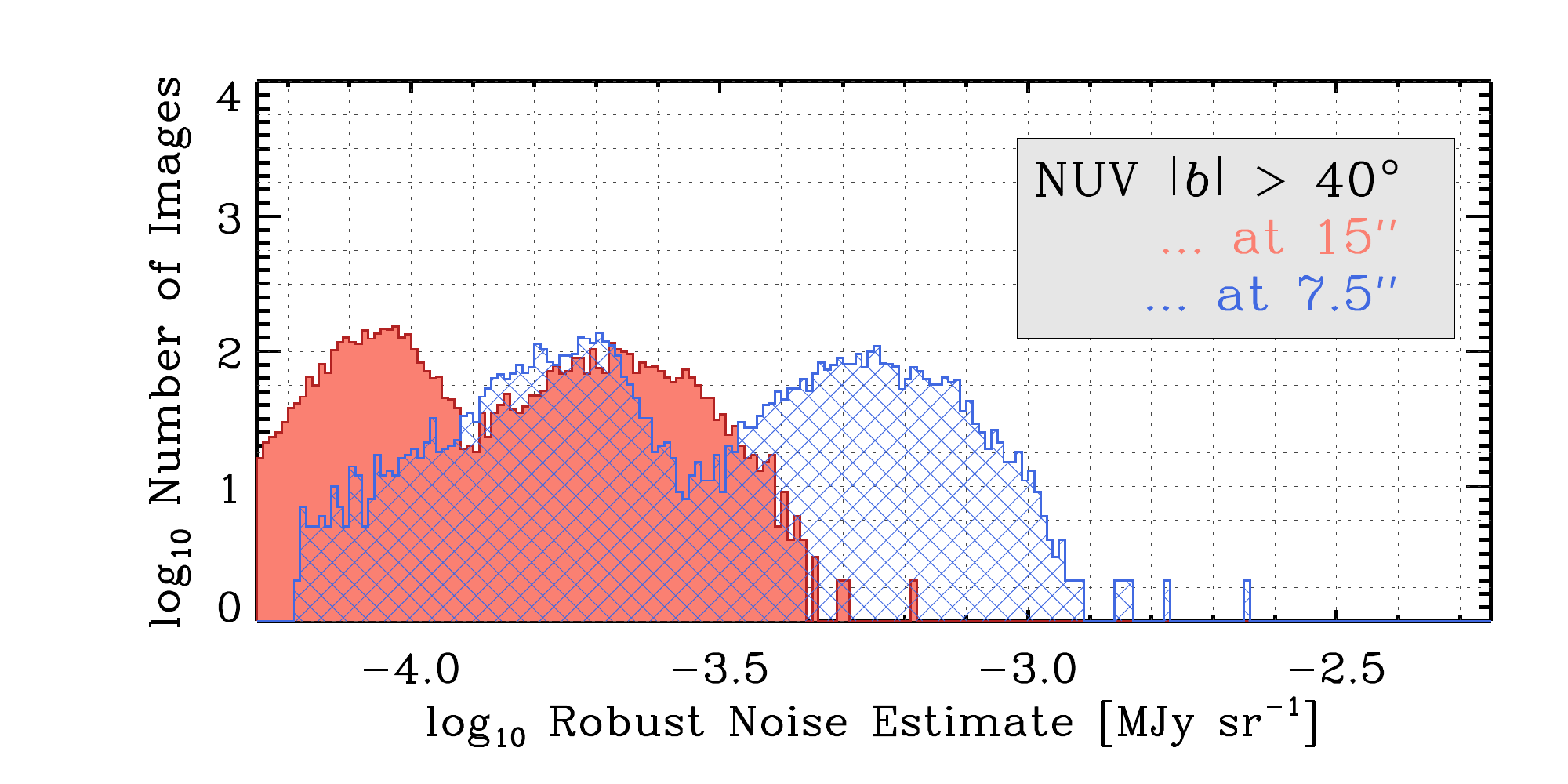}{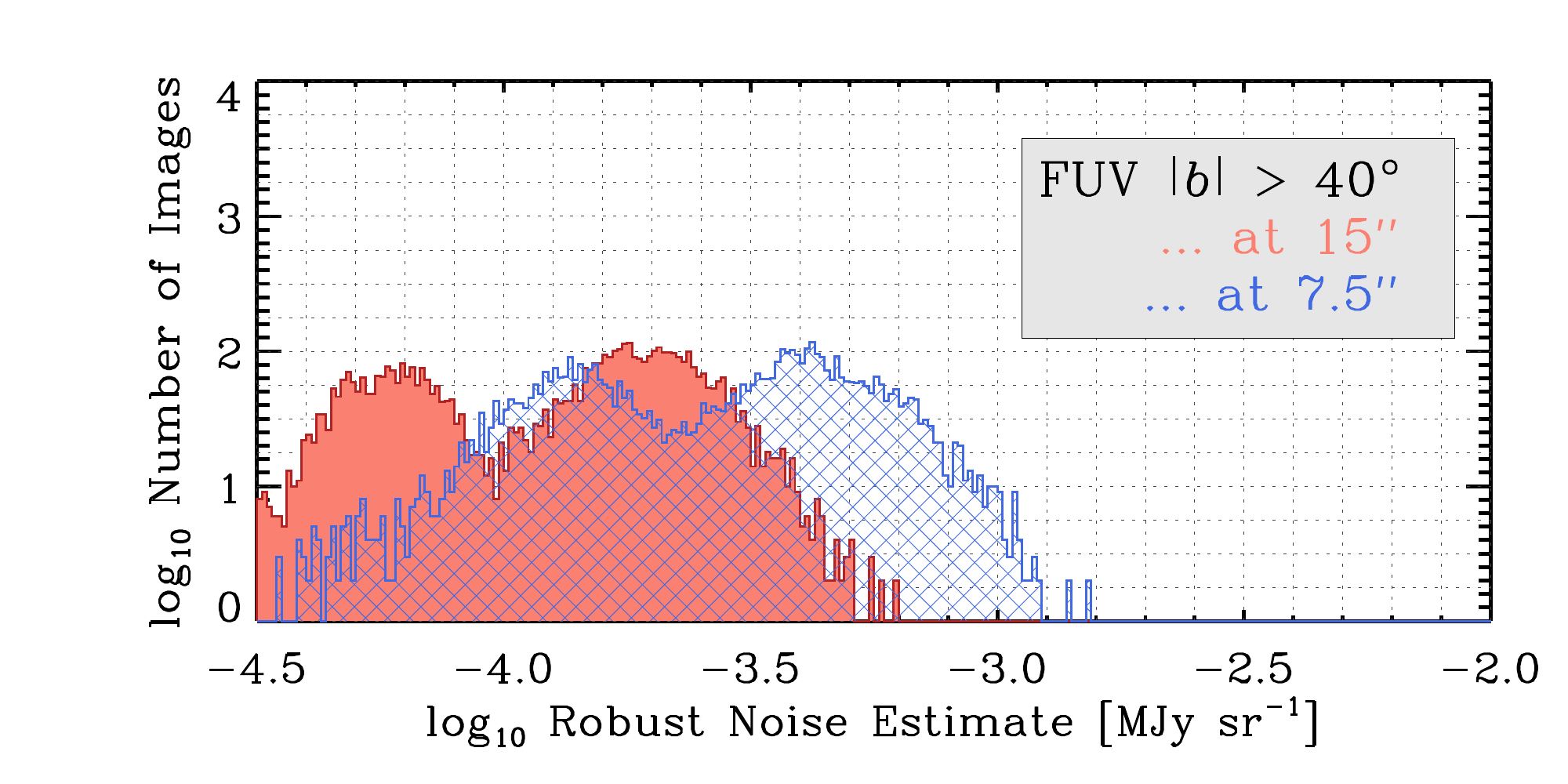}
\caption{{\bf Noise Distribution.} Median absolute deviation-based rms noise estimate in the $7.5\arcsec$ (blue) and $15\arcsec$ (red) resolution versions of the atlas. ({\em top and middle}:) The WISE atlas. Histograms show the distribution of robustly estimated RMS noise values for images at high Galactic latitude ($|b| > 40^\circ$). The bimodal histograms for the GALEX atlas ({\em bottom row}) reflect the variable exposure time that GALEX achieved across the nearby galaxy population (Figure \ref{fig:galex_time}).}
\label{fig:wisenoise}
\end{figure*}

\begin{figure*}
\centering
\plottwo{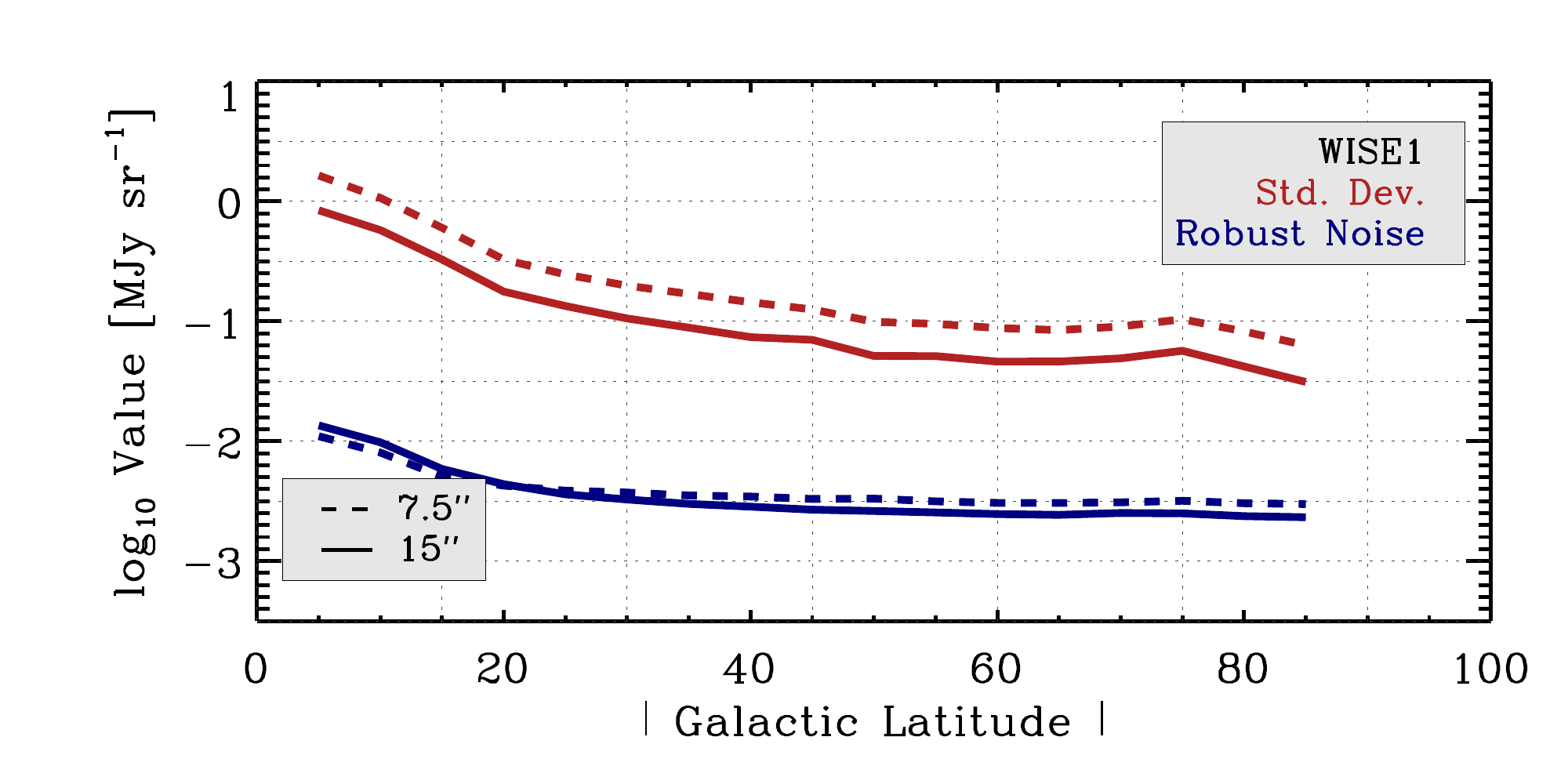}{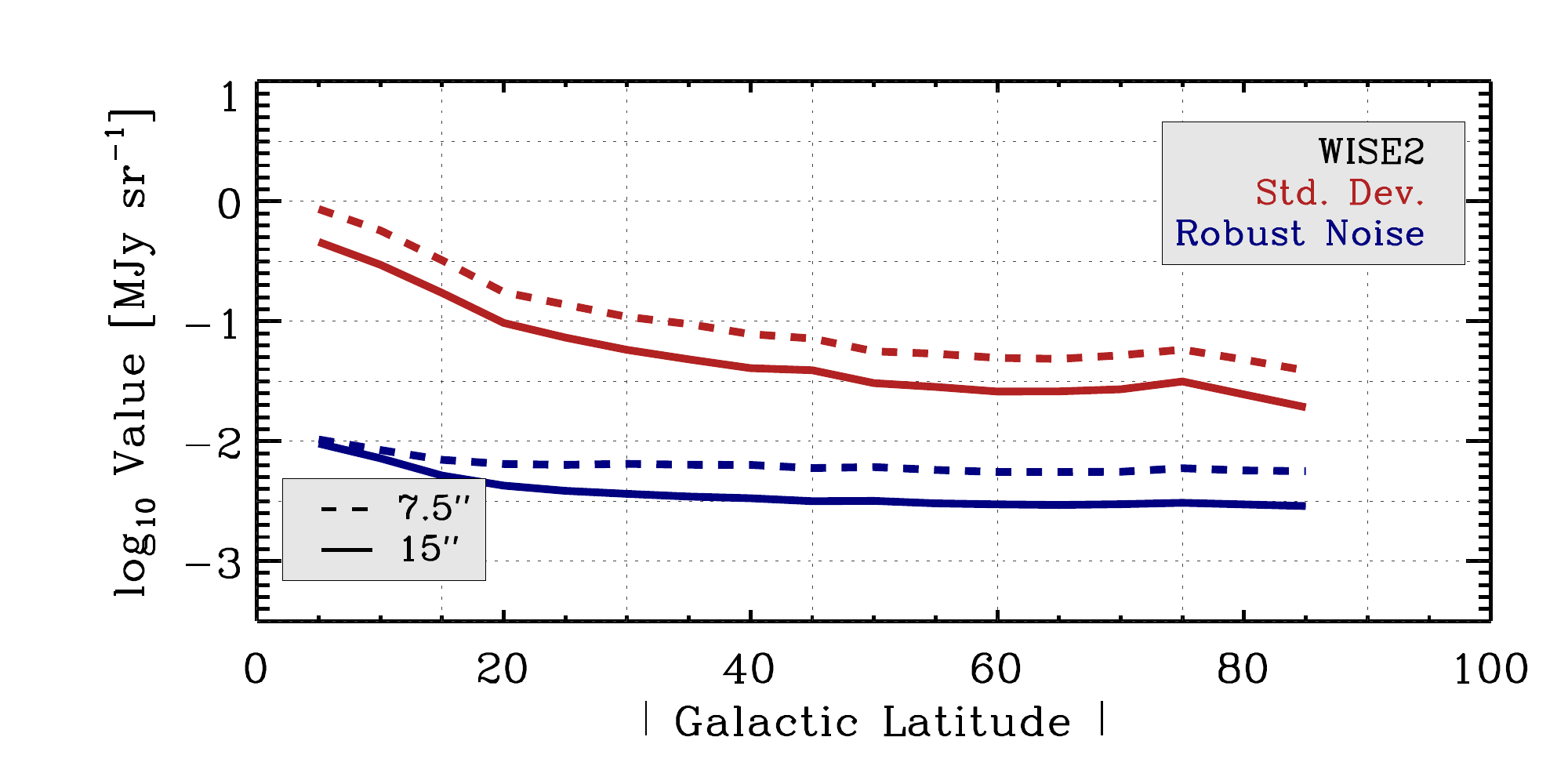}
\plottwo{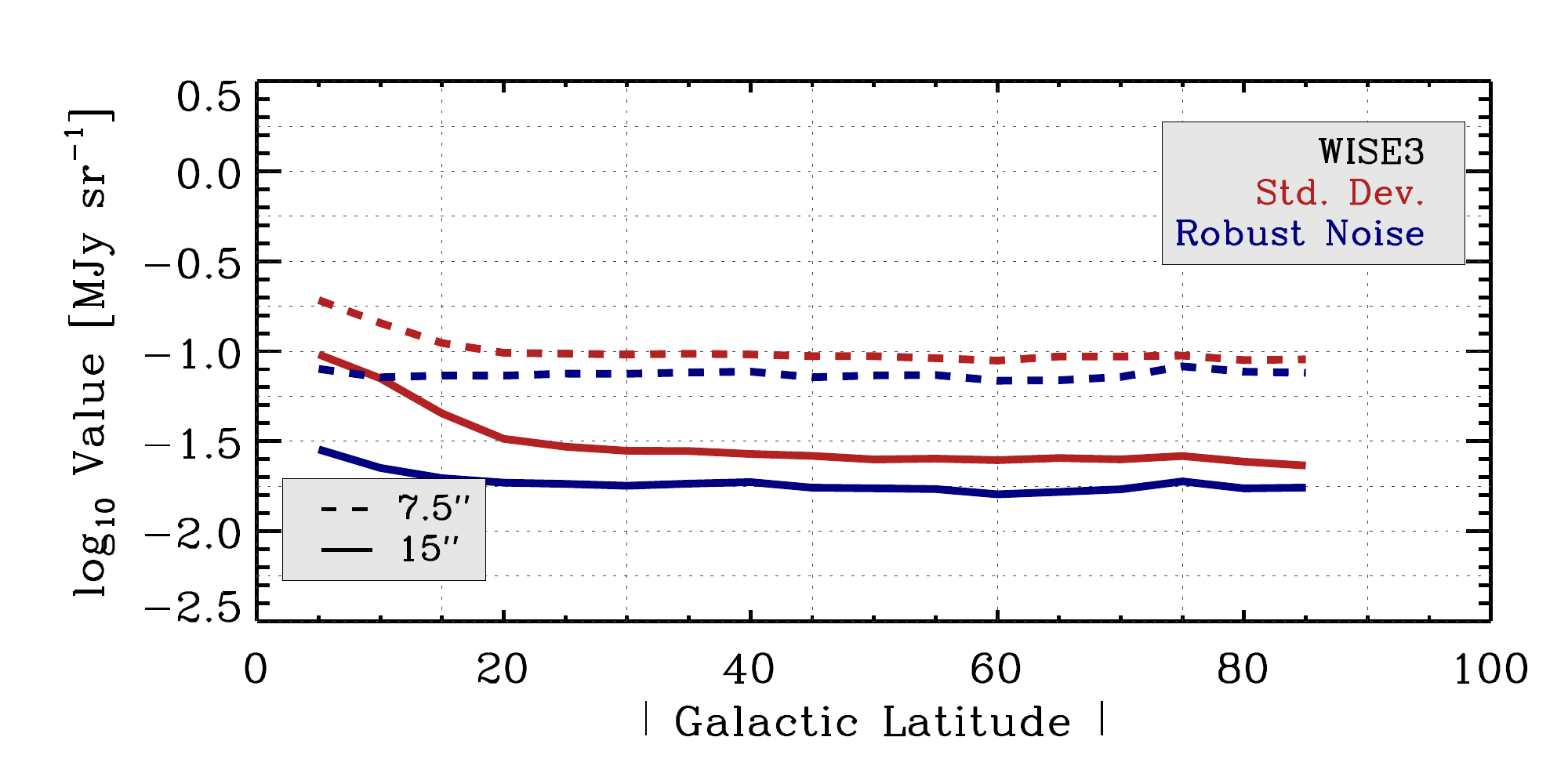}{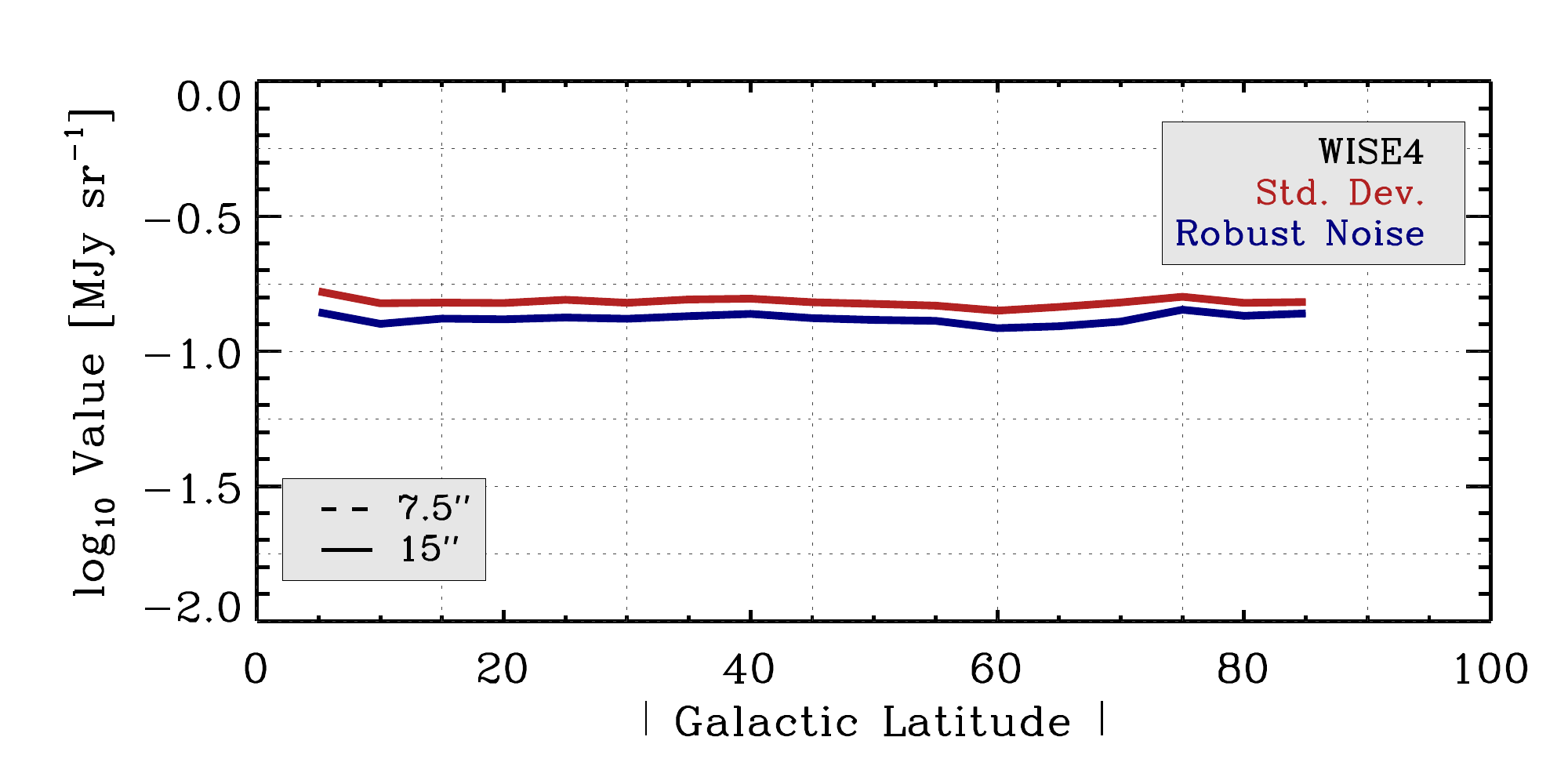}
\plottwo{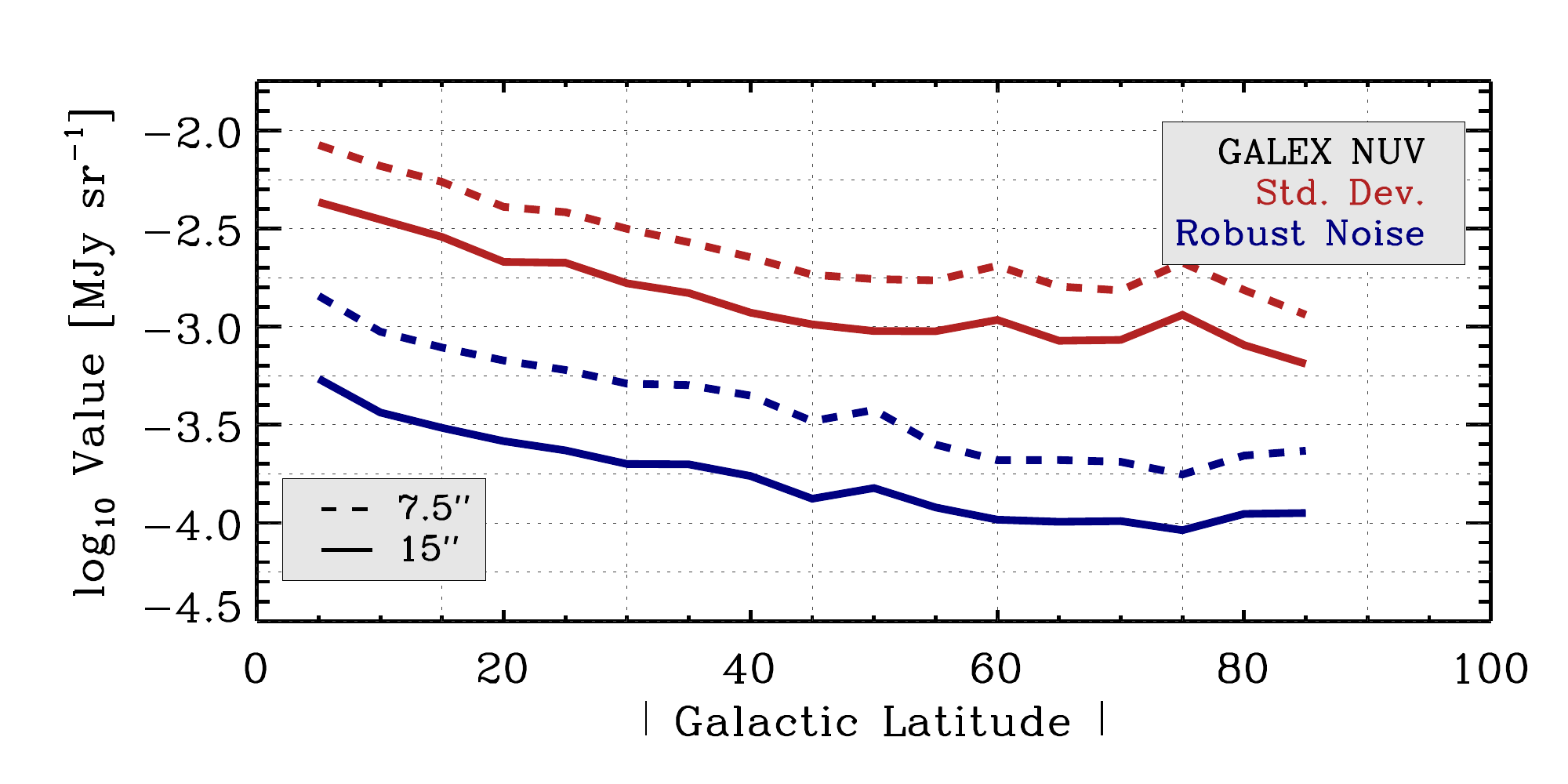}{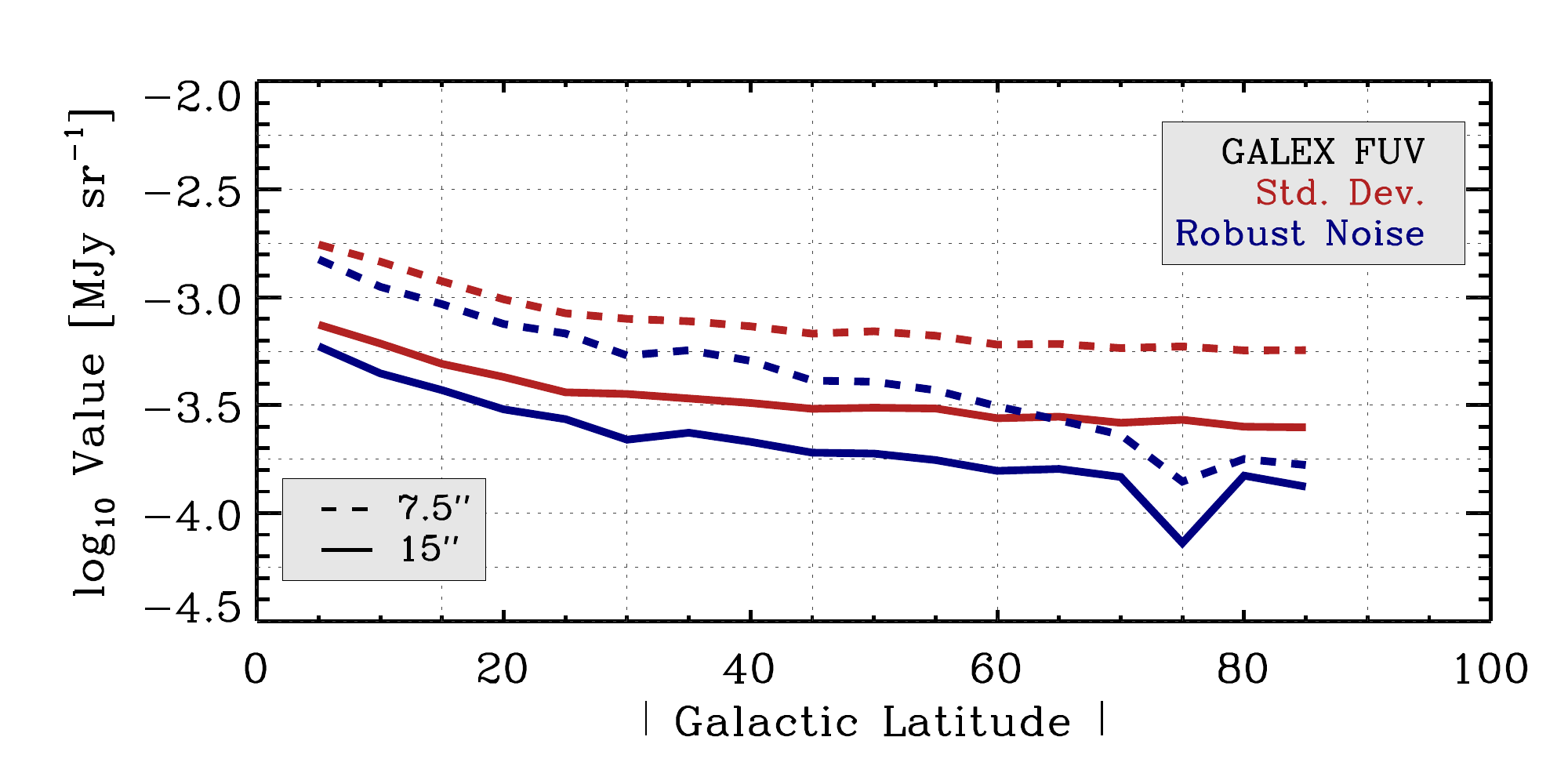}
\caption{{\bf Noise, Foreground Contamination, and Galactic Latitude.} Typical standard deviation-based noise (red) and robust noise (blue) in our atlas as a function of Galactic latitude, $|b|$. The much higher standard deviation (red) compared to the robust noise (blue) at WISE1, WISE2, and GALEX NUV demonstrates the dominant contribution of foreground stars to the image noise. The same effect is present, but weaker, at WISE3 and GALEX FUV, with Galactic cirrus acting as a major foreground for WISE3 \citep[e.g.,][]{MEISNER14}. The noise in WISE4 remains statistical at all but the lowest latitudes.}
\label{fig:wisenoise_vs_b}
\end{figure*}

Figure \ref{fig:wisenoise} shows our robust estimates of the noise in the WISE atlas. We plot the distribution of noise for each band and focus on targets at high Galactic latitude, $|b| > 40^\circ$. Because of the lower confusion due to foreground stars and cirrus, the high latitude histograms give our best estimate for the statistical noise in the atlas. 

The key results for users of our atlas are:

\begin{enumerate}
\item The statistical noise at high Galactic latitudes is uniform, with a $1\sigma$ range of $\pm 20\%$ across our sample. We report median values and the $16{-}84^{\rm th}$ percentile range for both working resolutions in Table \ref{tab:wisenoise}.

\item The statistical noise at WISE1 at $15''$ resolution, $\sim 3.2 \times 10^{-3}$~MJy~sr$^{-1}$, corresponds to a stellar surface density $\Sigma_\star \sim 1$~M$_\odot$~pc$^{-2}$. Surface densities of this magnitude are typical of the outer parts of galaxy disks. Thus, in the case of a filled beam and the absence of foreground stars (both significant caveats), the WISE1 images have surface brightness sensitivity appropriate to study the full area of galaxy disks.

\item The typical noise at WISE4 at $15''$ resolution corresponds to a star formation surface density $\sim 5 \times 10^{-4}$~\sigsfrunits . In the case of a well-resolved disk, we expect the maps to recover Solar Neighborhood star formation at a signal to noise ratio of a few.

\item Although less established as a SFR tracer, the $1\sigma$ noise in a typical WISE3 image at $15''$ resolution corresponds to $\sim 8 \times 10^{-5}$~\sigsfrunits . The WISE3 images are an order of magnitude more sensitive to recent star formation than the WISE4 images. However, given that PAH features dominate this part of the mid-IR spectrum and are known to have a strong environmental dependence, WISE3 also suffers from larger systematic uncertainties (we verify and quantify this in the Appendix). 

\item Typical M$_\star$-to-SFR ratios for star-forming galaxies (``main sequence'' galaxies) are $\sim 1{-}3 \times 10^{10}$~yr. Taking this as shorthand for a typical star-forming spectral energy distribution, the WISE1 and WISE3 images have roughly the same sensitivity to a star-forming disk. In this framework, the WISE4 images are an order of magnitude less sensitive (to star formation) than the WISE1 images (to stellar mass).
\end{enumerate}

The measured noise shows a strong Galactic latitude dependence. This reflects that for WISE bands WISE1 through WISE3, foreground contamination by the Milky Way, not instrumental noise, often represents the main source of uncertainty. The WISE depth of coverage does vary across the sky, but this pattern aligns with the ecliptic, with lower coverage near the ecliptic, and not the Milky Way.

Figure \ref{fig:wisenoise_vs_b} shows the importance of these foregrounds. For each band and each resolution, we plot the noise estimated from the median absolute deviation after blanking bright sources in blue. This represents our best estimate of the statistical noise. In red, we plot the standard deviation of the image away from the galaxy with no blanking applied. This calculation includes all outliers and does not reject bright sources and so captures either contaminating foreground emission or noise, whichever is dominant. In the limit of pure instrumental noise, we expect the two estimates to resemble one another. The lines in the figures show the median values among all images in bins of Galactic latitude.

Foreground stars drive the standard deviation for WISE1 and WISE2. The Figure shows that these represent the dominant source of scatter in the images, with an increasing contribution at low Galactic latitude. Even at high latitude, foreground stars are present and will introduce more scatter into the image than the instrumental noise. WISE3 also suffers from foregrounds, most notably Galactic cirrus, but the contamination is weaker compared to the statistical noise than at WISE1 and WISE2. WISE4 remains dominated by statistical noise at all but the lowest latitudes.

\section{GALEX Maps}
\label{sec:galex}

\subsection{GALEX Cutouts}

The sky coverage of GALEX is more uneven than that of WISE, and there is no public database analogous to unWISE. Therefore, we constructed our own cutouts.

To do this, we downloaded every FUV and NUV tile obtained by GALEX during its mission. Specifically, we downloaded the counts per second image (\texttt{int}), the sky background subtracted image (\texttt{intbgsub}), the pixel-by-pixel flags (\texttt{flags}), and the high resolution relative response image (\texttt{rrhr}).

For our atlas, we used the \texttt{int} version of the images, which provide the counts per second per pixel after the pipeline processing described by \citet[][]{MORRISEY05}. As with WISE, we avoided the background subtracted version because the sky estimate often includes a significant contribution from the galaxy itself. This can lead to bowling and negative fluxes. Instead, we fit our own background estimate outside the galaxy.

To construct the cutouts, we carried out the following steps:

\begin{enumerate}
\item We built a target astrometric grid for each galaxy. As for unWISE, we adopted a default image size of $1800''$ across. We increased this to six times the LEDA-recorded optical radius, $r_{25}$, for larger galaxies. We adopted the native GALEX pixel scale of $1.5''$ for our cutouts.

\item We identified all tiles that overlapped the footprint of each galaxy. For each overlapping tile, we read in the intensity (\texttt{int}) image, the relative response (\texttt{rrhr}), and the pipeline-produced flags.

\item We converted the intensity image from counts per second per pixel to Jy~pix$^{-1}$. To do this, we followed the GALEX documentation and adopted conversions of $1.07647 \times 10^{-4}$ Jy~pix$^{-1}$ (count~s$^{-1}$~pix$^{-1}$) for FUV and $3.37289 \times 10^{-5}$ Jy~pix$^{-1}$ (count~s$^{-1}$~pix$^{-1}$). After this, we converted from Jy~pix$^{-1}$ to Jy~sr$^{-1}$ using the pixel area in sr and then to the final units of MJy~sr$^{-1}$.

\item We blanked the portions of each image flagged as being near the edge of the field of view, where the sensitivity declines.

\item We aligned the intensity and relative response images to our target astrometric grid.

\item We summed the intensity times the relative response and the relative response for each pixel over all of the aligned tiles

\begin{equation}
I = \frac{\sum \texttt{int} \times \texttt{rrhr}}{\sum \texttt{rrhr}}~{\rm and}~w = \sum \texttt{rrhr}~.
\end{equation}

\noindent The pipeline processed \texttt{int} reflects the counts per second, while the \texttt{rrhr} captures the effective integration time times the effective area of the telescope for this observation. Thus the numerator corresponds the total counts observed, the normalization corresponds to the effective integration time, and the noise should be proportional to $w$, the square root of the total integration time. The final intensity in the cutout is $I$ and the weight image, $w$, captures the effective sensitivity. For each pixel in the final image, we expect the noise to be proportional to $w^{-0.5}$.

\end{enumerate}

We recorded our intensity, $I$, in MJy~sr$^{-1}$ and the relative response for each pixel, in units of seconds, as separate images on our new astrometric grid.

We attempted to construct cutouts for all of the images in the WISE atlas. We create empty or partially filled images where GALEX does not cover the galaxy. After this exercise, we considered the area of each image within the nominal optical radius, $r_{25}$, from LEDA. We recorded the mean effective integration time in each band and the fraction of pixels within the galaxy footprint covered by GALEX.

\subsection{Convolution and Alignment}

Before background subtraction, we convolved GALEX images to our common final PSF. At this stage, we dropped images in which the target galaxy is not covered. As with the WISE data, we produced two sets of images, one at resolution of $7.5''$ (FWHM) and the other at resolution $15''$. Again we use the kernels and approach described by \citet{ANIANO11} and the final PSF shape is a Gaussian. We also also convolved the weight (effective integration time) images. Finally, we constructed foreground star masks for the NUV and FUV bands based on the 2MASS and GAIA star catalogs. As noted above, these are less certain than the foreground star masks for the WISE bands because of the strong dependence of the UV-to-$K_S$ and UV-to-$G$ color on stellar temperature.

\subsection{Background Subtraction}

We fit and subtracted a background from the convolved images. We follow a procedure similar to that adopted for the unWISE images. We blanked all pixels within $2~r_{25}$ of the galaxy, adjusted slightly from case to case. We also blanked all nearby galaxies and stars. Then, we calculated the median intensity in the region around the galaxy. We blanked pixels that deviate from the median by two times the local noise value. Then we recalculated and subtracted a new median. If enough pixels remain unmasked at this stage, we also fit and subtracted a plane to the sky, again using iterative outlier rejection. Otherwise, we proceed with only the median background fit. As with the WISE data, after this stage, we applied a slight correction to shift the local background to the mode of the histogram. This was most important in NUV images at low Galactic latitude, where foreground stars skewed the noise distribution high at a low level.

There are two subtleties in the treatment of the GALEX data during background subtraction. First, in some cases the rms noise varies across our images due to variations in the exposure time across the field of view. To account for this, when we reject pixels during background fitting we use a signal-to-noise based cut in which the local noise scales with the local effective integration time $w^{-0.5}$.

Second, for short exposure FUV images at $7.5''$ resolution, Poisson statistics may be more appropriate than Gaussian statistics and our outlier rejection and use of the median may be inappropriate. To account for this, we fit the background only at $15''$ resolution. Then we subtracted the bakground fit at $15''$ resolution from the $7.5''$ resolution images. At $15''$ resolution, our tests on low exposure time images suggested that there are enough counts in almost all images to make Gaussian statistics a reasonable approximation for our application.

M31 and M33 are large on the sky. For these two galaxies, we manually restricted the field of view to the region near the galaxy, $\lesssim 2~r_{25}$ and fit a plane rather than a single value to characterize the background. For M33, this yields good results. An ideal background subtraction M31 remains a work in progress.

After background subtraction, we aligned the GALEX images onto the same astrometric grid used for the final WISE images and the masks. 

\subsection{Statistical Properties of the GALEX Atlas}

\begin{figure*}
\centering
\plottwo{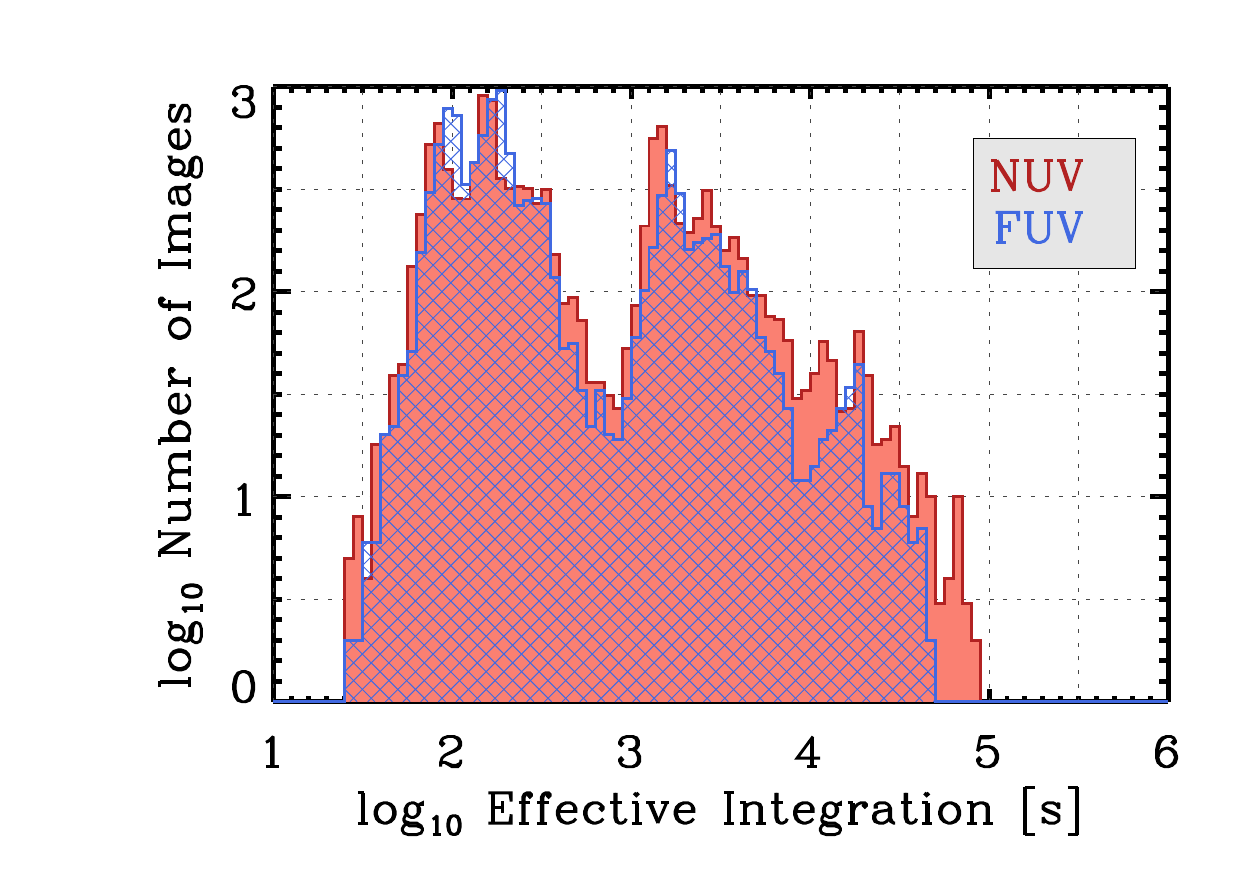}{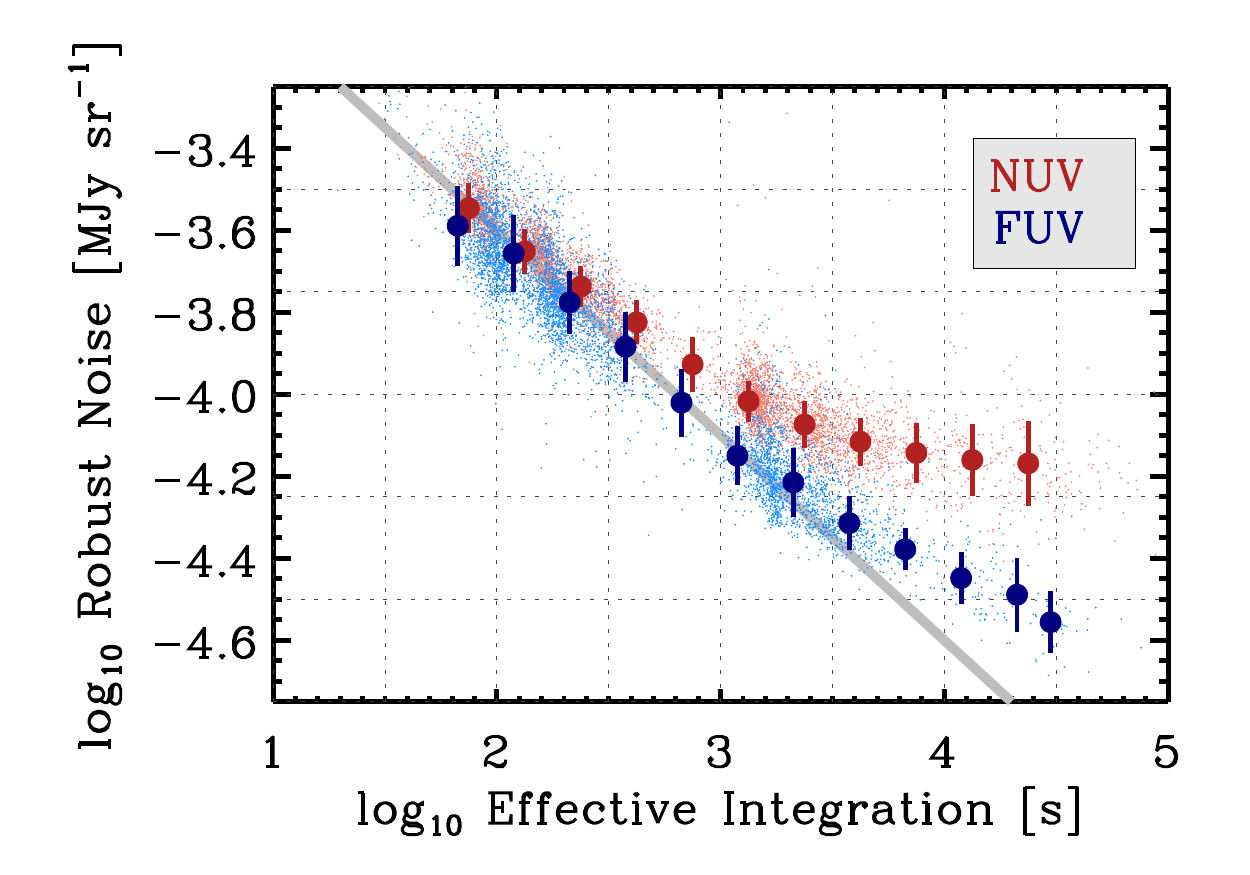}
\caption{{\bf GALEX Exposure Times and Noise at $15''$ Resolution vs. Exposure Time.} ({\em left}) Mean effective exposure time towards each target galaxy, in seconds, for the NUV (red, filled histogram) and FUV (blue, hatched histogram) parts of our atlas. ({\em right}) Robustly estimated noise as a function of mean effective integration time. The gray line shows the expected $\sigma \propto t^{-0.5}$ scaling. Deep NUV exposures tend to hit a noise floor, presumably due to the contribution of faint foreground sources to the noise. The GALEX images in both bands span at least a factor of $3$ in effective noise.}
\label{fig:galex_time}
\end{figure*}

Table \ref{tab:wisenoise}, Figure \ref{fig:wisenoise}, and Figure \ref{fig:wisenoise_vs_b} report the distributions of noise in the GALEX images and highlight the importance of Galactic foregrounds. These figures and the table show a bimodal distribution of noise values for the GALEX images, distinct from the uniform coverage in WISE. This reflects the highly varied integration times spent targeting local galaxies with GALEX. Table \ref{tab:wisenoise} also shows the incomplete sky coverage of GALEX. Out of our $\sim 15,750$ targets, which almost all have unWISE coverage, only $11,688$ have GALEX NUV coverage and only $10,754$ have GALEX FUV coverage.

Figure \ref{fig:galex_time} shows the distribution of effective integration time in NUV (red) and FUV (blue) for each galaxy in our atlas. The integration times show two broad peaks. The all sky surveys yield effective integration times $\sim 100$~s. Meanwhile targeted programs, such as the Nearby Galaxy Atlas \citep{GILDEPAZ07}, integrated for $\sim 10^3{-}10^4$~s and even up to $\sim 10^5$~s in a few cases. As a result of this factor of $\sim 100$ span in effective integration time, the effective noise varies by an order of magnitude across the GALEX atlas. The right hand panel of Figure \ref{fig:galex_time} shows that up to integration times $\sim 10^4$~s, the robustly estimated noise reflects the integration time in the expected way, with noise scaling as $\sim t^{-0.5}$. Above this value the NUV data appear to approach a noise floor, likely reflecting low level contamination by faint sources.

\subsection{Correction for Galactic Extinction}

\begin{figure}
\centering
\plotone{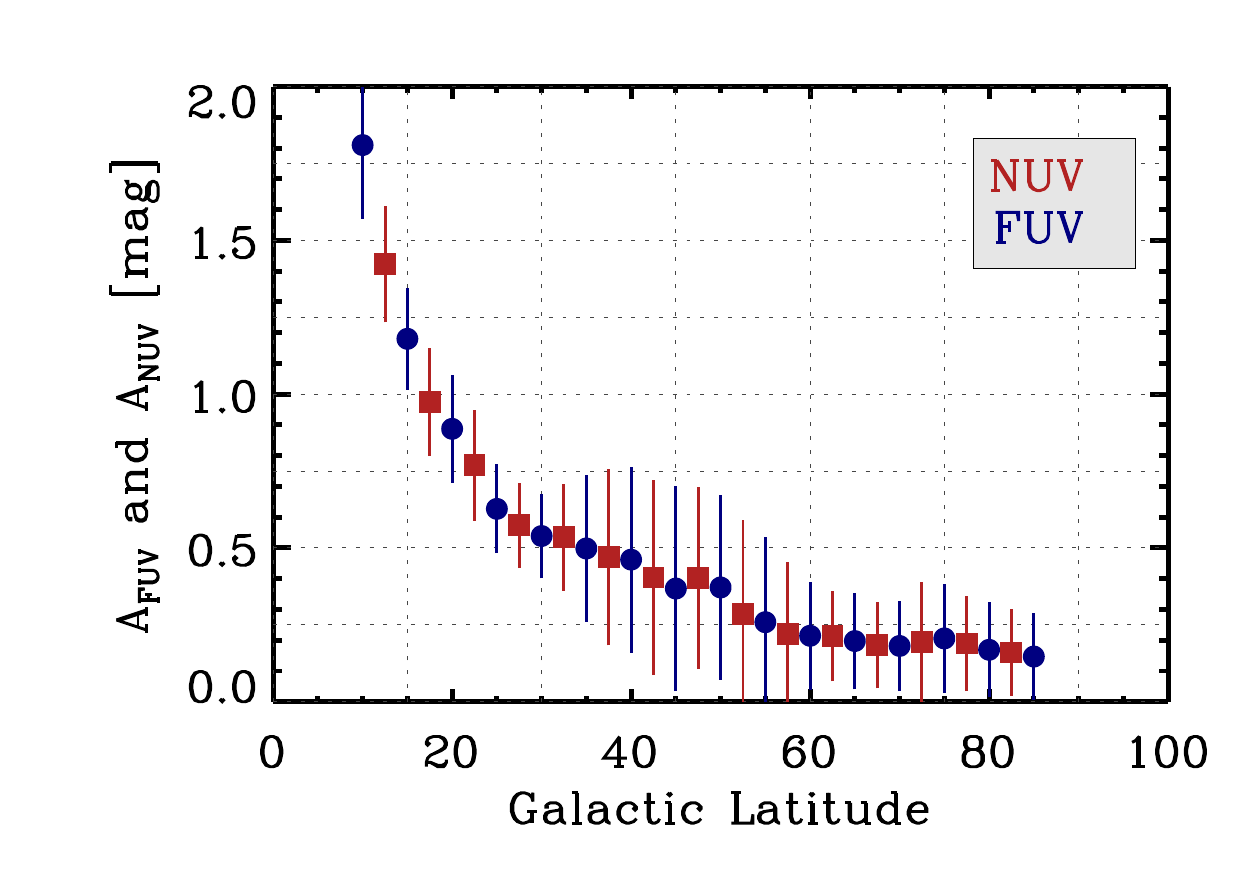}
\caption{{\bf Milky Way Extinction.} Median adopted corrections for Milky Way extinction in the FUV (blue) and NUV (red) bands as a function of Galactic latitude. We adopt $R_{\rm FUV}$ and $R_{\rm NUV}$ from \citet{PEEK13} and take $E(B-V)$ from \citet{SCHLEGEL98}.}
\label{fig:galex_afuv}
\end{figure}

Dust in the Milky Way attenuates the UV emission from our target galaxies. We expect the Milky Way foreground extinction to be relatively uniform across each target due to their small angular extent. The methods to account for this extinction are reasonably well understood. Therefore, we correct our GALEX atlas images for Galactic extinction. To do this, we adopt the prescription of \citet{PEEK13}, who used observations of the average colors of large numbers of galaxies to calibrate $A_{\rm FUV}$ and $A_{\rm NUV}$ as a function of the $E(B-V)$ predicted by the \citet{SCHLEGEL98} dust map. Motivating our choice, \citet{SALIM16} found the \citet{PEEK13} to give more self-consistent SED fitting results than competing prescriptions. We determine the $A_{\rm FUV} = R_{\rm FUV}~E(B-V)$ and $A_{\rm NUV} = R_{\rm NUV}~E(B-V)$ for each target using the \citet{SCHLEGEL98} $E(B-V)$, and $R_{\rm NUV}$ and $R_{\rm FUV}$ as follows:

\begin{eqnarray}
R_{\rm FUV} &=& 10.47+8.59~E(B-V)  \\
\nonumber && -82.8~E(B-V)^2  \\
\nonumber R_{\rm NUV} &=& 8.36+14.3~E(B-V)  \\
\nonumber && -82.8~E(B-V)^2  \\
\nonumber {\rm for}~E(B-V) &=& {\rm min} \left( E(B-V), 0.2 \right)~. \\
\nonumber \end{eqnarray}

\noindent The limit, $E(B-V) = 0.2$~mag represents approximately the maximum value for which \citet{PEEK13} derive $R_{\rm FUV}$ and $R_{\rm NUV}$. Above 0.2 mag, we fix the value of $E(B-V)$ to be 0.2 mag in calculating $R_{\rm FUV}$ and $R_{\rm NUV}$, although $A_{\rm FUV}$ and $A_{\rm NUV}$ continue to scale with the actual value of $E(B-V)$. Figure \ref{fig:galex_afuv} shows the typical extinction correction as a function of Galactic latitude for the atlas.

\section{Image Atlas}

\begin{figure*}
\centering
\includegraphics[width=0.85\textwidth]{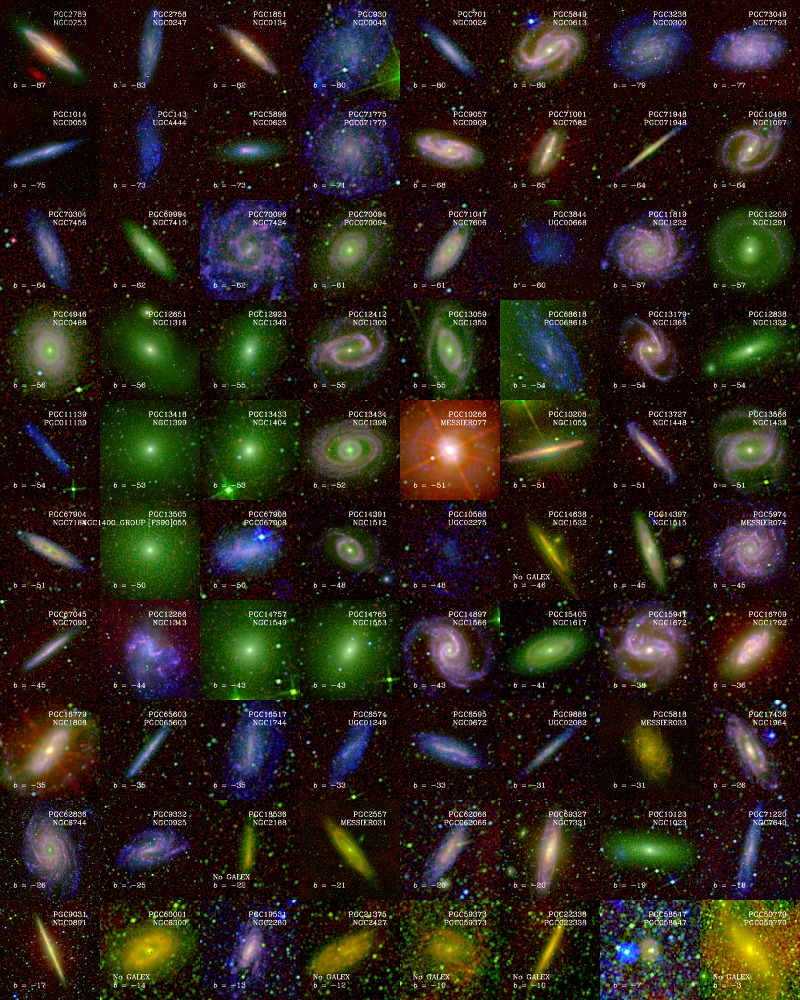}
\caption{{\bf Large Angular Size Targets 1.} Three color images of galaxies with $r_{25} > 150''$. Red shows WISE3 ($12\mu$m), mostly emission from small dust grains, blue shows NUV emission, mostly photospheric emission from relatively young stars, and green shows WISE1, mostly photospheric emission from older, evolved stars. We show the $7.5''$ resolution version of the axis. The images have full extent $2.5 r_{25}$, so that the cutouts vary in angular and physical size. The images all share a matched logarithmic stretch. In units of $\log_{10}$~MJy~sr$^{-1}$ they stretch from -2 to 1.5 (WISE3, red), -2 to 1.5 (WISE1, green), -3.0 to -0.25 (NUV, blue). We sort the images by Galactic latitude, so that those in the middle part of the sequence show the strongest contamination by foreground stars.}
\label{fig:images_1}
\end{figure*}

\begin{figure*}
\centering
\includegraphics[width=0.9\textwidth]{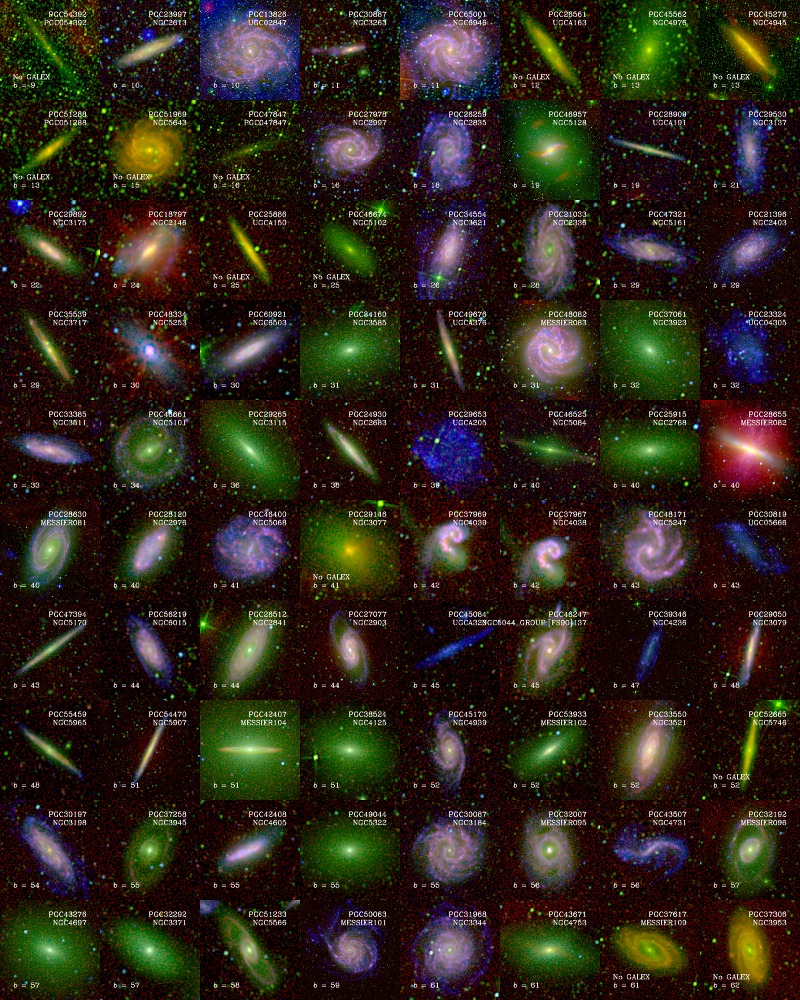}
\caption{{\bf Large Angular Size Targets 2.} Figure \ref{fig:images_1} continued.}
\label{fig:images_2}
\end{figure*}

\begin{figure*}
\centering
\includegraphics[width=0.9\textwidth]{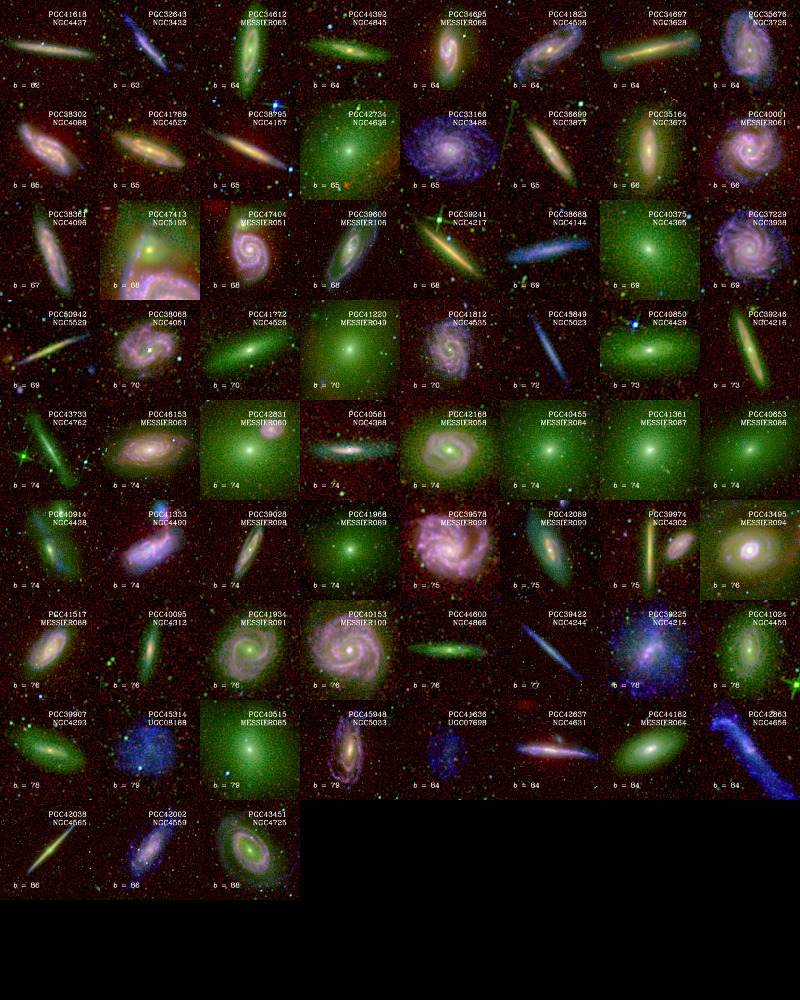}
\caption{{\bf Large Angular Size Targets 3.} Figure \ref{fig:images_1} continued.}
\label{fig:images_3}
\end{figure*}

Following the previous sections, our image atlas consists of six-band imaging for $\sim 15,750$ galaxies. For each galaxy, the image atlas consists of the following:

\begin{enumerate}
\item FUV, NUV, and WISE1, WISE2, WISE3, and WISE4 intensity images in units of MJy~sr$^{-1}$ at $15''$ resolution. FUV, NUV, and WISE1, WISE2, and WISE3 intensity images at $7.5''$ resolution.
\item Weight images to pair with the GALEX FUV and NUV images. These have the sense that the local noise is expected to be proportional to $w^{-0.5}$ and have units of seconds, capturing the effective exposure time.
\item Images for each resolution that give the galactocentric radius, in decimal degrees, used to fit the background and carry out the photometry. As described above, we cap the inclination used to calculate galactocentric radius at $60^\circ$.
\item Masks for each resolution identifying the likely footprint of other galaxies in the field.
\item Masks for each resolution and band that identify the expected footprint of bright stars.
\end{enumerate}

The atlas images have been resampled to have pixel sizes of $2.75\arcsec$ at $7.5''$ resolution and $5.5\arcsec$ at $15''$ resolution. 

An accompanying electronic ``index'' table summarizes the contents of the atlas. This index reports the presence of data for each band, the integration time for each GALEX band, the measured noise, the adopted Milky Way extinction, photometry in each band, several inferred physical properties of the galaxy (adopted distance, $M_\star$, SFR, \mtolwise, etc.) and flags identifying potential issues with the data (see below).

Figures \ref{fig:images_1} through \ref{fig:images_3} show examples of the image atlas. We show three color images combining the $7.5''$ resolution WISE3 (red), WISE1 (green), and GALEX NUV (blue) for all targets with isophotal radius $r_{25} > 150''$, i.e., 20 resolution elements.  We extract a region $\pm 1.25~r_{25}$ on a side for each galaxy, so that the cutouts show different sizes. The stretch for all images is the same: a logarithmic scaling between $10^{-2}$ and $10^{1.5}$~MJy~sr$^{-1}$ for WISE3 and WISE1 and between $10^{-3}$ and $10^{-0.25}$~MJy~sr$^{-1}$ for NUV.

The images show several important limitations of the atlas. We sort the images by Galactic latitude, $b$, and do not apply any masks to the images. Thus images with large contamination by foreground stars appear together in the middle of the sequence. These low $b$ images show that foreground stars represent an important source of contamination for WISE1 and NUV. 

Several of the brightest, closest targets also show evidence for saturation and non-Gaussian PSFs; e.g., see NGC 253, M77 (NGC~1068), NGC 1808, and M82 in WISE3 or note the bright stars near several galaxies in WISE1. Section \ref{sec:starstacks} quantifies these effects using stacks of bright stars known from 2MASS. 

To capture these issues, the index table and image headers include flags that indicate whether we expected contamination by other galaxies, issues with saturation, or significant contamination by foreground stars. We flag confusion due to galaxies when $> 10\%$ of the area associated with the target is also associated with the footprint of another galaxy. We note potential saturation issues when any pixel in the image exceeds intensity thresholds of $100$~MJy~sr$^{-1}$ at WISE1 and WISE2 or $300$~MJy~sr$^{-1}$ at WISE3 and WISE4 (see below). We flag contamination by stars as an issue when $> 20\%$ of the flux within the galaxy footprint lies inside the bright star mask. We repeat this calculation for each band. 

Table \ref{tab:flags} summarizes the fraction of galaxies affected by each type of flag. As the plots in the previous section show, contamination by foreground stars represents a serious problem at WISE1 and WISE2 for a significant fraction of our targets. Saturation and overlapping galaxies affect far fewer galaxies, though as Figures \ref{fig:images_1} --- \ref{fig:images_3} show, saturated targets are often among the nearest, brightest, and best-studied galaxies.

Although we label bright images with a ``saturation'' flag, a similar cut would identify the cases where we expect the PSF of the images to become significantly non-Gaussian. In Section \ref{sec:starstacks}, we estimate the average PSF by stacking stars, and show that despite our convolution, there is still significant non-Gaussianity at signal to noise $\gtrsim 1000$.

\begin{deluxetable}{lcc}
\tabletypesize{\scriptsize}
\tablecaption{Flags in the Atlas at $7.5''$ \label{tab:flags}}
\tablewidth{0pt}
\tablehead{
\colhead{Flag Type} & 
\colhead{Number Affected} &
\colhead{Fraction Affected}
}
\startdata
Overlapping galaxy & $456$ & $0.03$ \\
Heavy star contamination & & \\
$\ldots$ WISE1 & $5,013$ & $0.32$ \\
$\ldots$ WISE2 & $4,534$ & $0.29$ \\
$\ldots$ WISE3 & $1,502$ & $0.096$ \\
$\ldots$ WISE4\tablenotemark{a} & $100$ & $0.006$ \\
$\ldots$ NUV & $1,907$ & $0.16$ \\
$\ldots$ FUV\tablenotemark{b} & $31$ & $0.003$ \\
Saturation & & \\
$\ldots$ WISE1 & $322$ & $0.020$ \\
$\ldots$ WISE2 & $207$ & $0.013$ \\
$\ldots$ WISE3 & $71$ & $0..0045$ \\
$\ldots$ WISE4\tablenotemark{a} & $74$ & $0.005$ 
\enddata
\tablenotetext{a}{At $15''$ resolution.}
\tablenotetext{b}{Color information not included in star masking. FUV masks likely an underestimate.}
\tablecomments{``Overlapping galaxy'' means that $>10\%$ of the target galaxy footprint is also associated with another galaxy. ``Heavy star contamination'' means that $>20\%$ of the flux in the galaxy footprint lies inside the star mask. ``Saturation'' indicates that part of the image exceeds the threshold for likely saturation effects.}
\end{deluxetable}

\subsection{Validation Against Previous Resolved Mapping Surveys}

\begin{figure*}
\centering
\epsscale{0.925}
\plottwo{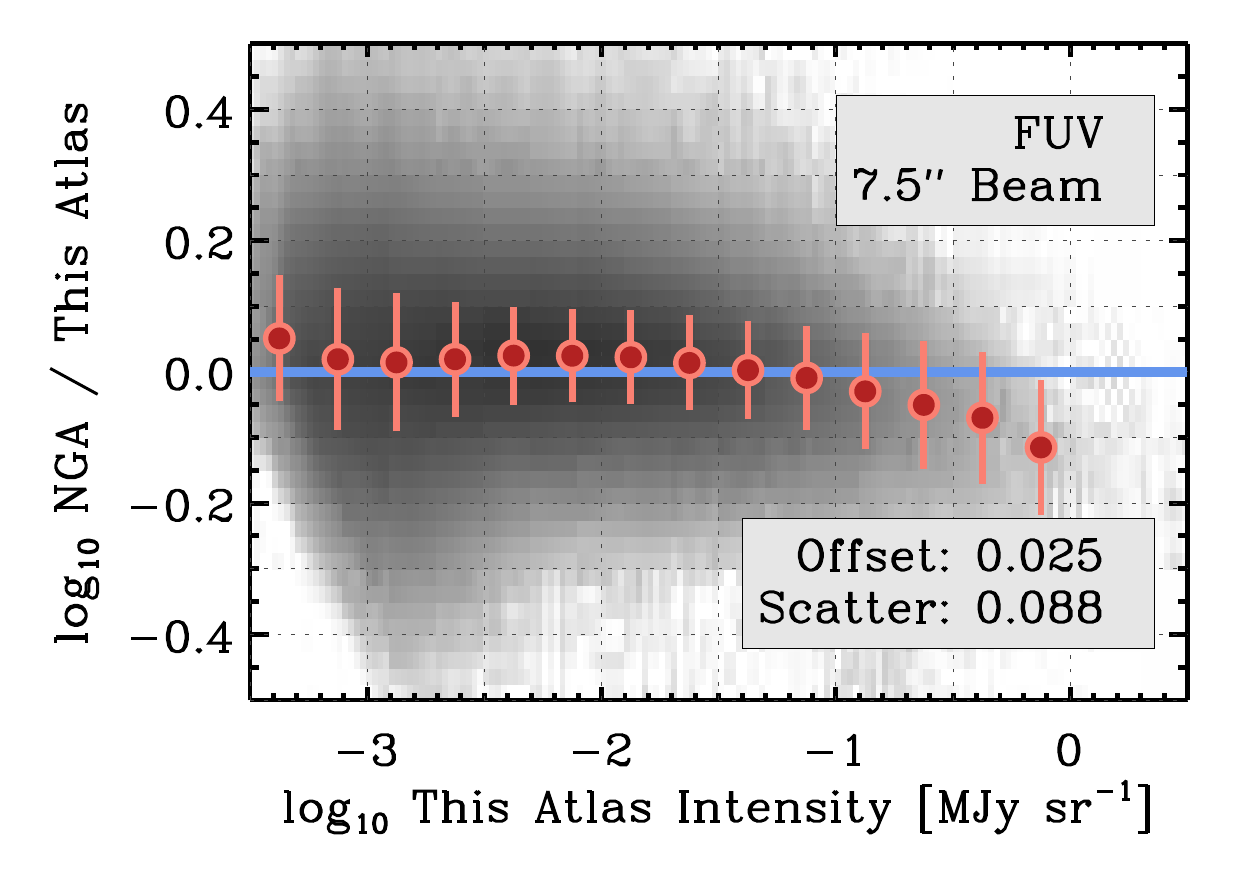}{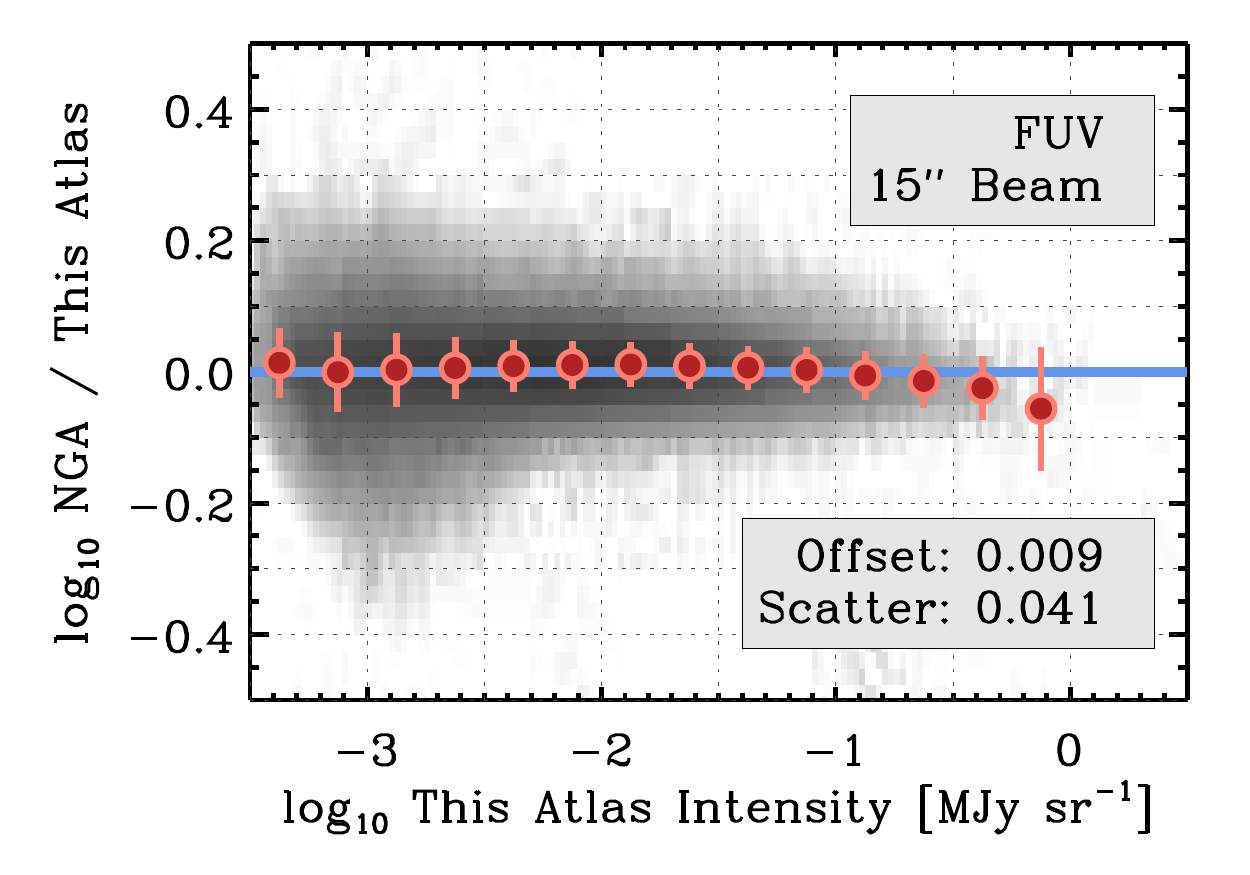}
\plottwo{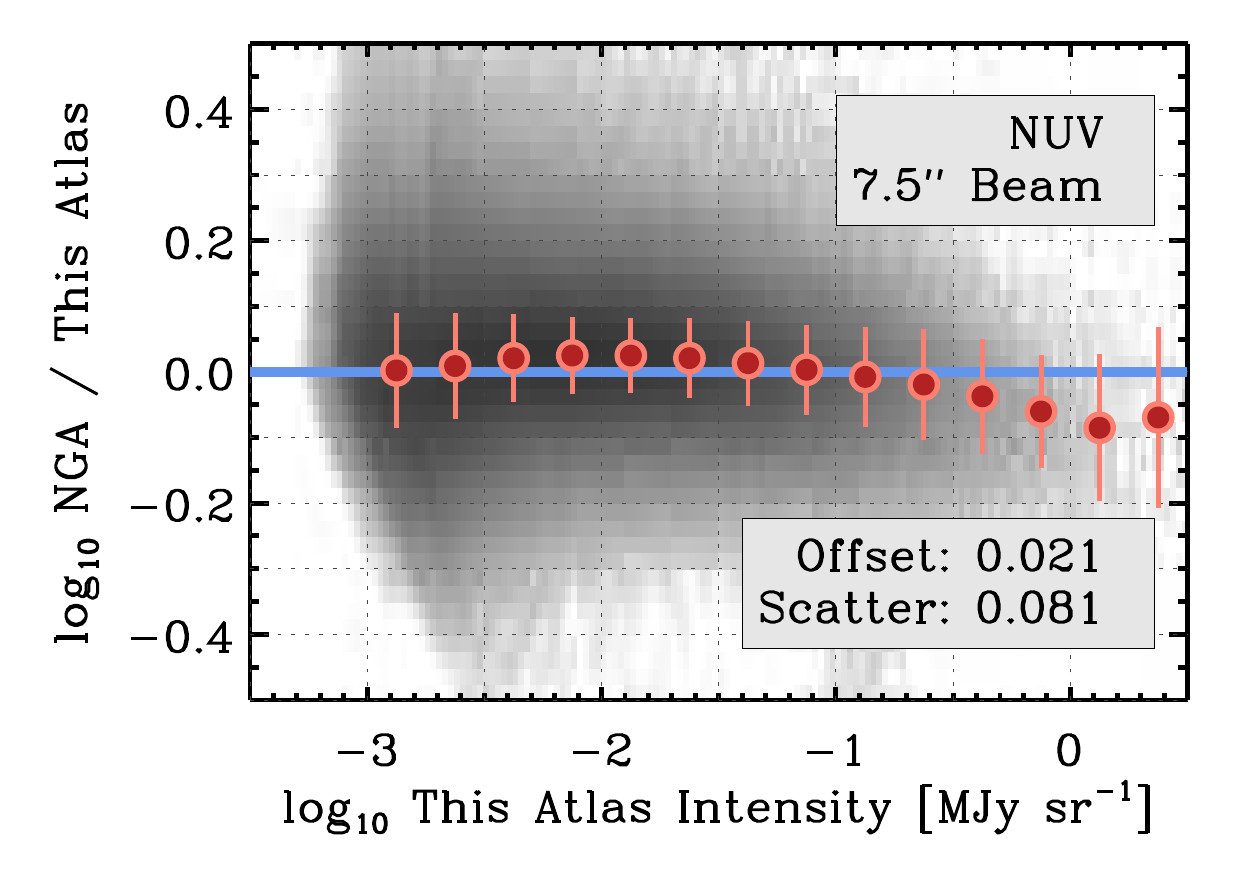}{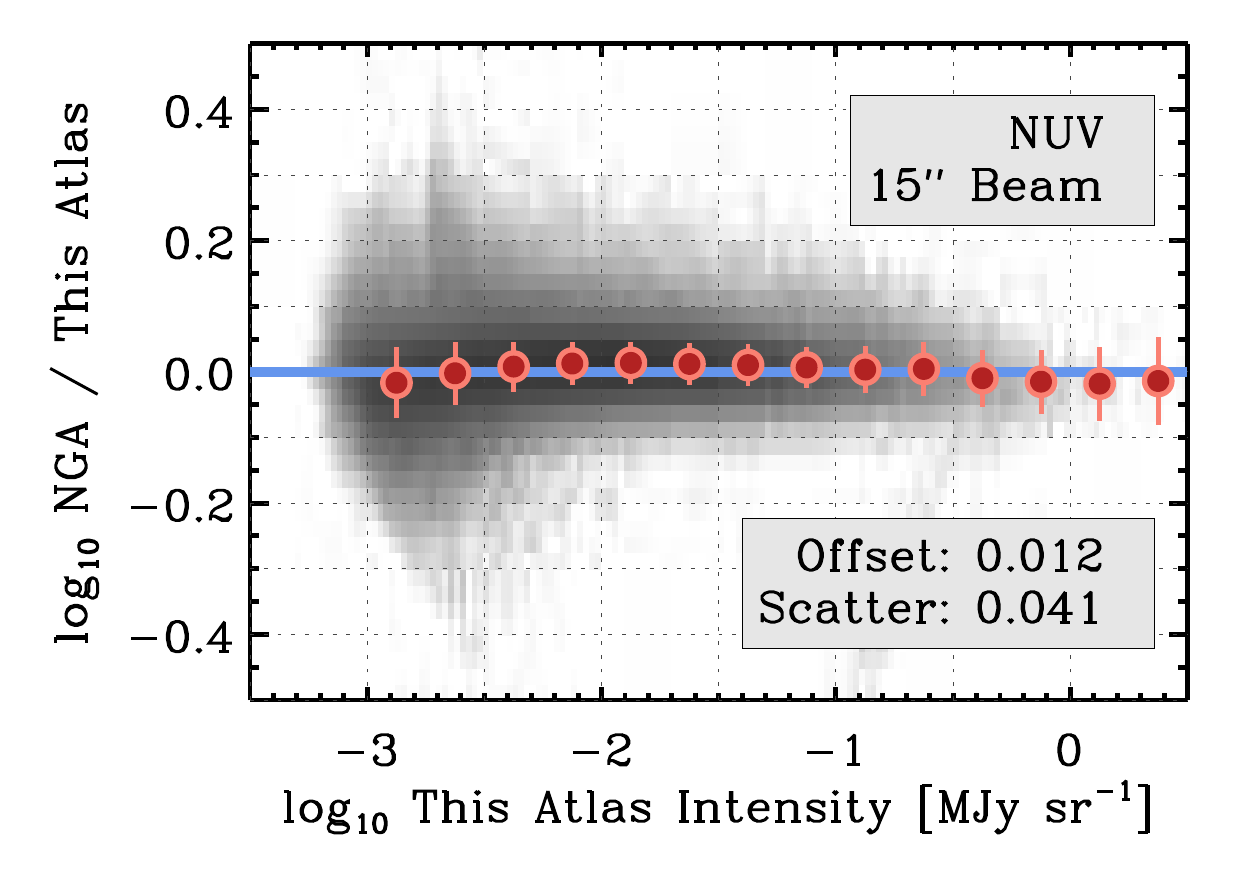}
\plottwo{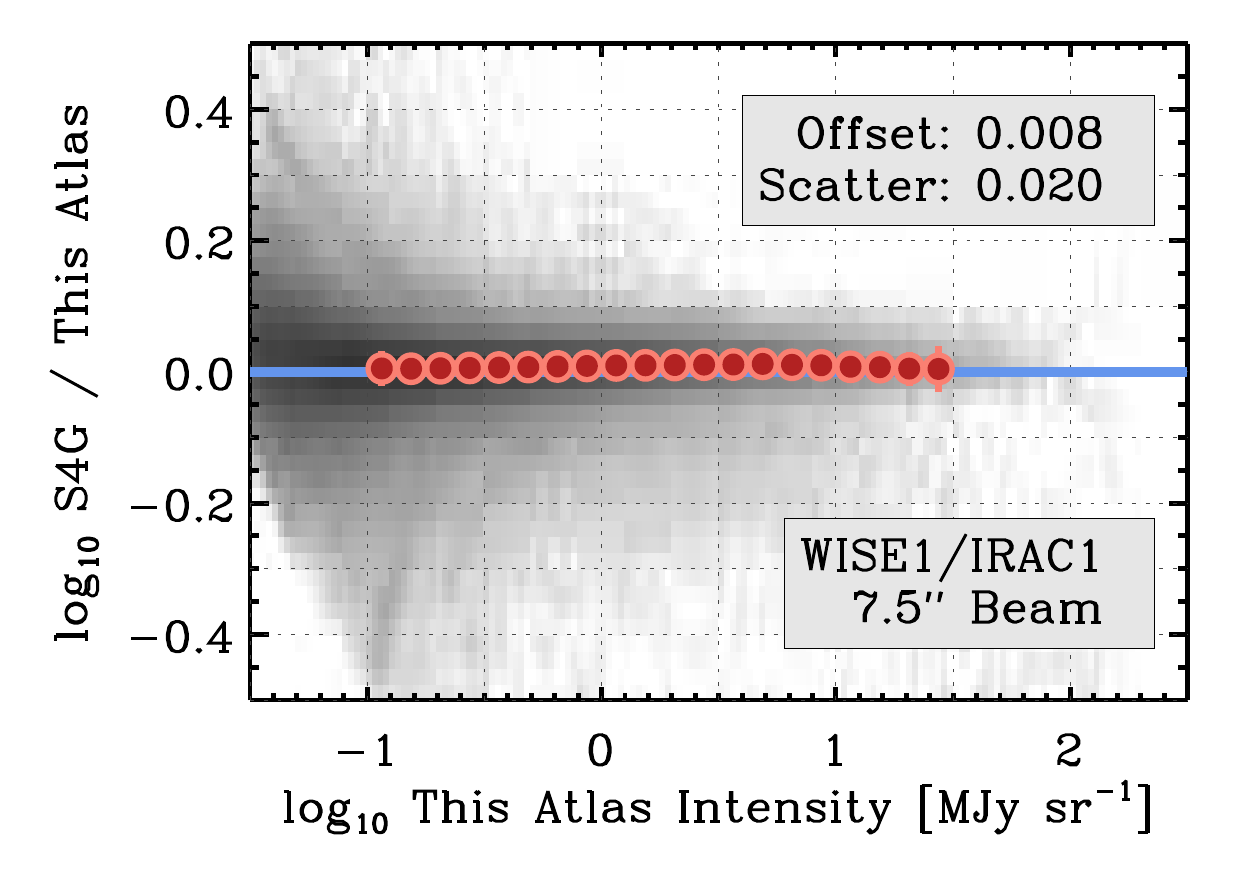}{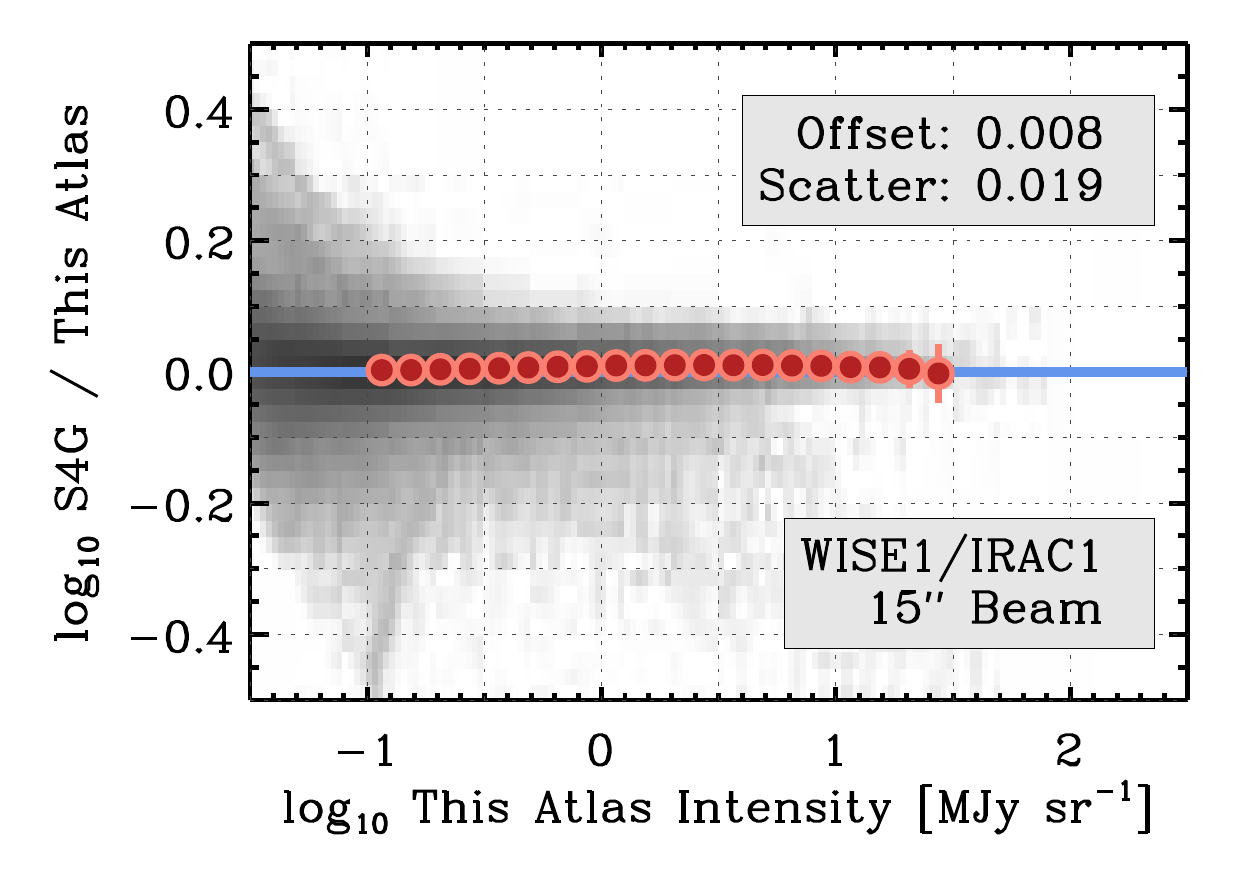}
\plottwo{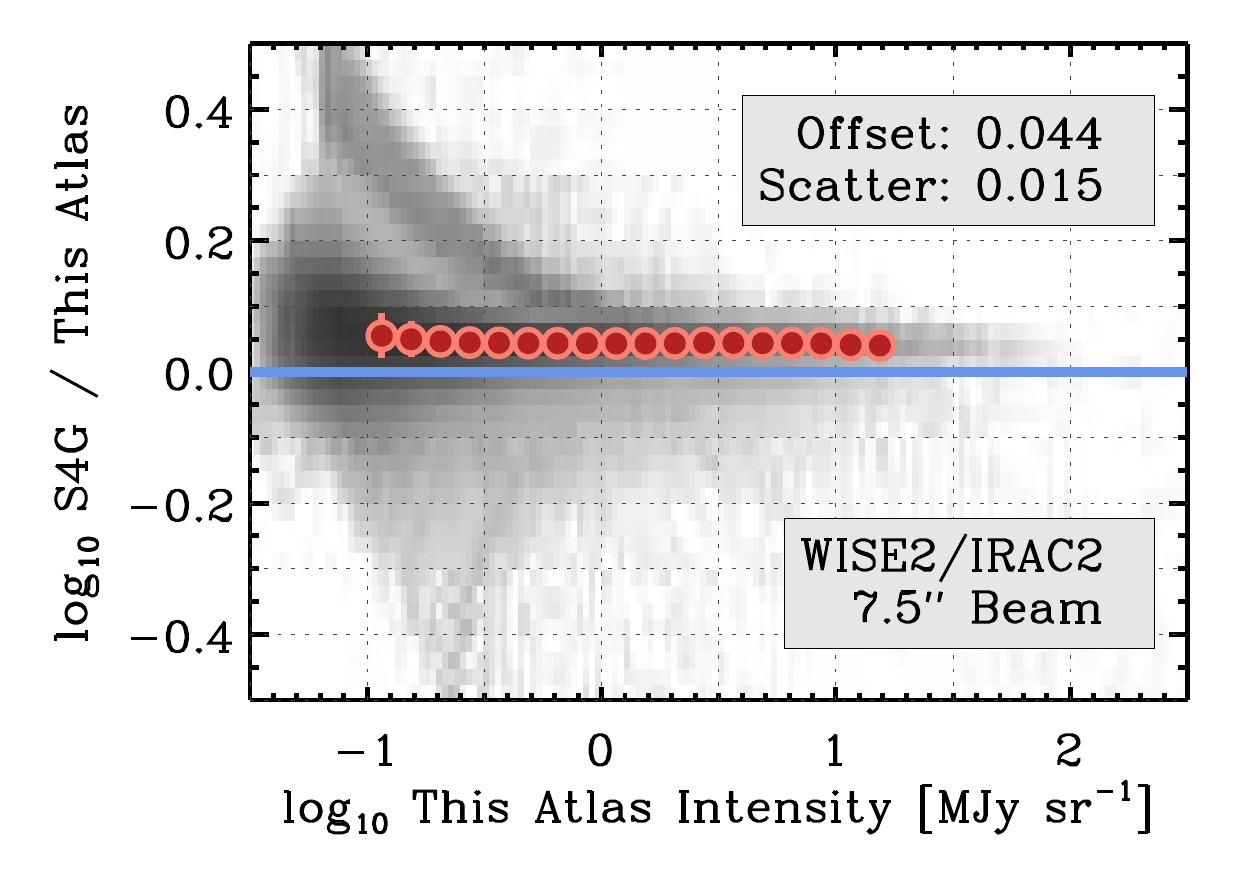}{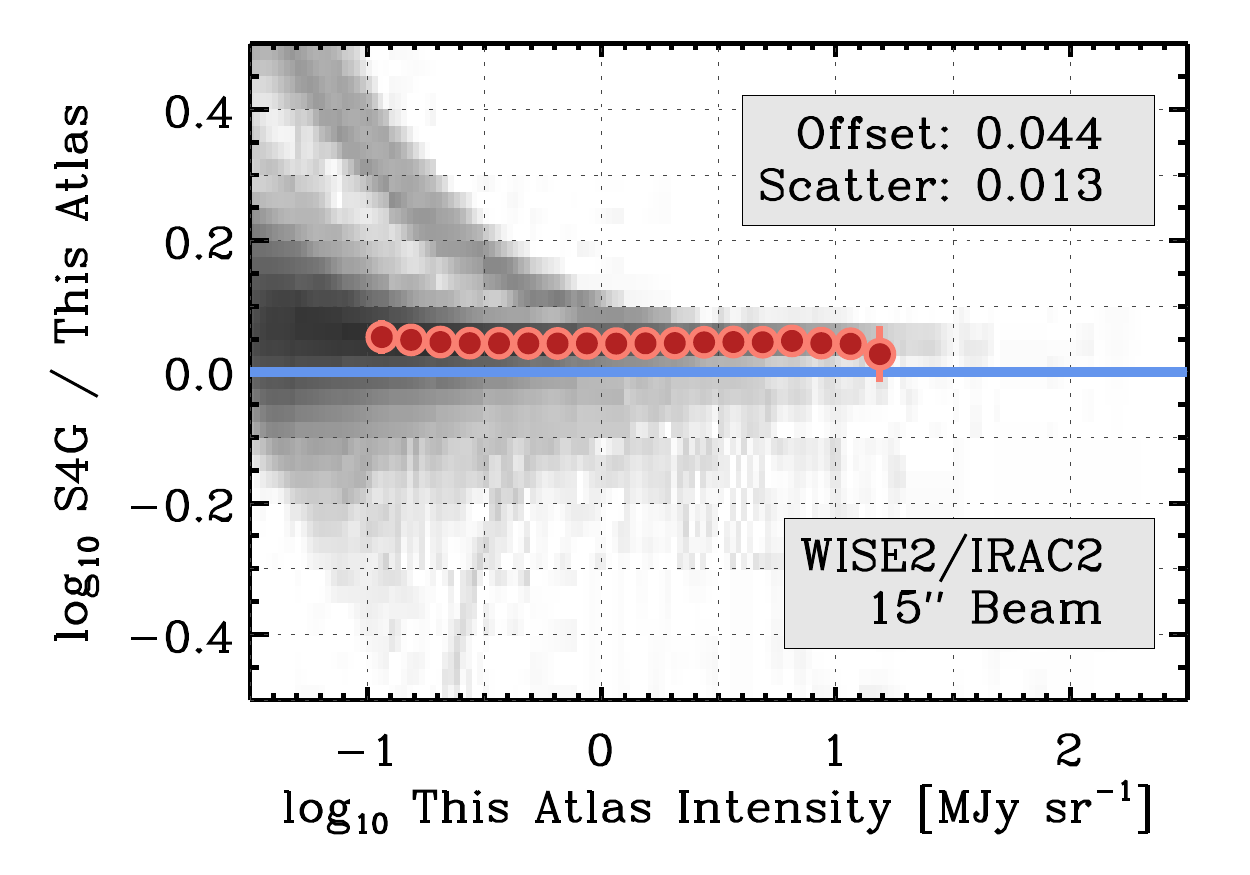}
\caption{{\bf Validation of the Image Atlas Against Literature Maps.} Comparison of our atlas to the GALEX Nearby Galaxies Atlas \citep[{\em top},][]{GILDEPAZ07} and {\em Spitzer} S4G \citep[{\em bottom},][]{SHETH10}. Each panel shows intensity from the other survey divided by intensity from our atlas ($y$-axis, left panels) as a function of intensity from our atlas ($x$-axis). This comparison contains only lines of sight with signal to noise $>10$. Grayscale shows density of data in this space with 0.025~dex cells and a logarithmic stretch. Red points show the median and median absolute deviation-based scatter in $0.25$~dex wide bins. The panels report the median offset from unity and rms scatter about the median offset, both in dex.}
\label{fig:validation}
\end{figure*}

We validate our images by comparison to the GALEX Nearby Galaxy Atlas \citep[NGA,][]{GILDEPAZ07} and S4G \citep{SHETH10}. These surveys produced large sets of images for overlapping samples of galaxies using the same (NGA) or similar (S4G) bands. They allow us the opportunity to check the correctness of our images line-of-sight by line-of-sight across a large part of the atlas. We carry out similar benchmarks using integrated photometry below.

We convolve the publicly available GALEX NGA images to share our $7.5\arcsec$ and $15\arcsec$ resolution, align them to our astrometric grid, subtract a local background determined outside $r_{25}$, and then apply our correction for Milky Way extinction to those data. We perform the same steps, except for the extinction correction, for the S4G IRAC1 and IRAC2 images.

For both comparison surveys, the area to fit the local background is limited, restricted by the provided image size for the NGA and by the IRAC field of view for S4G. We use an outlier-rejecting median to fit the background, and note the mismatch between the aperture used to fit the background in our atlas and the comparison as a main source of uncertainty.

After these steps, we compare the other surveys to our atlas. We consider all pixels within $r_{25}$ with S/N greater than $10$ in our our atlas. Then for every pixel we calculated the ratio of intensity in the other atlas to intensity in our atlas. This gives us a point-by-point measurement of how well we reproduce the intensities measured by other programs.

We plot the results of this comparison in Figure \ref{fig:validation}. We omit M31 and M33 from the comparison to the NGA because their large area translates to an overwhelming number of pixels, which would dominate the comparison. Each panel shows the ratio of intensity in the other atlas to our atlas for individual pixels. The grayscale shows data density on a logarithmic stretch. The blue points show the median and robustly-estimated $1\sigma$ scatter. The arc-like features visible in the S4G comparisons represent cases where the background does not perfectly agree between the two images. 

The diagonal cut through the data in the bottom left part of the plot reflects the signal-to-noise cut applied to our atlas. Near this cut we expect a modest bias towards high values, i.e., our atlas will appear high relative to the comparison, due to the signal-to-noise cut applied to our data. The comparison has a large dynamic range and we expect this bias to play only a minor role.

Overall the figure demonstrates a good match between our images and previous work at both resolutions. Our images match the GALEX NGA images in both bands with mean offset $\sim 5\%$. We find point-to-point scatter $\sim 20\%$ at $7.5\arcsec$ resolution and $10\%$ at $15\arcsec$ resolution. 

Our WISE Band 1 and 2 images also track the IRAC 1 and IRAC 2 images from S4G well, with less than $\sim 5\%$ scatter at both resolutions. The small median offset between WISE2 and IRAC2 reflects differences in the calibration (IRAC is calibrated using point sources and $14''$ apertures), bandpass, and imperfect knowledge of the PSF. In fact, WISE1 and IRAC1 should not agree as well as they do, given that the two instruments have different bandpasses and central wavelengths. In both cases, we view the small scatter as the key metric. Our atlas tracks the IRAC intensities measured by S4G very well.

The prominence of the arcs in the lower panels of Figure \ref{fig:validation} show that mismatched background subtraction causes much of the scatter that we do observe.

\subsection{Validation Against Known Stars: Saturation Limits and PSF Shape}
\label{sec:starstacks}

\begin{figure*}
\centering
\includegraphics[width=0.45\textwidth]{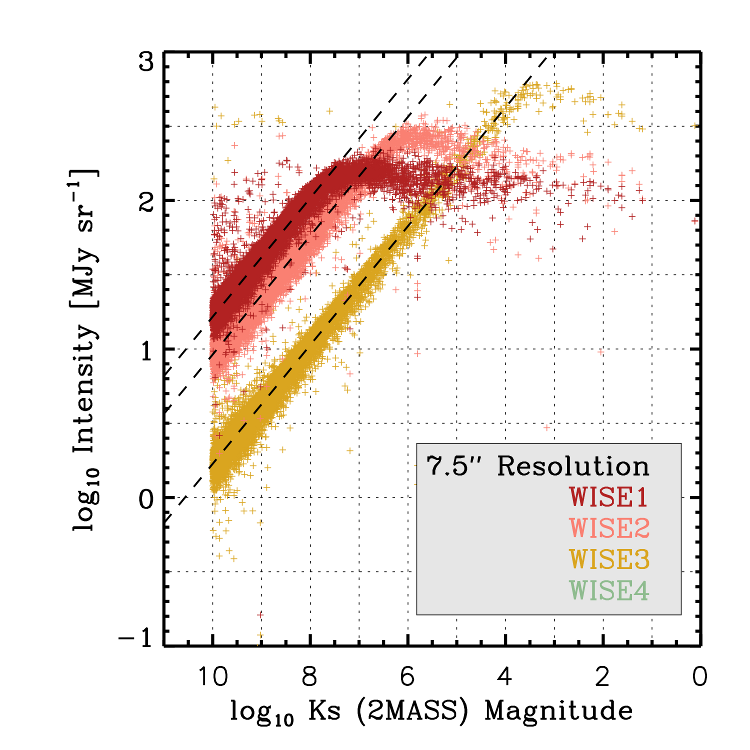}
\includegraphics[width=0.45\textwidth]{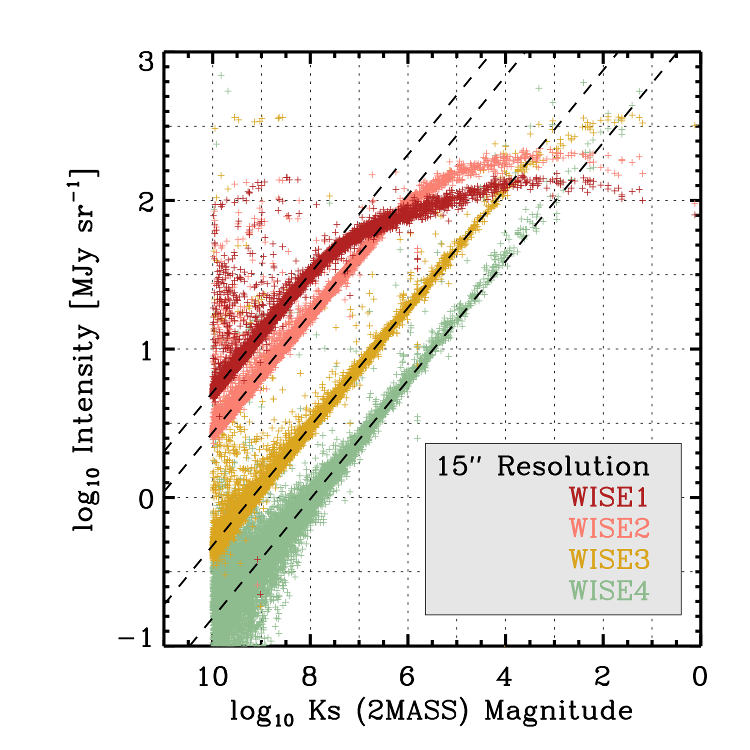}
\includegraphics[width=0.45\textwidth]{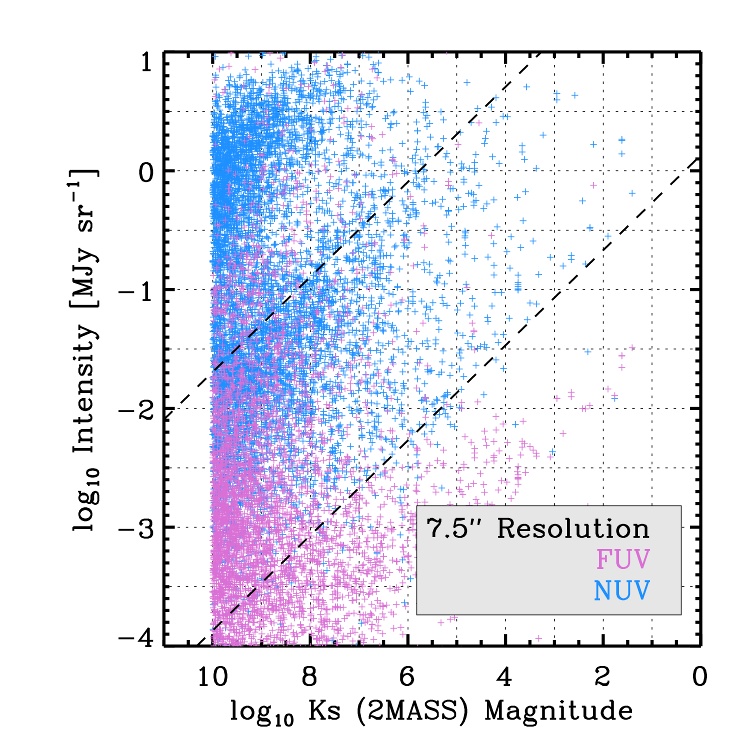}
\includegraphics[width=0.45\textwidth]{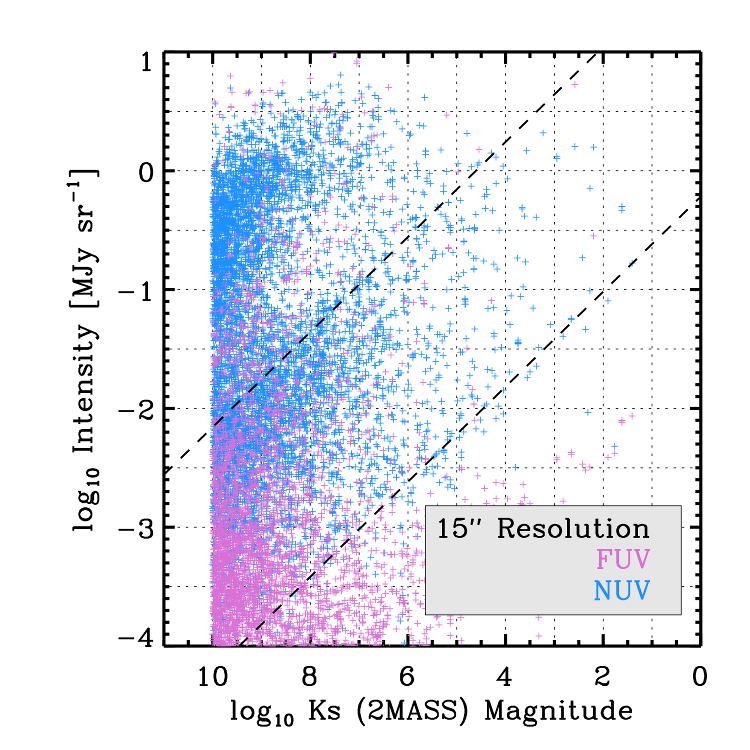}
\caption{{\bf Bright Stars in the Atlas.}  intensity measured in our atlas ($y$-axis) at $7.5''$ (\textit{left}) and $15''$ (\textit{right}) resolution in the WISE (\textit{top}) and GALEX (\textit{bottom}) as a function of the 2MASS $K_S$ magnitude for bright stars. We plot values for stars brighter than $10$~mag in the first $1,000$ images in our atlas. Dashed lines show the expectation for a fixed color stars (meaning fixed $K_S$-to-WISE or GALEX color). Saturation effects are evident in WISE1, WISE2, WISE3 at high intensities. The highly variable UV-to-near-IR color of stars is clear from the bottom panels.}
\label{fig:stars_unwise}
\end{figure*}

\begin{figure*}
\centering
\plottwo{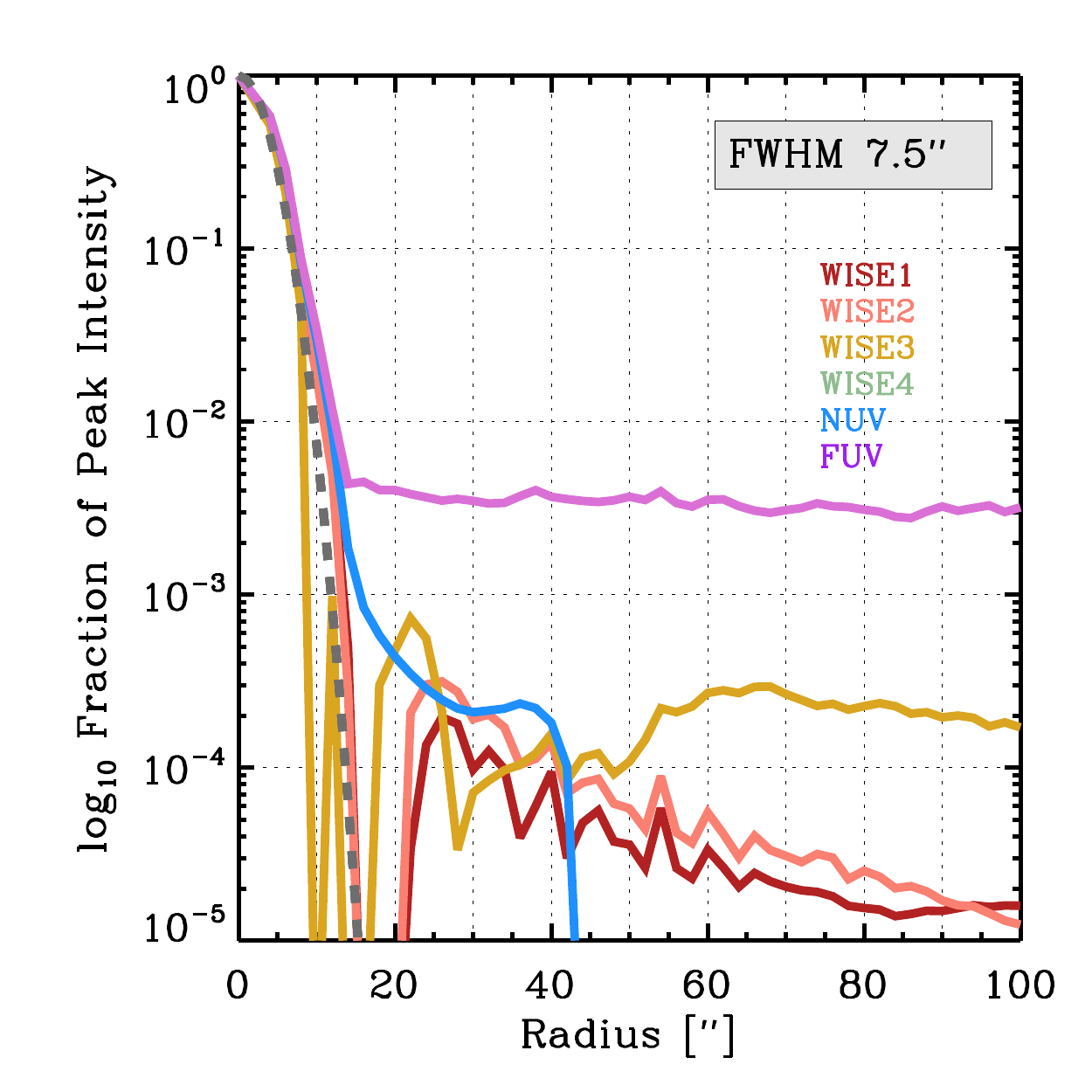}{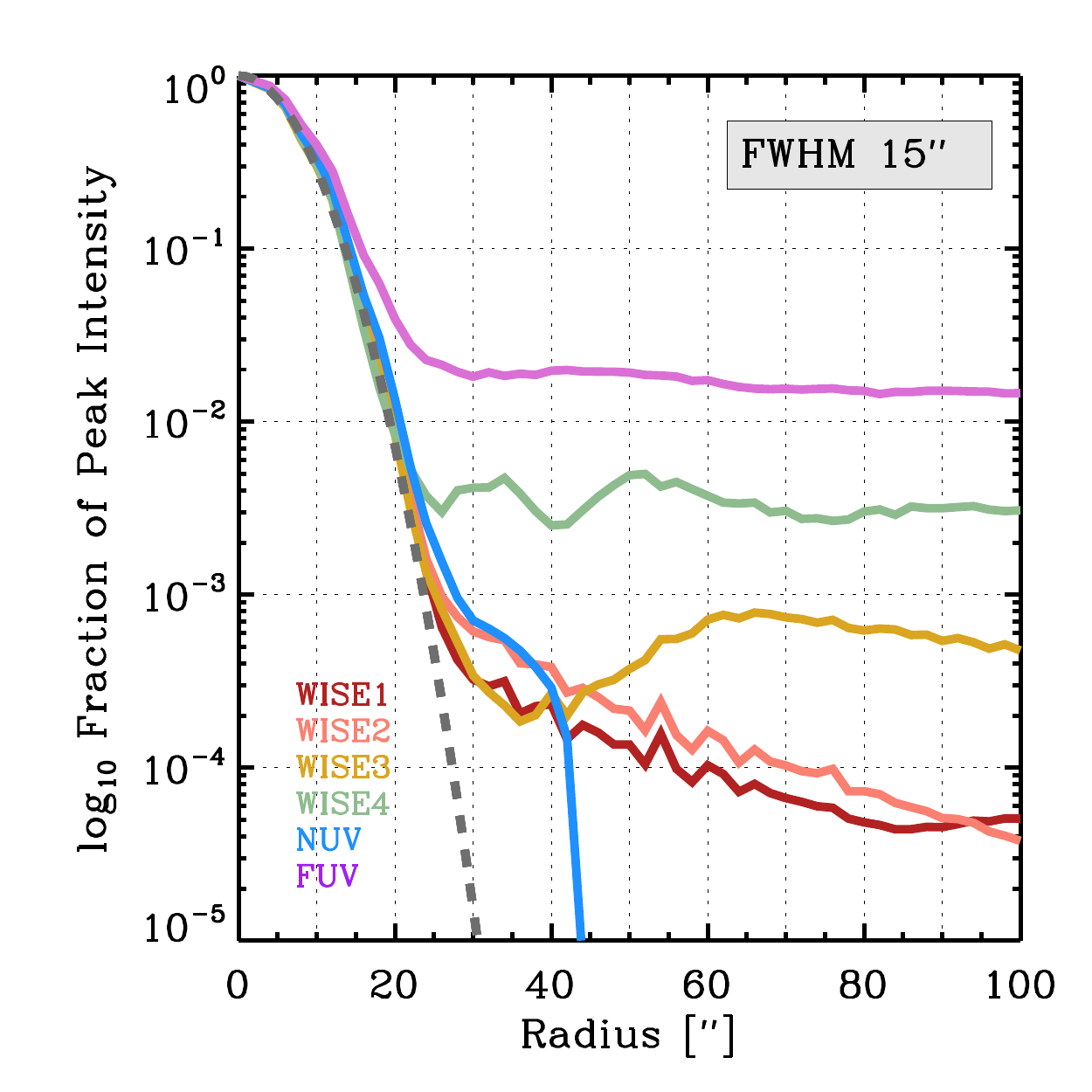}
\caption{{\bf Beam Profiles Based on Stacks of Bright Stars.} The median profile  of bright stars in the first 1000 images of the atlas. We normalize each cutout star image by the peak value, so that the profiles should reflect the average shape of a point source in the image. Gray dashed lines show the expected Gaussian PSF, which is a good description down to $\sim 10^{-3}$ of the peak. Below this artifact, including diffraction spikes and negative bowls in the NUV, dominate the shape of the PSF. The cutouts are extracted from the wide area images, which are not background subtracted. This affects the profiles in the FUV and WISE4 bands, where we can only say that the PSFs appear Gaussian down to the background level.}
\label{fig:beams}
\end{figure*}

We check the brightness and profile of known, bright stars in our atlas. For this exercise, we consider 2MASS stars with $K_S$ magnitude brighter than $10$~mag. We extract cutouts around each star from the full area images (i.e., before clipping to a smaller cutout around the galaxy) and before background subtraction, which occurs after the images are trimmed to a smaller size. The full field of view offers access to more stars and a wider view around each star.

\textbf{Intensity scaling and saturation:} Figure \ref{fig:stars_unwise} shows how the peak intensity at the location of a star in our atlas ($y$-axis) relates to the 2MASS catalog magnitude ($x$-axis). We plot results for stars from the first thousand images in our atlas, with each point representing one bright star. Dashed lines show the expected behavior if all stars have the same color (e.g., WISE1-to-$K_S$). This should be a reasonable approximation for WISE (top panels) but not GALEX (bottom panels).

The figure shows good agreement between WISE and 2MASS at low intensities. We see the expected behavior that stars are fainter at WISE2, much fainter at WISE3, and very faint at WISE4. For stars fainter than $\sim 7$~mag in $K_S$ the relationship between $K_S$ brightness and our atlas appears quite linear. 

Above this value, we see saturation effects. The plot appears consistent with the observatory-provided saturation limits (in Vega magnitudes) of $8.1$~mag and $6.7$~mag for WISE1 and WISE2. In our atlas, we expect saturation effects above $\sim 100$~MJy~sr$^{-1}$ in WISE1 and WISE2 at $7.5''$ resolution, a bit below this value at $15''$ resolution. Saturation appears at $\sim 300$~MJy~sr$^{-1}$ at WISE3. We do not detect clear saturation for stars at WISE4, likely because stars appear faint at this band. We also adopt the WISE3 $300$~MJy~sr$^{-1}$ saturation limit for WISE4. 

The atlas index includes and image headers include a flag indicating whether any pixel intensities in the image exceed these saturation threshold.

The figure also shows that for GALEX, the brightness of foreground stars tends to be low in the FUV and highly variable in the NUV.

\textbf{Stacked PSF:} We also stack the cutouts around bright stars to check the shape of the beam in the atlas. To do this, we normalize the cutout image around each star by the peak value, calculated at the location of the star. Then we take the median over all bright stars ($\sim 8,000$) in the first $1,000$ targets to check the shape of our PSF. The resulting profiles appear in Figure \ref{fig:beams}.

The figure shows that our data are reasonably described by a Gaussian PSF down to a contrast of $\sim 1000$-to-$1$. Signal to noise of $\gtrsim 1000$ does occur in the atlas and for these brighter sources, asymmetries in the true PSF (we assume a symmetric PSF in the convolution) and shortcomings of the convolution kernel begin to emerge. Among the most notable features are diffraction spikes in the WISE1 and WISE2 images and a modest amount of negative bowling around bright NUV sources.

\subsection{Notes on the Atlas}

Figures \ref{fig:images_1} through \ref{fig:beams} illustrate a few key points for users of the image atlas:

\begin{enumerate}

\item Our atlas matches previous work well when compared at matched resolution with approximately matched background treatment. Based on this comparison, we suggest to adopt a multiplicative uncertainty of a few percent ($\sim 0.05$~dex). The zero point of the images still represents a significant uncertainty, which will be correlated across the whole galaxy. We suggest to treat the zero point as systematically uncertain by $\pm 1{-}2 \times 10^{-4}$~MJy~sr$^{-1}$ for GALEX and $\pm 2 \times 10^{-3}$~MJy~sr$^{-1}$ for WISE.

\item The PSFs are reasonably Gaussian and the images match 2MASS catalog magnitudes well for bright stars. Saturation effects are visible at about the expected levels, which translate to $\sim 100{-}300$~MJy~sr$^{-1}$ at our resolution. Non-Gaussianity in the PSFs is visible at levels $\sim 10^3$ lower than the peak. This can be significant in bright galaxies --- e.g., such effects are visible in M77 (NGC 1068) and M82 in Figures \ref{fig:images_1} --- \ref{fig:images_3}. This can also be important when bright stars are near the galaxy.

\item Foreground stars are visible in all images at WISE1. At low $|b|$ they contribute a major fraction of the light in the images. We provide masks indicating the likely footprint of bright stars, but these should be applied with care to avoid masking bright regions in the target galaxy.
\end{enumerate}

Bearing these caveats in mind, Figures \ref{fig:images_1} -- \ref{fig:validation} also highlight the remarkable legacy of GALEX and WISE. We have matched resolved infrared and ultraviolet maps covering the full extent of thousands of galaxies. From only these images, one can pick out stellar bulges, nuclear starbursts, bars, star-forming spiral arms, and variable levels of dust obscuration and star formation activity.

\section{Integrated Photometry}

\begin{figure*}
\centering
\plottwo{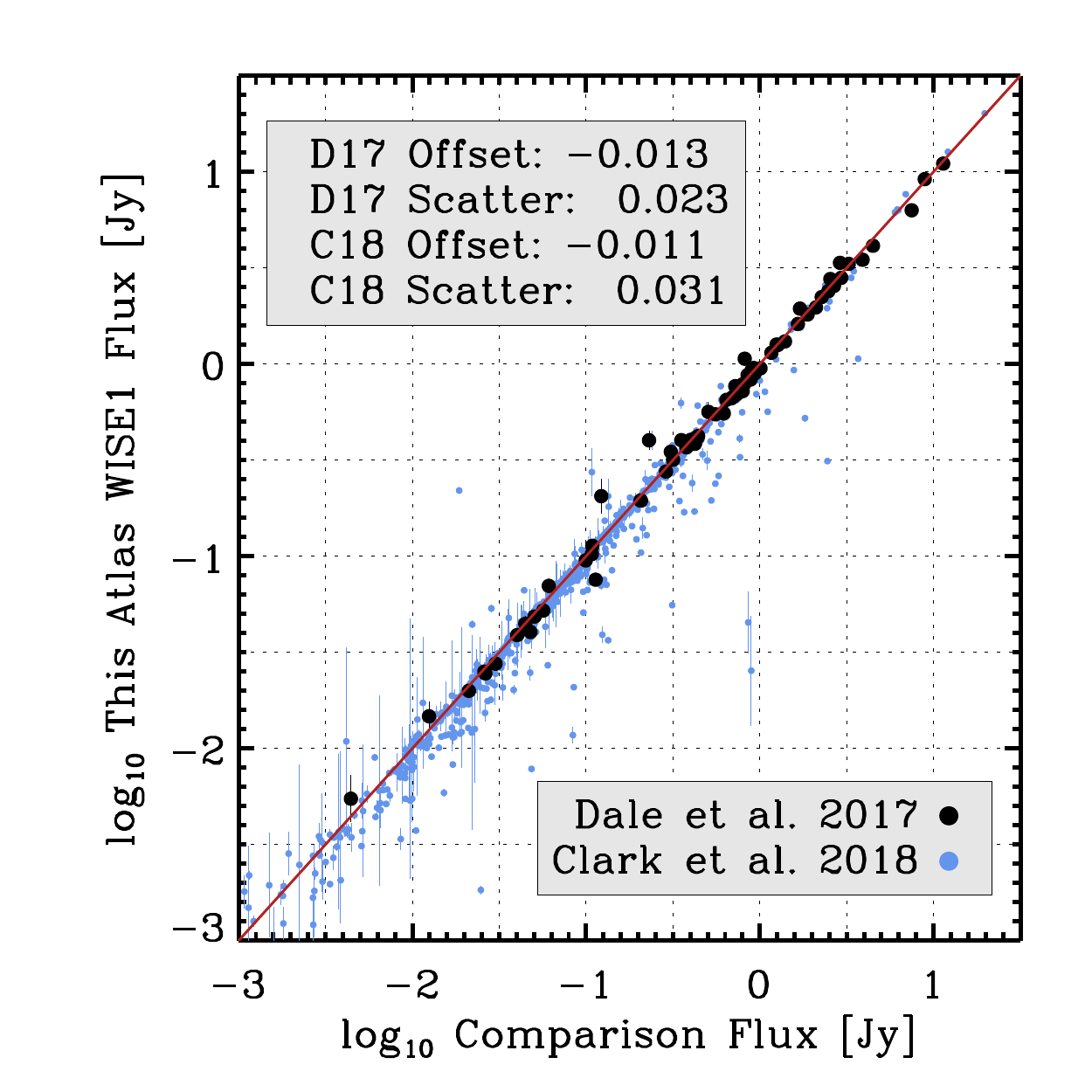}{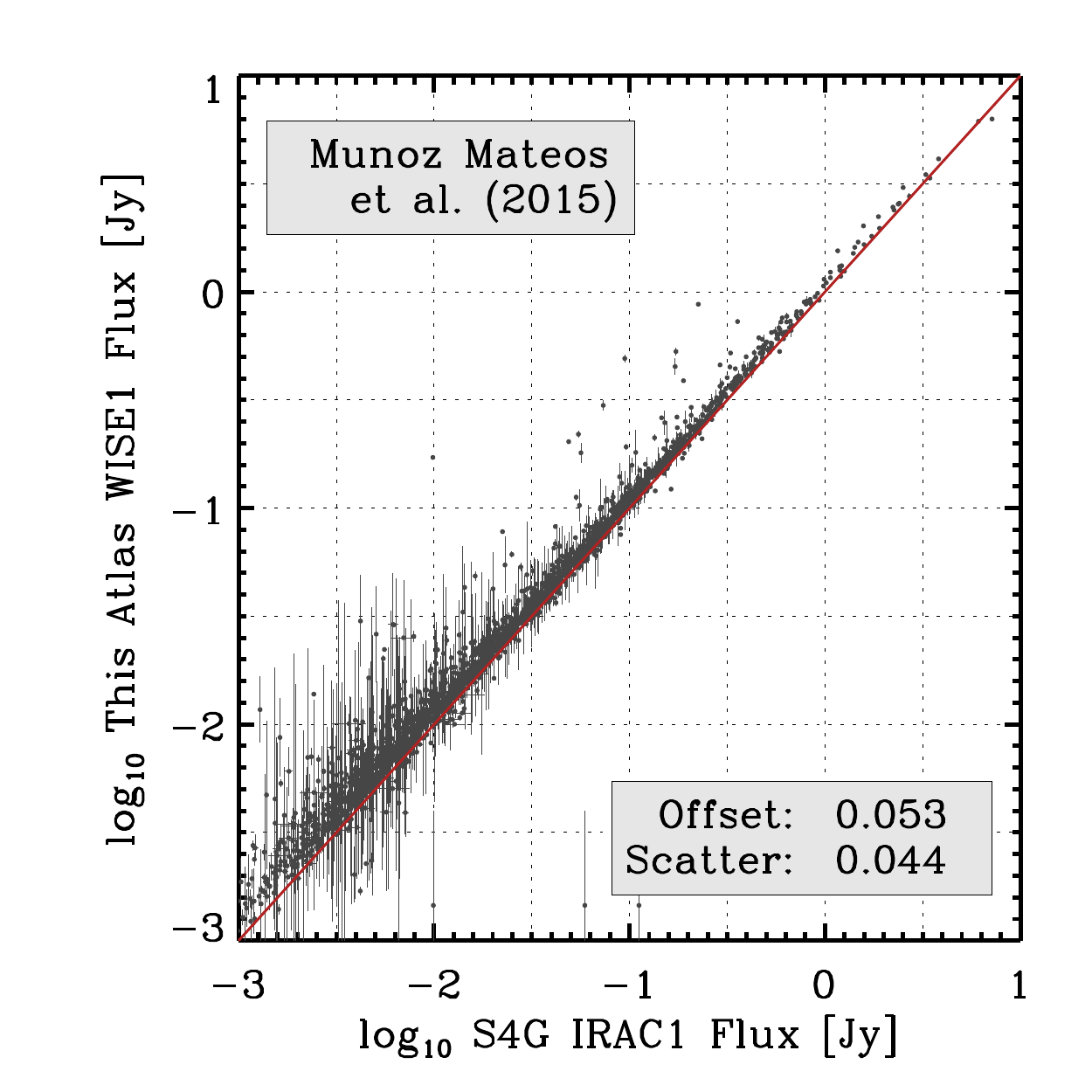}
\plottwo{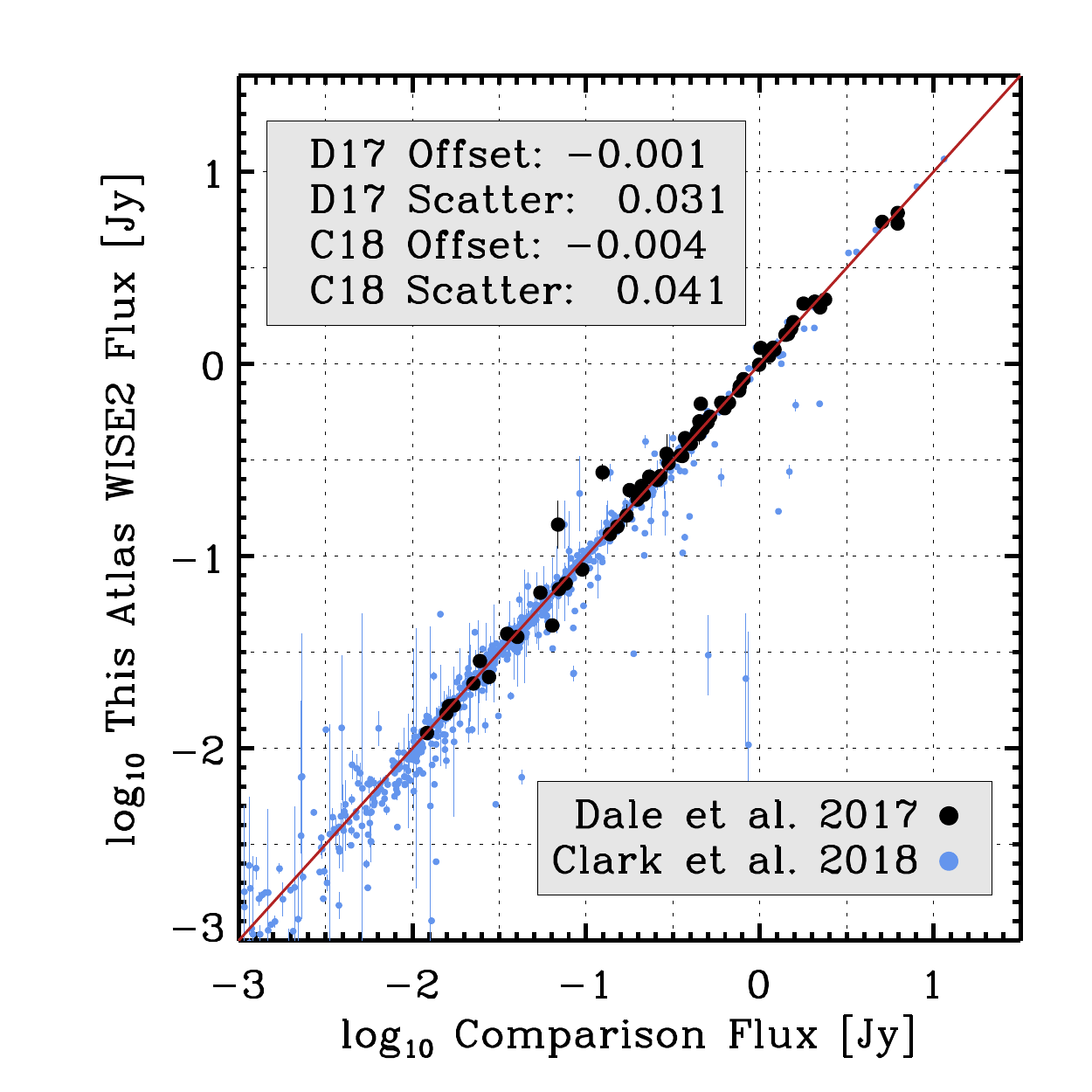}{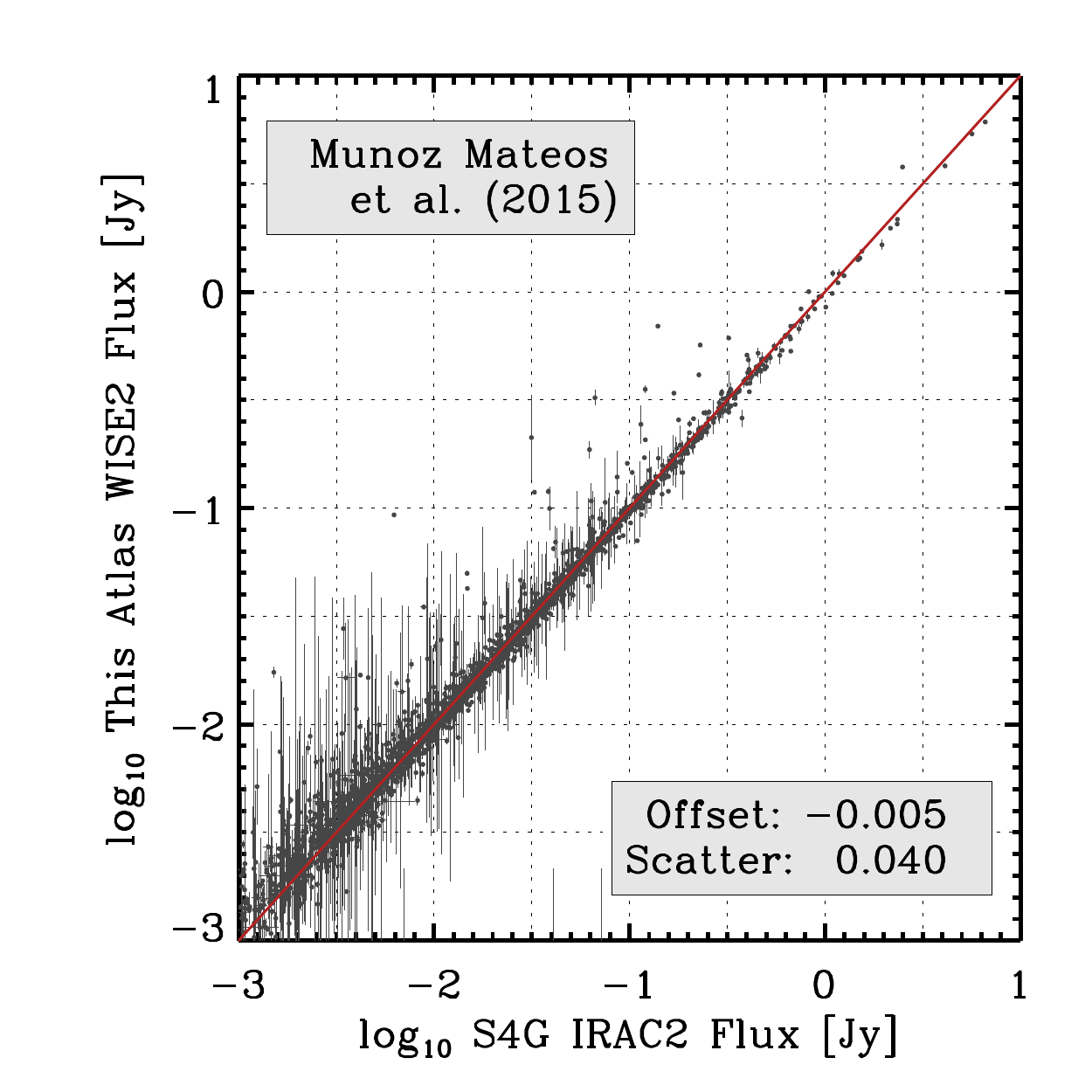}
\caption{{\bf Integrated Photometry vs. Literature Measurements 1.} Our WISE1 ({\em top}) and WISE2 ({\em bottom}) flux measurements ($y$-axis) compared to those from ({\em left}) \citet[][black]{DALE17} and \citet[][color]{CLARK18} and ({\em right}) \citet{MUNOZMATEOS15}. In each panel, we report the median ratio (in dex) and robust robustly estimated scatter (also in dex) dividing our flux by the literature flux.}
\label{fig:phot_check_1}
\end{figure*}

\begin{figure*}
\centering
\plottwo{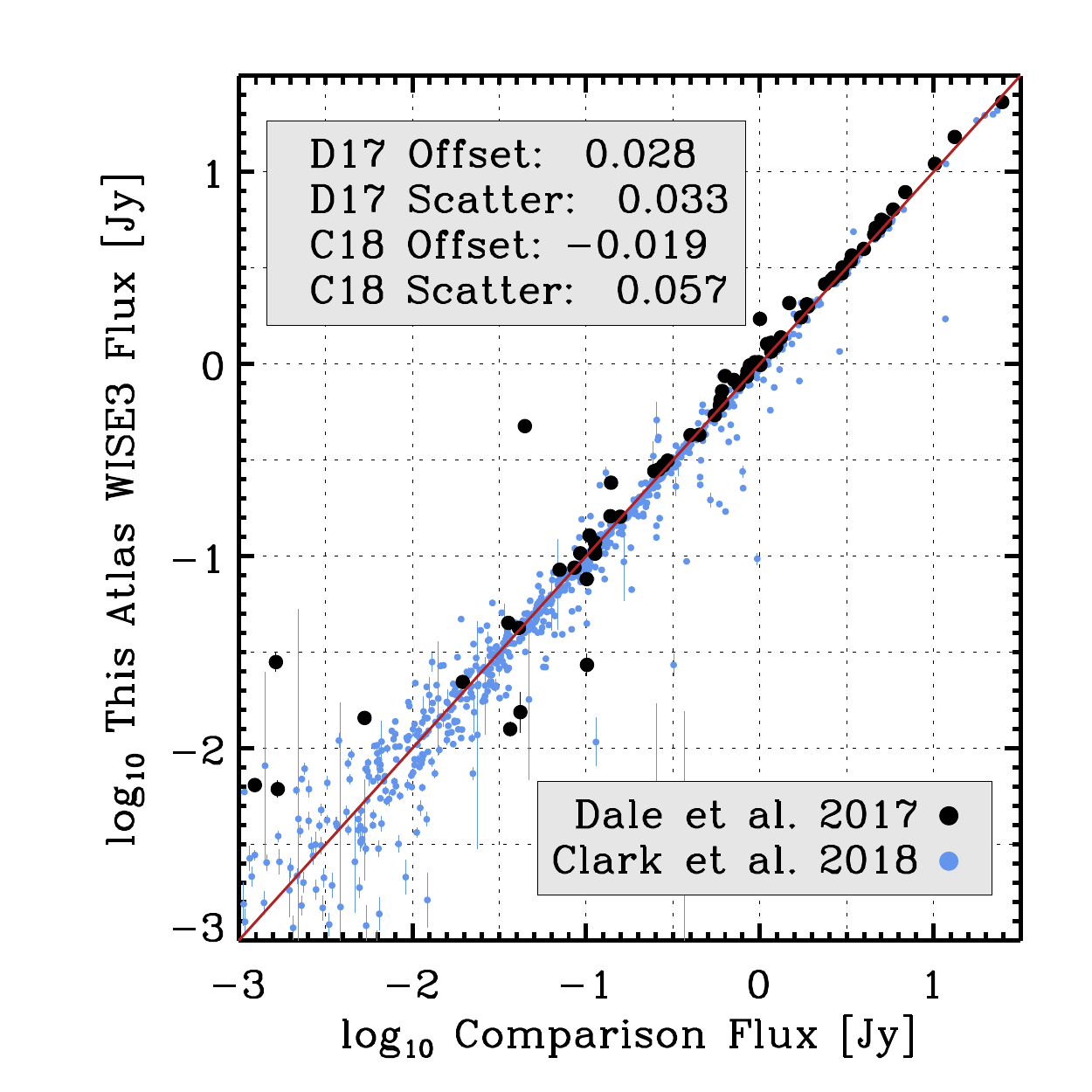}{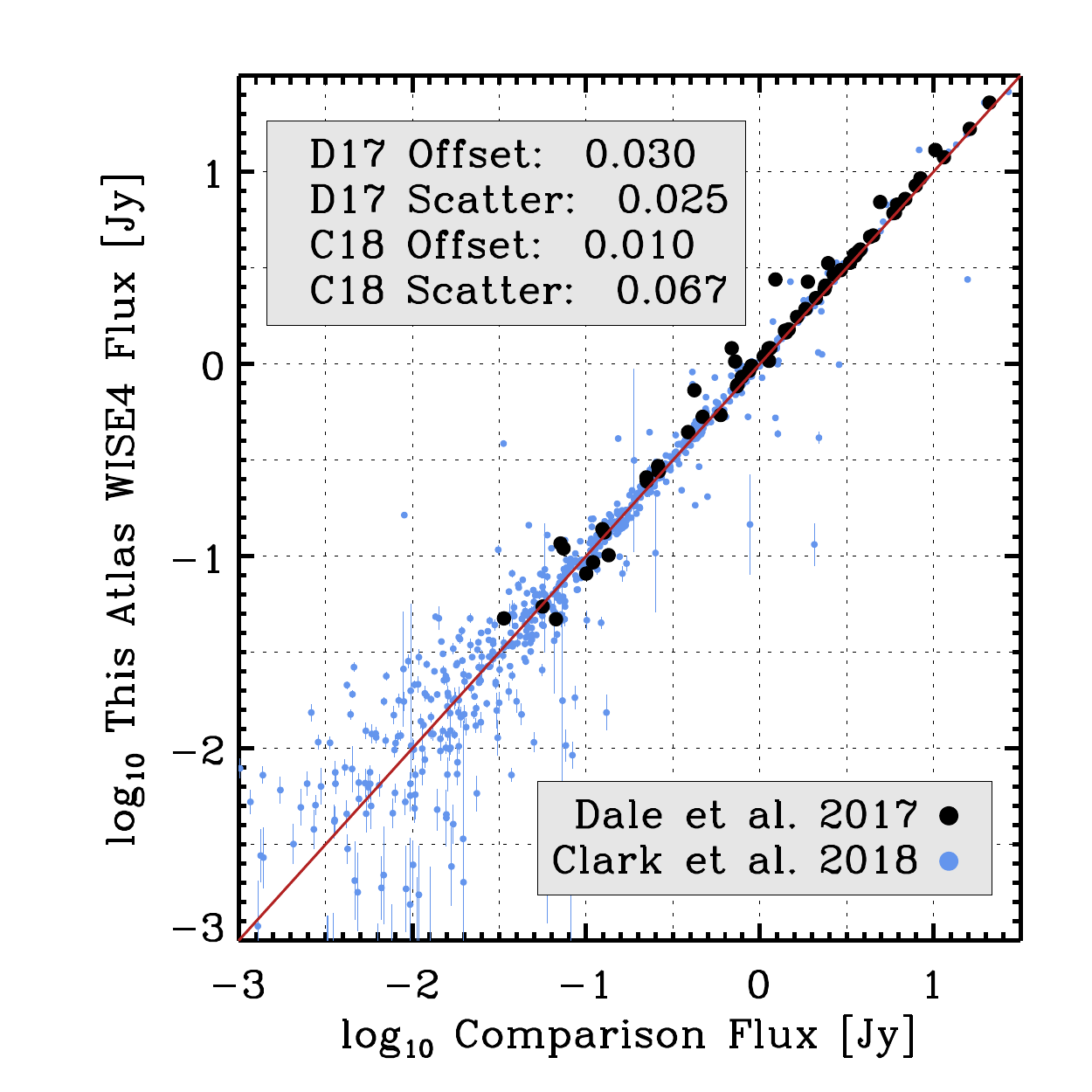}
\plottwo{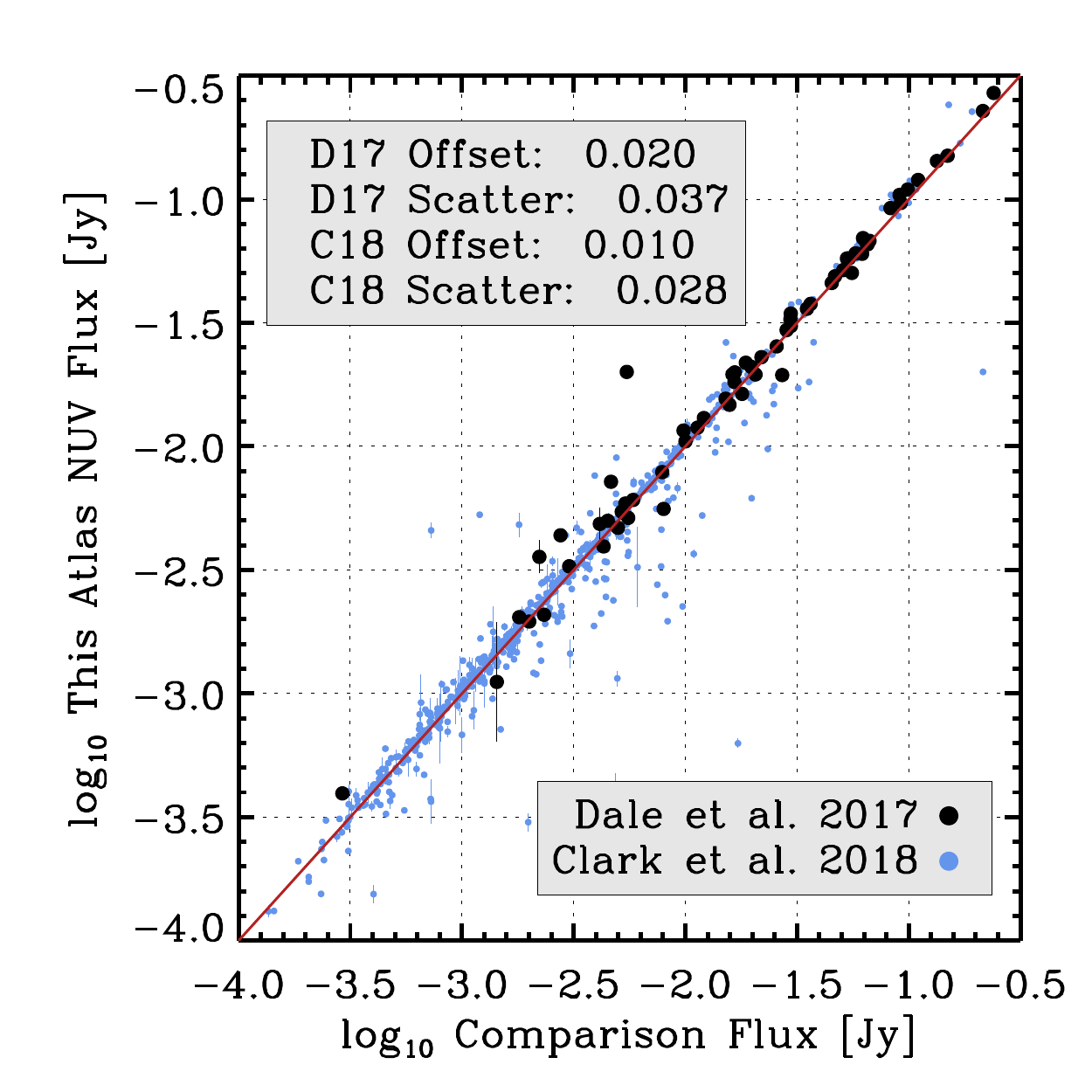}{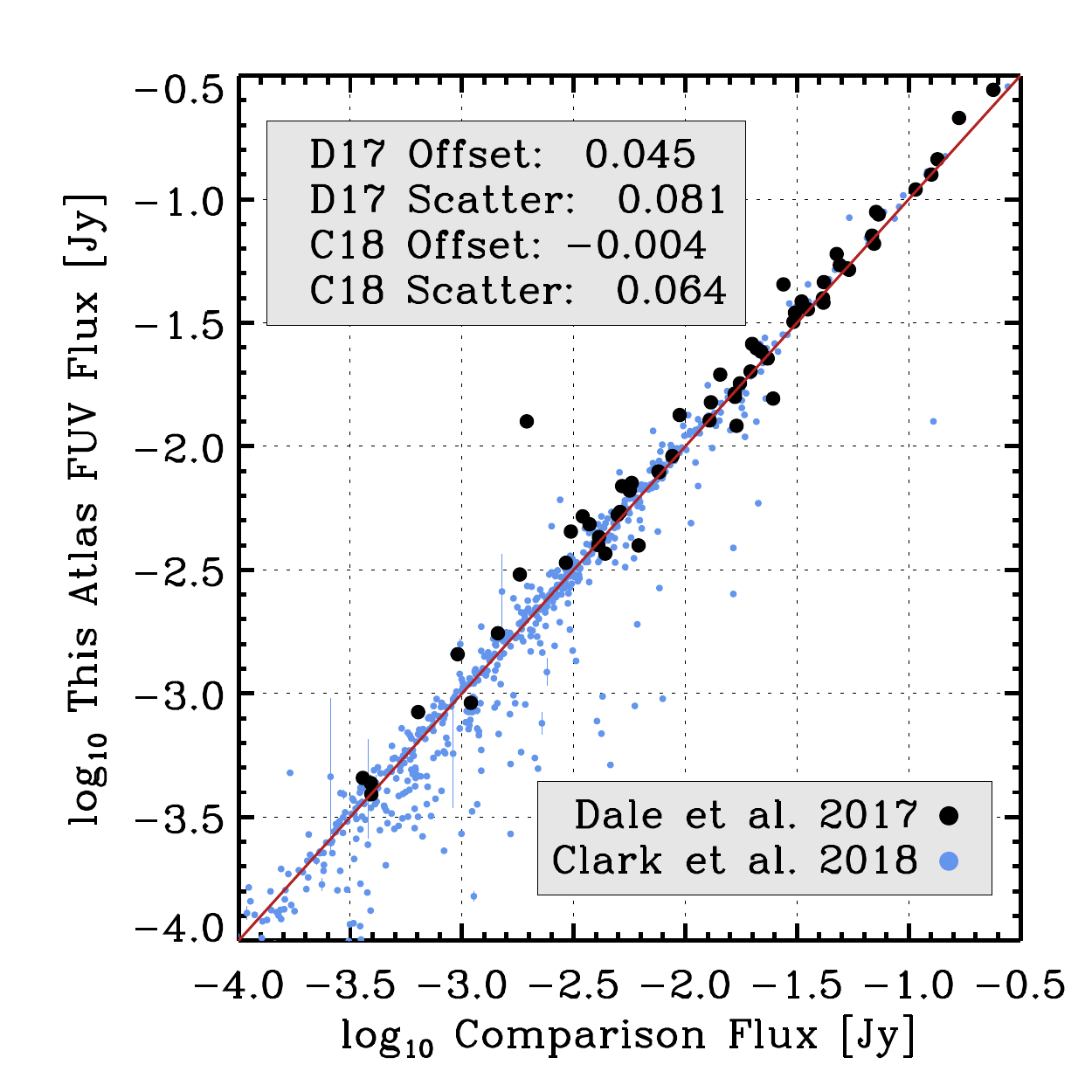}
\caption{{\bf Integrated Photometry vs. Literature Measurements 2.} Our WISE3 ({\em top left}), WISE4 ({\em top right)}), FUV ({\em bottom left}), and NUV ({\em bottom right}) flux measurements ($y$-axis) compared to those from \citet[][black]{DALE17} and \citet[][color]{CLARK18}. In each panel, we report the median ratio (in dex) and robust robustly estimated scatter (also in dex) dividing our flux by the literature flux}
\label{fig:phot_check_2}
\end{figure*}

From the image atlas, we estimate the integrated the flux from each target at each band. In order to minimize the effect of foreground stars, we carry out our fiducial photometry at $7.5''$ resolution for FUV, NUV, WISE1, WISE2, and WISE3. We carry out the WISE4 photometry on the $15''$ resolution data. To do this, we:

\begin{enumerate}
\item {\bf Construct ellipsoidal apertures.} We consider a series of ellipsoidal apertures. Our fiducial semimajor axis is the $25^{\rm th}$ magnitude $B$-band isophote, $r_{25}$, or $30''$, whichever is greater. As discussed above, we adopt the LEDA position and inclination angle, but set the inclination angle to $60^\circ$ in cases where the galaxy is more inclined than this. If either the inclination or the position angle is unknown (true for $\sim 10\%$ of galaxies), we adopt $0^\circ$ for both.

\item {\bf Apply masks.} We apply our masks to the images. We blank regions outside one-half the fiducial radius, i.e., $0.5~r_{25}$, that we identified as stars or overlapping galaxies. We do not blank regions inside this radius, because we found that the star identification, even in 2MASS, often breaks down and identifies nuclear features associated with the galaxy as stars. We replace the blanked values with the mean or median value at that radius when calculating the flux.

\item {\bf Build binned profiles.} We calculate the mean and median intensity in bins of fixed galactocentric radius. These bins have width $3.75''$, i.e., half the beam size for the $7.5''$ version of the atlas. We calculate binned intensities out to four times the fiducial semimajor axis. We exclude masked pixels from the calculation of the profiles.

\item {\bf Integrate the profiles out to $2~r_{25}$. For WISE1 and WISE2 we use the median profile instead of the mean from $0.75{-}2~r_{25}$.}  To calculate the flux, we sum the binned profile, equivalent to integrating the image within $2~r_{25}$. That is, in each ring over this range, we take the mean intensity, multiply it by the ring area, and add this to the flux. In this step, we assume that all pixels, even masked pixels, have the mean or median intensity in the bin. For WISE1 and WISE2, the galaxies appear smooth while contributions from foreground stars remain a concern. To suppress the effects of these stars, we switch to use the median profile outside $0.75~r_{25}$ for these two bands. We expect this to bias us moderately low in galaxies with asymmetric extended structure or edge on galaxies; in exchange, contamination from stars will be dramatically suppressed. For WISE3, WISE4, NUV, and FUV, we use the mean profile throughout.

\item {\bf Apply extended source corrections to WISE.} The WISE calibration scheme requires a modest aperture correction to account for the fact that the default calibration scheme is optimized for point sources. Following the WISE documentation, we multiply the calculated fluxes by $0.97$, $0.96$, $1.03$, and $0.97$ at WISE1, WISE2, WISE3, and WISE4.
\end{enumerate}

\textbf{Validation:} We compare our results to the photometry by \citet{MUNOZMATEOS15}, \citet{DALE17}, \citet{CLARK18}. \citet{MUNOZMATEOS15} determined apparent magnitudes for the S4G sample. Although the bandpass does not match WISE perfectly, as above this provides a large sample to benchmark against. \citet{DALE17} conducted careful, by-hand aperture definition and removal of foreground stars to derived integrated photometry for the KINGFISH galaxies \citep[see also][]{CLUVER17}. \citet{CLARK18} adopt a more automated approach, similar to what we do here. They target a much larger set of galaxies than \citet{DALE17}. \citet{DALE17} and \citet{CLARK18} target the same bands that we use, though they employ different versions of the maps.

Figures \ref{fig:phot_check_1} and \ref{fig:phot_check_2} shows a good match between our photometry all three literature sources for WISE1 and WISE2. Because the bandpasses differ, a mild average offset from the {\em Spitzer} data can be expected. The scatter in the ratio of our measured flux to the literature value appears small, $\lesssim 10\%$, for all comparison samples. We quote the robustly estimated scatter, which does suppress a few strong disagreements. In the case of the S4G comparison, these almost all reflect bright stars overlapping the galaxy.

We find similar good agreement between our measurements, \citet{DALE17}, and \citet{CLARK18} for all bands. We find systmematic offsets less $\sim 10\%$ ($0.045$~dex) for all bands and often $\lesssim 5\%$ ($0.02$~dex). For individual galaxies, we find robustly estimated scatter $0.03{-}0.1$~dex depending on the band. For both this comparison and the check against S4G, the residuals correlate across bands for individual galaxies. This indicates that methodology drives the low level differences among the measurements.

Overall, Figures \ref{fig:phot_check_1} and \ref{fig:phot_check_2} provide another validation of the image atlas, and confirms that our integrated photometry yields a robust set of host galaxy properties to place measurements for individual lines of sight in context.

\section{Local Galaxies on the Star Forming Main Sequence}\label{sec:sfms}

\subsection{Estimating M$_\star$ and SFR}

We aim to place local galaxies in the context of the full galaxy population, as studied at larger distances using bigger samples. As a first step, we use our integrated photometry to estimate the stellar mass and star formation rate of our targets.

The Appendix describes how we do this. In brief, we leverage the GALEX-SDSS-WISE Legacy Catalog \citep[GSWLC][]{SALIM16,SALIM18}. That project combined GALEX and WISE photometry with SDSS observations. They carried out population synthesis modeling using the CIGALE code \citep{BOQUIEN19}. This yields high quality SFR and M$_\star$ estimates along with matched GALEX and WISE photometry for a sample of $\sim 130,000$ galaxies. In the Appendix, we use these data to calibrate recipes for estimating the stellar mass and SFR in our sample using GALEX and WISE data. These recipes have the advantage of a much larger training set compared to previous work on the topic. They also place our local targets on the same system as the GSWLC, and so also the SDSS main galaxy sample. This allows a ready comparison of local galaxies to the full galaxy population.

Following the recipes in the Appendix, we estimate M$_\star$ and SFR for galaxies in our sample. We note several key points from the appendix here:

\begin{enumerate}
\item We adopt a variable WISE1 mass-to-light ratio, \mtolwise . The GSWLC and other population synthesis results strongly suggest a close link between \mtolwise\ and the age of the stellar population. Mechanically, this amounts to a close dependence of \mtolwise\ on specific star formation rate, SFR/M$_\star$ in our recipes. This variable \mtolwise\ is crucial to reproduce the SFR-M$_\star$ scaling relations seen in SDSS.
\item We strongly prefer linear hybrids of FUV and WISE4 or NUV and WISE4 to estimate SFR. The translation of WISE3 to SFR shows strong dependence on stellar mass and specific star formation rate, consistent with the heavy contribution of PAHs to that band. Meanwhile, using only WISE without GALEX leads to significant underpredictions of the SFR in low mass galaxies.
\item The GALEX-WISE hybrid SFR estimators work well for star forming galaxies, but different calibrations are required to treat quiescent galaxies. We do not focus on quiescent galaxies here, but caution that our SFR estimates for those galaxies will be overestimates. This manifests as an apparent, but somewhat artificial pile up of quiescent galaxies at $\log_{10}$ SFR/M$_\star \approx -11.5$~yr$^{-1}$.
\end{enumerate}

\noindent We justify these points and present related calculations in more detail in the Appendix. 

Table \ref{tab:z0mgs_sfr_mstar} presents physical parameter estimates for our sample. Following discussions throughout this paper, these estimates have several associated caveats. Distance estimates, masking of stars and galaxies, and physical parameter estimation all represent our ``best effort'' as of this publication but are likely to improve. 

\begin{figure*}
\plottwo{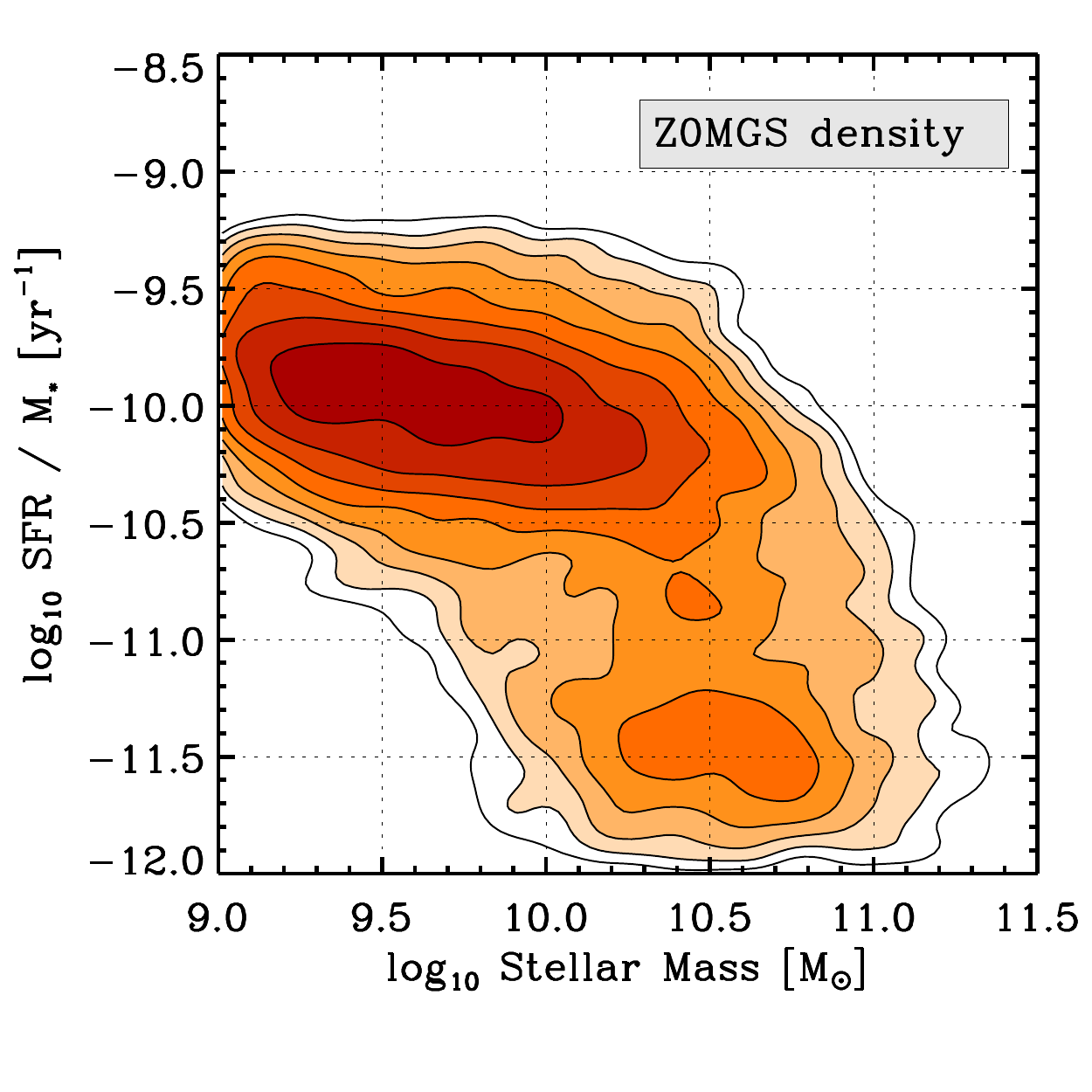}{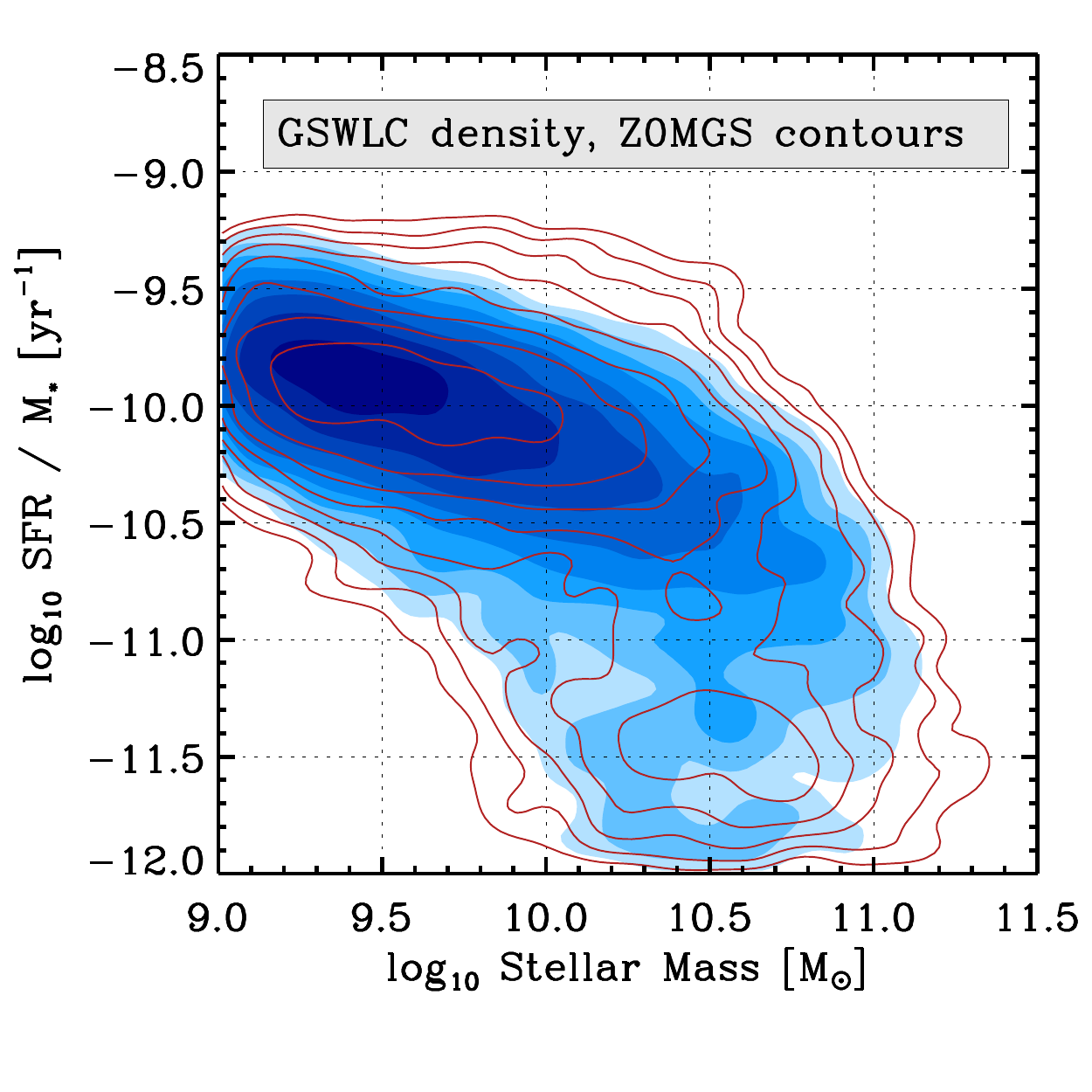}
\plottwo{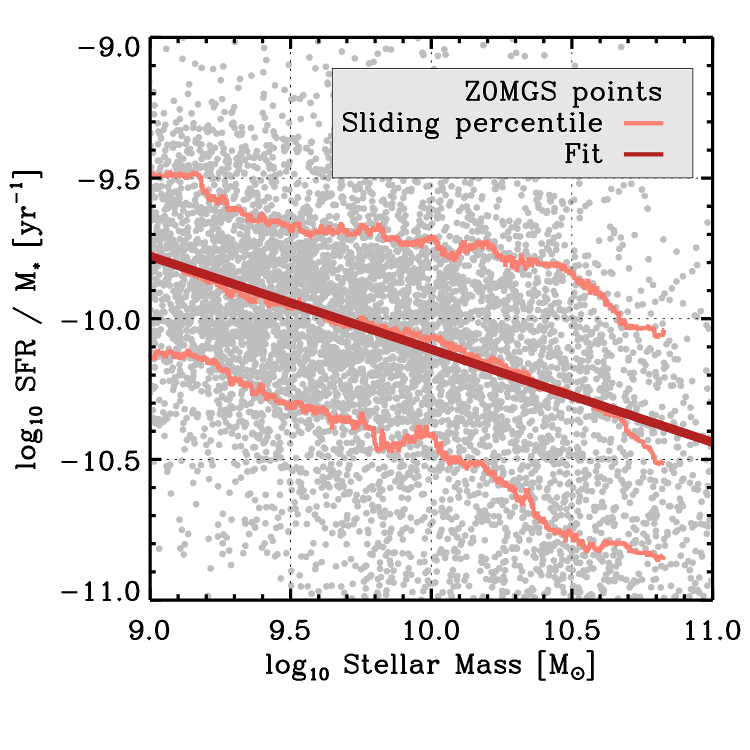}{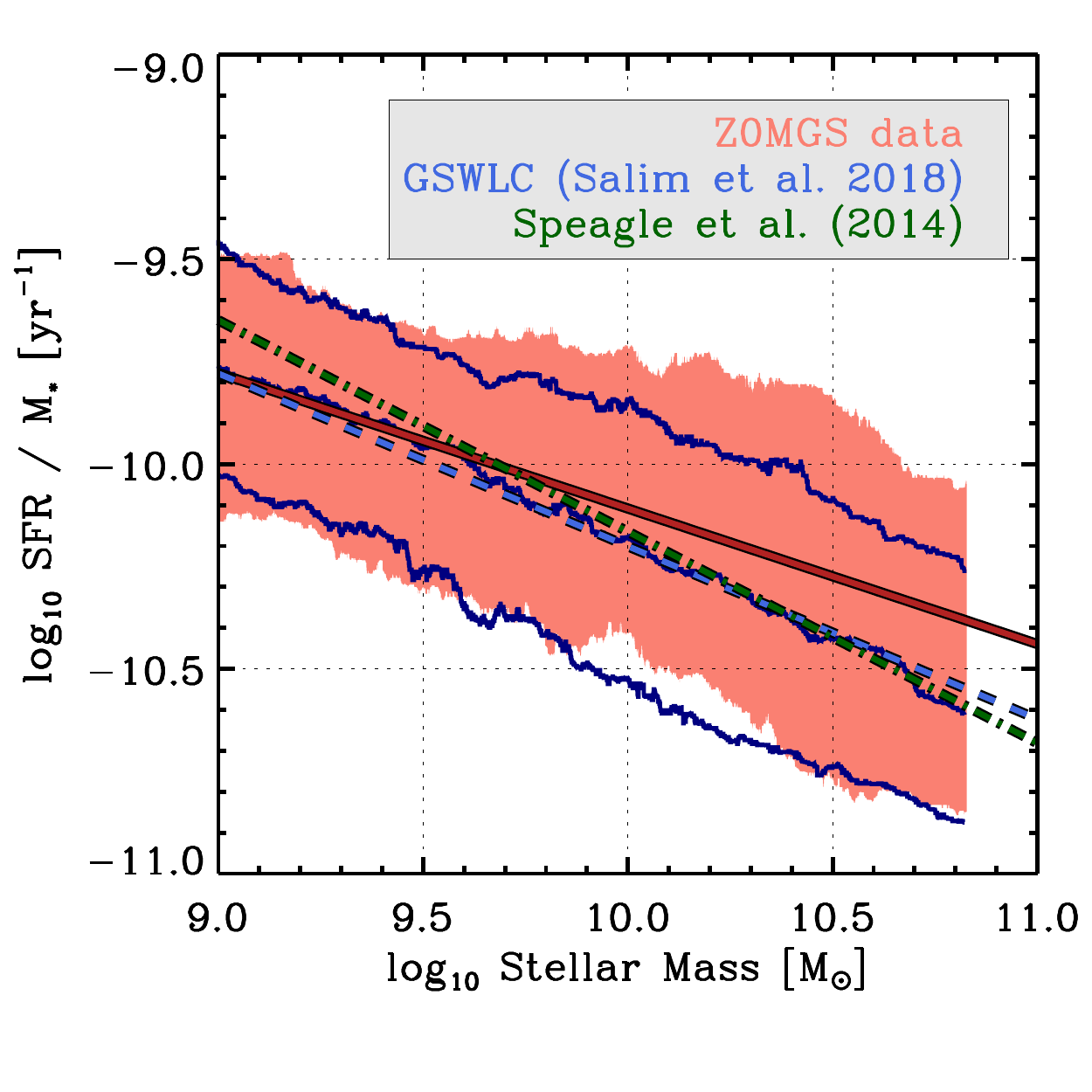}
\caption{{\bf Local Galaxies on the ``Star Forming Main Sequence.''} Galaxies from our atlas and the GSWLC \citep{SALIM16,SALIM18} in SFR/M$_\star$ vs. M$_\star$ space. ({\em top left}) Data density for the $\sim 11,000$ galaxies from our sample. Contours are spaced logarithmically and stepped by a factor of $2$ in data density. ({\em top right}) Data density for $\sim 90,000$ $z < 0.05$ galaxies from the GSWLC in the same parameter space. Contours show the local galaxy population from the left panel. Our estimates show a well defined star forming main sequence that agrees well with that seen in the larger SDSS (GSWLC) sample. We also see a significant population of quiescent galaxies, and an intermediate population of ``green valley'' targets. Our GALEX+WISE estimates tend to overestimate the SFR in quiescent ($\log_{10}$ SFR/M$_\star < -11$) galaxies, which are not the focus of this analysis. ({\em bottom left}) Individual points for star-forming galaxies in our our sample in gray, with running values of the $16$, $50$, and $84^{\rm th}$ percentile in 250 galaxy wide bins (light red). A solid red line shows our best fit to the median ($50^{\rm th}$ percentile) trend. This line represents our estimate of the local star forming main sequence. ({\em bottom right}) The $16$ to $84^{\rm th}$ percentile range for our sample (shaded light red) and the GSWLC (dark blue, also showing the median trend). We plot our fit to the star forming main sequence (dark red) along with fits to the GSWLC (blue, dashed) and the \citet{SPEAGLE14} redshift-dependent star forming main sequence extrapolated to $z=0$ (green dash-dotted).}
\label{fig:mainseq_wise}
\end{figure*}

About $1/3$ of our sample also lacks GALEX measurements. In these cases the WISE-only SFR represents an underestimate. As we show in the Appendix, this preferentially affects low mass galaxies, leading to significant underestimates of the SFR. We drop galaxies without GALEX from comparisons to the GSWLC, do not use them to fit the ``star forming main sequence'' below, and do not provide offsets from the star forming main sequence. In future work, we will use the calculations in the Appendix, along with other multiwavelength data, to improve the WISE-only SFR estimates. For now, we suggest to treat those estimates as incomplete and so lower limits. Based on the appendix and Figure \ref{fig:mainseq_wise}, we also suggest to treat estimates for quiescent galaxies as likely upper limits.

\subsection{The Local Main Sequence of Star Forming Galaxies}

Figure \ref{fig:mainseq_wise} shows the SFR and $M_\star$ estimates for our sample and compares them to $z < 0.05$ galaxies in the GALEX-SDSS-WISE Legacy Catalog \citep{SALIM16,SALIM18}. We show both samples in SFR/$M_\star$ vs. $M_\star$ space, a key parameter space for classifying galaxies \citep[e.g., see][]{BLANTON09}. We restrict this comparison to our targets with both WISE and GALEX data and to the $\sim 90,000$ GSWLC galaxies with good SED fits at redshift $< 0.05$. This redshift cut reduces bias in the GSWLC comparison sample, but selection effects may still play an important role.

The top left panel shows the data density of our targets in SFR/M$_\star$ vs. M$_\star$ space. The top right panel shows our results as contours over the same plot for the GSWLC comparison sample. Overall we find good agreement. Our sample shows a well-defined ``star-forming main sequence,'' a significant population of quiescent galaxies, and an intermediate green valley. We agree well with the distribution in the GSWLC for star-forming galaxies. As expected, using GALEX+WISE tends to overestimate SFR in quiescent galaxies \citep[see the appendix and, e.g.,][]{DAVIS14,UTOMO14,SIMONIAN17}. This leads to the pile-up of quiescent galaxies in our data near $\log_{10}$~SFR/M$_\star \sim -11.5$~yr$^{-1}$. We will return to this in future work. For now, we suggest only to label galaxies with $\log_{10}$~SFR/M$_\star \lesssim -11$~yr$^{-1}$ as quiescent \citep[see more in][]{SALIM16}.

The bottom two panels focus on star-forming galaxies. Points in the bottom left panel show our individual targets. Light red lines show running values for the 16, 50, and 84$^{\rm th}$ percentile, i.e., the median $\pm1\sigma$. We calculate these in 250-wide galaxy bins after sorting our targets by stellar mass and considering only galaxies with $\log_{10}$~SFR/M$_\star > -11$~yr$^{-1}$. The dark red line shows a power law fit to star-forming galaxies with $\log_{10} M_\star = 9.5{-}11$~M$_\odot$:

\begin{equation}
\label{eq:sfmainseq}
\log_{10} {\rm SFR} / M_\star \left[ {\rm yr}^{-1} \right] = (-0.32) \left( \log_{10} \frac{M_\star}{10^{10} {\rm M}_\odot} \right) - 10.17
\end{equation}

\noindent Star forming galaxies show $\sim \pm 0.36$~dex scatter about this line in our data. This number includes a contribution from statistical noise; i.e., the physical scatter will be moderately less than this value. We tested the robustness of Equation \ref{eq:sfmainseq} by imposing distance cuts, galactic latitude cuts, and varying the fitting range. These shifted the power law index by $\lesssim 0.04$ and the intercept by a similar small amount.

The bottom right panel compares these results to the GSWLC and the literature synthesis of \citet[][$t=13.7$~Gyr in their formula]{SPEAGLE14}. Red again shows our results. In blue, we show the same running percentiles applied for the GSWLC SED-fitting based masses and SFRs. Fitting the GSWLC, we find a slightly steeper slope, $-0.42$, and similar normalization, $-10.2$, compared to Equation \ref{eq:sfmainseq}. GSWLC galaxies exhibit modestly lower residuals than our targets, $\sim 0.31$~dex compared to $0.36$~dex. Compared to our value, \citet{SPEAGLE14} find a steeper slope, $0.48$. At $\log_{10} M_\star \sim 10$~M$_\odot$, \citet{SPEAGLE14} predict $\log_{10} {\rm SFR}/M_\star = -10.17$~yr$^{-1}$, similar to what we find. 

Overall, the bottom right panel shows good agreement between our local galaxy measurements and results for larger samples. We find a moderate excess in SFR/M$_\star$ galaxies near $\log_{10}$ M$_\star \sim 10.5$~M$_\odot$, i.e., near the knee in the star-forming galaxy galaxy mass function. The differences among the three sets of results gives some sense of the systematic uncertainties affecting our SFR and $M_\star$ estimates, distance estimates, and sample selection. The main issues appears to be that we find a high abundance of high specific star formation rate galaxies compared to the GSWLC at $\log_{10}$M$_{*} \sim 10.5$~M$_{\odot}$. We proceed using Equation \ref{eq:sfmainseq} as a reference for local galaxies, and report offsets from this relation in Table \ref{tab:z0mgs_sfr_mstar}.

\begin{deluxetable*}{lcccccccccccc}[t!]
\tabletypesize{\scriptsize}
\tablecaption{SFR and M$_\star$ Estimates for Local Galaxies \label{tab:z0mgs_sfr_mstar}}
\tablewidth{\textwidth}
\tablehead{
\colhead{PGC \#} &
\colhead{NGC name} & 
\colhead{UGC name} &
\colhead{IC name} & 
\colhead{$d$} &
\colhead{$\delta d$} &
\colhead{$\log_{10}$ M$_\star$} & 
\colhead{\mtolwise } & 
\colhead{\mtolwise Method} & 
\colhead{$\log_{10}$ SFR} & 
\colhead{SFR Method} &
\colhead{$\Delta$~MS} & 
\colhead{Flags}  \\
\colhead{} &
\colhead{} & 
\colhead{} &
\colhead{} & 
\colhead{(Mpc)} &
\colhead{(dex)} &
\colhead{(M$_\odot$)} & 
\colhead{(M$_\odot$/L$_\odot$)} & 
\colhead{} & 
\colhead{(M$_\odot$~yr$^{-1}$} & 
\colhead{} &
\colhead{(dex)} &
\colhead{}
\\
\colhead{(1)} &
\colhead{(2)} &
\colhead{(3)} & 
\colhead{(4)} &
\colhead{(5)} & 
\colhead{(6)} &
\colhead{(7)} &
\colhead{(8)} & 
\colhead{(9)} & 
\colhead{(10)} & 
\colhead{(11)} & 
\colhead{(12)} &
\colhead{(13)} \\
}
\startdata
       4 &            &            &            & $ 65.9$ & $ 0.10$ & $ 9.28$ $\pm$ $ 0.45$ & $ 0.46$ &   SSFRLIKE & $-1.28$ $\pm$ $ 0.21$ &  FUV+WISE4 & $-0.70$ &       \\
      38 &            &   UGC12893 &            & $ 17.2$ & $ 0.37$ & $ 9.04$ $\pm$ $ 0.10$ & $ 0.50$ &   SSFRLIKE & $-1.66$ $\pm$ $ 0.20$ &  FUV+WISE4 & $-0.92$ &     S \\
      43 &            &            &            & $ 43.4$ & $ 0.15$ & $ 9.59$ $\pm$ $ 0.10$ & $ 0.29$ &   SSFRLIKE & $-0.32$ $\pm$ $ 0.20$ &  FUV+WISE4 & $ 0.06$ &       \\
      55 &            &   UGC12898 &            & $ 71.2$ & $ 0.09$ & $ 9.15$ $\pm$ $ 0.20$ & $ 0.25$ &   SSFRLIKE & $-0.60$ $\pm$ $ 0.20$ &  FUV+WISE4 & $ 0.07$ &       \\
      94 &            &   UGC12905 &            & $ 63.5$ & $ 0.10$ & $ 9.30$ $\pm$ $ 0.10$ & $ 0.34$ &   SSFRLIKE & $-0.81$ $\pm$ $ 0.20$ &  FUV+WISE4 & $-0.23$ &     S \\
     102 &            &   UGC12909 &     IC5376 & $ 74.8$ & $ 0.08$ & $10.70$ $\pm$ $ 0.10$ & $ 0.50$ &   SSFRLIKE & $-0.10$ $\pm$ $ 0.20$ &  FUV+WISE4 & $-0.45$ &       \\
     109 &    NGC7805 &   UGC12908 &            & $ 71.8$ & $ 0.09$ & $10.75$ $\pm$ $ 0.10$ & $ 0.50$ &   SSFRLIKE & $-0.88$ $\pm$ $ 0.21$ &  FUV+WISE4 & $  NaN$ &     G \\
     112 &    NGC7806 &   UGC12911 &            & $ 71.2$ & $ 0.09$ & $10.65$ $\pm$ $ 0.10$ & $ 0.50$ &   SSFRLIKE & $-0.21$ $\pm$ $ 0.20$ &  FUV+WISE4 & $-0.53$ &     G \\
     120 &            &   UGC12914 &            & $ 55.4$ & $ 0.11$ & $10.86$ $\pm$ $ 0.10$ & $ 0.45$ &   SSFRLIKE & $ 0.35$ $\pm$ $ 0.20$ &  FUV+WISE4 & $-0.11$ &     G \\
     129 &            &   UGC12915 &            & $ 64.7$ & $ 0.10$ & $10.56$ $\pm$ $ 0.10$ & $ 0.28$ &   SSFRLIKE & $ 0.70$ $\pm$ $ 0.20$ &  FUV+WISE4 & $ 0.44$ &     G \\
\enddata
\tablecomments{This table is a stub. The full version appears as an online table. Columns: (1) PGC number, (2) NGC name, (3) UGC name, (4) IC name, (5) adopted distance in Mpc, (6) adopted logarithmic uncertainty on the distance in dex, (7) $\log_{10}$ of inferred stellar mass, in M$_\odot$ and associated uncertainty (not including distance uncertainty), (8) adopted WISE1 mass to light ratio, (9) method used to calculate WISE1 mass-to-light ratio, (10) $\log_{10}$ of inferred star formation rate, in M$_\odot$~yr$^{-1}$ and associated uncertainty (not including distance uncertainty), (11) method used to calculate star formation rate, (12) offset from the best-fit ``main sequence'' of star forming galaxies in dex, not provided for quiescent galaxies or galaxies without hybrid SFR tracers, (13) flags, S: heavy star contamination, G: galaxy overlap, A: saturation a concern.}
\end{deluxetable*}

\section{Resolved Intensities}

\begin{figure*}
\plottwo{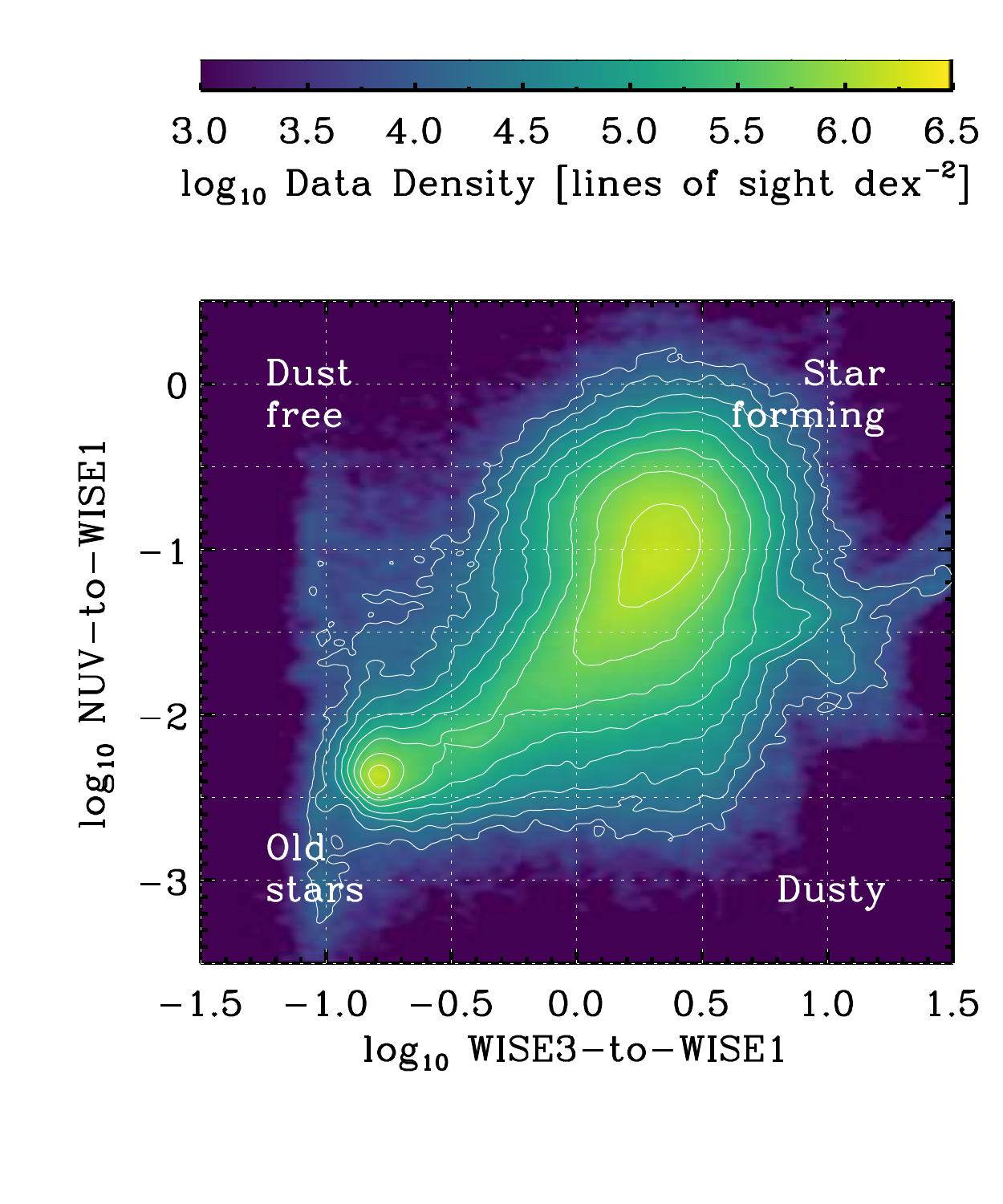}{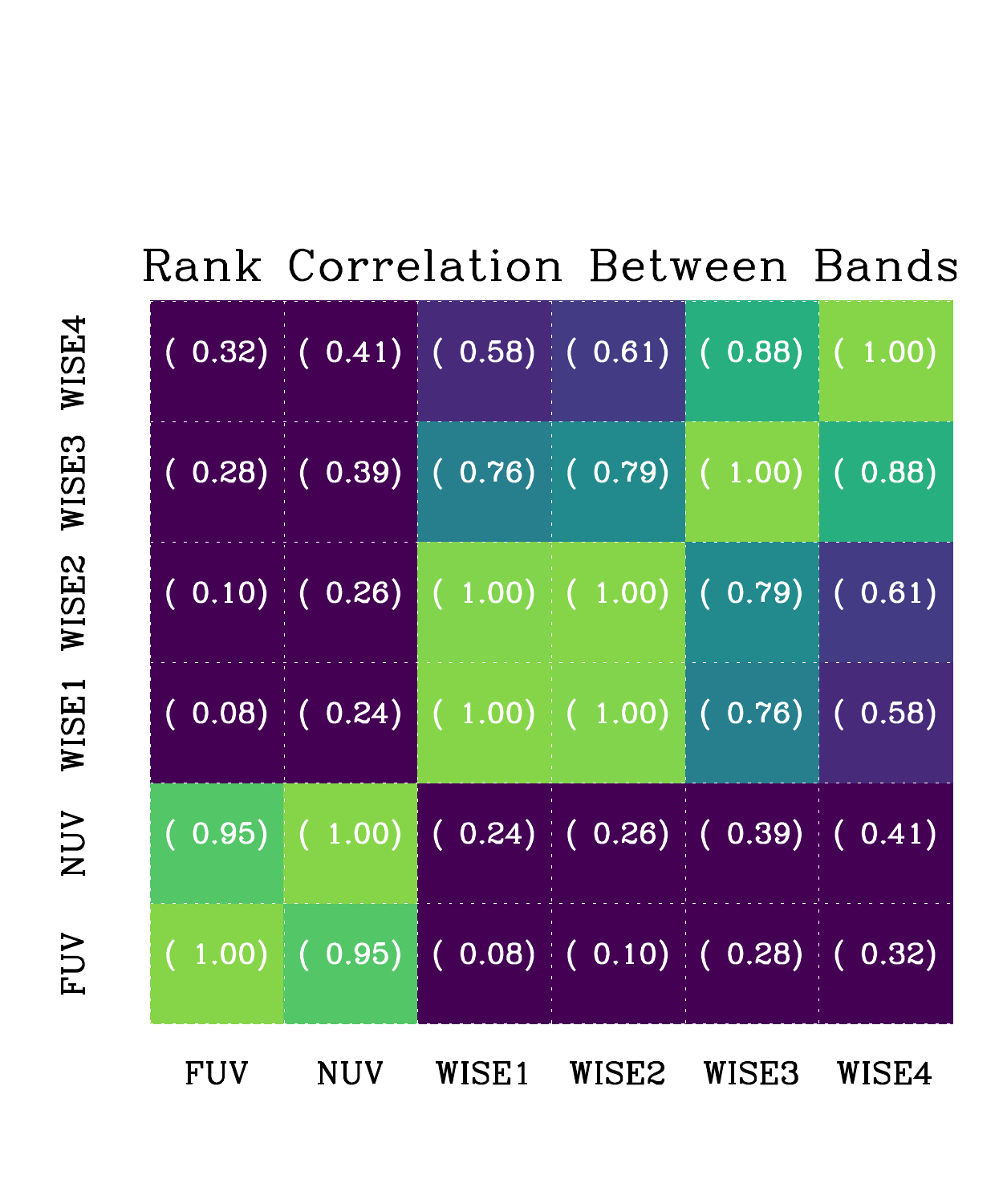}
\caption{{\bf Resolved GALEX and WISE Intensities} Illustration of a resolved application of the atlas. ({\em left}) Data density for lines of sight in our $15''$ atlas in NUV-to-WISE1 vs WISE3-to-WISE1 space (all ratios compare intensities in MJy~sr$^{-1}$). ({\em right}) Band-to-band rank correlation among lines of sight where all six bands are detected at high significance. Numbers report the rank correlation between the bands indicated on the axes. The similarity of the NUV and FUV bands, WISE1 and WISE2, and, to a lesser extent, WISE3 and WISE4 stand out, motivating our choice of parameter space in the left panel.}
\label{fig:intensities}
\end{figure*}

We also aim to provide resolved maps of galaxies, not only integrated measurements. In Figure \ref{fig:intensities}, we illustrate this aspect of the atlas. We show the distribution of data in the key NUV-to-WISE1 vs. WISE3-to-WISE1 parameter space. Here, these colors express  $\log_{10}$ of the ratio of intensities in MJy~sr$^{-1}$).

Figure \ref{fig:intensities} shows the distribution of individual lines of sight, taken from the whole $15''$ resolution atlas (except M31), in this space. We plot this as a data density plot, with successive contours spaced by a factor of two. We show only lines of sight with significant NUV, WISE3, and WISE1 detections. In total, a few million lines of sight contribute to the figure. 

Although this space does not capture all of the key physics in the six bands, it does show much of the variation. FUV and NUV correlate tightly, as do WISE1 and WISE2, and to a slightly lesser extent WISE3 and WISE4. The sense of the space is that NUV-to-WISE1 and WISE3-to-WISE1 both capture the degree to which a region is gas rich and has recent star formation, while NUV-to-WISE3 captures the dustiness of the region. In detail, the picture can be much more complicated than this, with dust physics and several timescales including the age of the stellar population playing important roles. 

The figure already shows interesting geography. In the bottom left, we see a high density of points at low NUV- and WISE3-to-WISE1 indicative of old stellar populations and perhaps remaining contaminating foreground stars. The complementary region to the upper right shows the area occupied by star-forming disks. This covers a much broader area in all directions, a dex in NUV-to-WISE1 and half a dex in WISE3-to-WISE1. This indicates a wide range of dustiness and degree of star formation. If we were to treat each galaxy equally instead of treating each line of sight equally, we might expect an even wider distribution as dust poor but small dwarf galaxies receive more weight.

The right panel of the figure shows the covariance among our six bands, measured via the rank correlation coefficient, for lines of sight where all six bands are well ($\gtrsim 10\sigma$) detected. The numbers report the rank correlation coefficient for each pair of bands. They demonstrate the points we made in choosing to plot NUV-to-WISE1 vs WISE3-to-WISE1. FUV and NUV intensities are highly correlated (bottom left). WISE1 and WISE2 track one another very closely (middle). And WISE3 and WISE4 also correlate tightly, though not as tightly as the other two pairs. The variations in the WISE3-to-WISE4 ratio partially track physics related to PAH abundance \citep[e.g.,][]{ENGELBRACHT05}; see the Appendix for more details.

\section{Summary}

As part of the $z=0$ Multiwavelength Galaxy Synthesis (\zomgs ) project, we present a public atlas of ultraviolet, near-infrared, and mid-infrared images built from observations by NASA's GALEX and WISE satellites. We construct the atlas on a common astrometric grid with two common angular resolutions, $7.5''$ and $15''$ (FWHM of a Gaussian beam). We build images for $\approx 15,750$ nearby galaxies, selected to have $>10\%$ chance of lying within $50$~Mpc with absolute $B$-band magnitude $\lesssim -18$~mag. This is about the distance out to which WISE can resolve a star-forming galaxy and roughly the magnitude above which IR detections become common.

This paper described the selection of targets and construction of the atlas. We show that our measurements reproduce previous measurements and other surveys well. In the overlap of the samples, the beam-by-beam intensities in our atlas match those measured from {\em Spitzer}'s S4G Survey \citep{SHETH10} and the GALEX Nearby Galaxy Atlas \citep{GILDEPAZ07}. 

We present integrated photometry from our sample, using an algorithm that matches the by-hand results from \citet{DALE17}. Our photometry also agrees with the recent measurements by \citet{CLARK18} and the integrated S4G photometry by \citet{MUNOZMATEOS15}.

In the appendix, we use the GALEX-SDSS-WISE Legacy Catalog \citep{SALIM16,SALIM18} to calibrate GALEX- and WISE-based estimators of the star formation rate (SFR) and stellar mass (M$_\star$). With a training set of detailed population synthesis fits to $> 100,000$ galaxies, these prescriptions should represent an improvement over previous literature work on integrated galaxies. Perhaps more important, they allow us to place local galaxies on the same measurement framework as the full SDSS sample considered by the GSWLC. We present SFR and M$_\star$ estimates for our targets and use these to estimate a local ``main sequence'' of star forming galaxies. This agrees well with previous estimates and results from applying a matched methodology to the GSWLC.

This atlas allows local, high-detail case studies to be placed in the context of the larger galaxy population. It provides a ``finding chart'' to plan more detailed observations of these key nearby systems. And it provides resolved, IR-based M$_\star$ and SFR estimates for a large sample of galaxies. Due to the limited resolution of IR telescopes to date, such measurements are still not possible at much greater distances. We also anticipate that it will play a key role in other aspects of the \zomgs\ project, providing a point of comparison for far-IR-based dust modeling, {\sc Hi} mapping, and CO mapping. We also anticipate using the atlas on its own to revisit the demographics and structure of star formation and stellar content in local galaxies.

\acknowledgments We thank the anonymous referee for a constructive, detailed, and timely report. We thank the GALEX and WISE teams for their hard work and excellent public data archives. This project benefited from support from the helpdesks for both facilities. We also thank Matthew Penny and Jiayi Sun for helpful consultation. We also acknowledge heavy use of the HyperLEDA database, NASA/IPAC Extragalactic Database (NED), and Extragalactic Distance Database (EDD). KS thanks Josh Peek for useful discussions regarding extinction corrections for GALEX. AKL acknowledges helpful correspondence with the MAST and IPAC helpdesks. AKL acknowledges support by the OSU Center for Cosmology and Astroparticle Astrophysics that made this work possible.

The work of AKL, KS, AL, IC, JC, MJG, SK, and DU is supported by NASA ADAP grants NNX16AF48G and NNX17AF39G and National Science Foundation grant No.~1615728. The work of AKL, MJG, SK, and DU is partially supported by the National Science Foundation under Grants No.~1615105, 1615109, and 1653300.

This work uses observations made with the \textit{Spitzer} Space Telescope, which is operated by the Jet Propulsion Laboratory, California Institute of Technology under a contract with NASA. This research has made use of NASA's Astrophysics Data System. This research has made use of the NASA/IPAC Extragalactic Database (NED) which is operated by the Jet Propulsion Laboratory, California Institute of Technology, under contract with the National Aeronautics and Space Administration.

This work has made use of data from the European Space Agency (ESA) mission {\it Gaia} (\url{https://www.cosmos.esa.int/gaia}), processed by the {\it Gaia} Data Processing and Analysis Consortium (DPAC, \url{https://www.cosmos.esa.int/web/gaia/dpac/consortium}). Funding for the DPAC has been provided by national institutions, in particular the institutions participating in the {\it Gaia} Multilateral Agreement.

\begin{appendix}

\section{Stellar Mass and Star Formation Rate Estimates Anchored to the GALEX-SDSS-WISE Legacy Catalog}

This project aims to place local galaxy population in the context of large surveys of more distant galaxies, e.g., SDSS, and observations of galaxies at higher redshifts. To do this, we need to translate our measured luminosities into estimates of the stellar mass, $M_\star$, and star formation rate, SFR, in a way that is consistent with the treatment of these more distant samples.

We discuss existing prescriptions to translate UV and mid-IR light into $M_\star$ and SFR in Section \ref{sec:conv}. These have mostly been developed for smaller samples of local galaxies, often focusing on cases where ``gold standard'' estimates of M$_\star$ or SFR are available \citep[e.g.,][]{CALZETTI07,MURPHY11,MEIDT14,HUNT19}. Other studies have built on these works and considered larger populations of local galaxies, aiming for internal consistency among the available tracers \citep{CALZETTI10,HAO11}. \citet{KENNICUTT12} and \citet{CALZETTI13} provide excellent reviews of the topic, and \citet{JARRETT13} extended this work to the WISE bands.

In this appendix, we adopt a more ``top down'' approach, following, e.g., \citet[][]{SALIM07}. We benchmark our estimates of $M_\star$ and SFR to the GALEX-SDSS-WISE Legacy Catalog \citep{SALIM16,SALIM18}. That project combined GALEX and WISE photometry with SDSS observations in order to estimate the integrated mass and SFR for $\sim 600,000$ galaxies. To do this, they carried out population synthesis modeling using the CIGALE code \citep{BOQUIEN19}. Their combination GALEX and WISE photometry with and physical parameter estimates offers an enormous reference data set. We use this to calibrate prescriptions that place local galaxies onto the same framework as this very large comparison sample using the available WISE and GALEX fluxes.

\citet{SALIM16} showed good agreement among SFRs estimated from emission lines, mid-IR emission, and their population synthesis fits in star-forming galaxies. \citet{SALIM18} include infrared emission (WISE4 when available, WISE3 in other cases) in the fitting, and so impose energy balance constraints \citep[e.g., following][]{DACUNHA08}. This should further improve the fidelity of their results. Thus, we expect that the GSWLC represents an excellent training set, with good accuracy and number statistics exceeding any local reference sample. A more conservative way to view the exercise is that regardless of any systematic effects in the GSWLC, this exercise places our measurements on an almost-identical scale to this large comparison data set that extends to $z\sim 0.3$.

For reference, \citet{SALIM18} use CIGALE \citep{BOQUIEN19} to fit a star formation history that combines an old population, an exponentially decaying star formation history, and a younger population with nearly constant SFR. Their fits include energy balance \citep[e.g.,][]{DACUNHA08} with the IR luminosity extrapolated from the 22$\mu$m band (or $12\mu$m if 22$\mu$m is not detected) via the IR templates of \citet{CHARY01}. They also account for contamination by emission lines. \citet{SALIM16} shows good agreement, on average, between their approach and emission line-based or IR-based methods. This should yield accurate average SFRs over the $\sim 100$~Myr timescales to which UV light is sensitive \citep[e.g., see][]{KENNICUTT12}. This smooth parameterization of the SFR does mean that recent non-smooth features in the star formation history, e.g., triggered by interactions, gas flows through the galaxy, will only be reflected on average.

\subsection{Selection}

We work with the GSWLC version 2 ``X'' catalog \citep{SALIM18}, which includes the results of broadband SED fitting spanning from the UV to $22\mu$m. For each galaxy, we also consider the matched photometry used by \citet{SALIM16} and \citet{SALIM18} in the FUV, NUV, WISE1, WISE2, WISE3, and WISE4 bands. Following \citet{SALIM16} we calculate luminosity distances assuming $H_0 = 70$~km~s$^{-1}$~Mpc$^{-1}$ and a flat cosmology with $\Omega_m = 0.27$.

We consider all galaxies with S/N$>3$ detections in WISE1, WISE3, WISE4, NUV, and FUV that have a ``good fit'' flag. This yields $\sim 101,000$ galaxies. We experimented with selections that considered only the more sensitive NUV, WISE3, and WISE1 bands and with varying the S/N cut. This can yield up to $\sim 400,000$ galaxies and the results appear qualitatively similar. Given the importance of WISE4 and FUV to trace SFR and our reliance on band ratios, which require good S/N, we focus on this selection only. We do caveat that this leads us to miss low mass quiescent galaxies and may somewhat bias our results for more massive quiescent galaxies.

We carry out no correction for completeness or Malmquist bias. For this exercise, we assume that once the colors and luminosity of a galaxy are specified it will represent a reasonable point of comparison for local galaxies with the same luminosity and colors. But we do caution that this GSWLC selection does not represent a volume limited sample. High luminosity galaxies will be over-represented and low luminosity galaxies with little or no star-formation will be under-represented.

\subsection{Calculation of Mass-to-Light Ratios and SFR Calibration Pre-Factors} 

We use the population synthesis-based stellar masses, SFRs, and photometry from \citet{SALIM18} to estimate the WISE1 mass-to-light ratio, \mtolwise, and pre-factors on several SFR calibrations involving the UV and mid-IR.

We calculate \mtolwise\ following Section \ref{sec:conv}. That is, we divide the CIGALE population synthesis-based $M_\star$ from \citet{SALIM18} by the WISE1 luminosity, expressed in units of the Sun's luminosity at $\lambda = 3.4\mu$m. The result typically lies in the range $\sim 0.1{-}0.7$~M$_\odot$~L$_\odot^{-1}$, consistent with previous work (see Section \ref{sec:conv}).

We also consider the ``pre-factors'' on several SFR estimators. We follow a modified version of the notation used by \citet{KENNICUTT12}. We define $C$ as

\begin{equation}
\label{eq:cdefn}
C = \frac{{\rm SFR} [{\rm M_\odot~yr}^{-1}]}{\nu L_\nu [{\rm erg~s}^{-1} ]}~.
\end{equation}

\noindent That is, $C$ represents the translation from luminosity to SFR. In CGS units, with $\nu L_\nu$ in erg~s$^{-1}$ and SFR in M$_\odot$~yr$^{-1}$, $\log_{10}$ C has typical magnitude $-43.5 \pm 0.5$ for the UV and $-42.5 \pm 0.5$ for WISE \citep[e.g.,][]{KENNICUTT12}.

We also calculate $C$ for linear ``hybrid'' tracers. We consider each linear combination of NUV and FUV with WISE3 and WISE4, so that:

\begin{equation}
\label{eq:sfr_hybrid}
{\rm SFR} = C_{\rm UV} \nu_{\rm UV} L_{\rm \nu, UV} + C_{\rm WISE} \nu_{\rm WISE} L_{{\rm \nu, WISE}}~.
\end{equation}

\noindent We follow the literature in treating $C_{\rm UV}$, the ``unobscured'' term, as fixed and set by the underlying stellar physics \citep[e.g., see][]{SALIM07}. We check this below by using the best-fit GSWLC2 FUV extinctions, $A_{\rm FUV}$, so that:

\begin{equation}
\label{eq:cfuv}
C_{\rm FUV} = \left( \frac{{\rm SFR} (\rm{Total})}{\nu_{\rm FUV} L_{\rm \nu,FUV} \times 10^{A_{\rm FUV}/2.5} } \right)~.
\end{equation}

Here SFR~(Total) refers to the total SFR from the GSWLC catalog, $L_{\rm \nu, FUV}$ is the observed FUV luminosity, and $A_{\rm FUV}$ is the population synthesis-based extinction estimate.

Again following the literature, we treat the pre-factor on the WISE terms as empirical, depending on the choice of WISE band, UV hybrid, and --- we will see below --- the properties of the galaxy. This reflects that the WISE bands do not trace the bolometric luminosity and that there remain significant uncertainties related to dust composition (especially for WISE3), geometry, and heating sources. For each galaxy, we calculate the appropriate $C_{\rm IR}$ via:

\begin{equation}
\label{eq:cwise}
C_{\rm IR} = \left( \frac{{\rm SFR} (\rm{Total}) - {\rm SFR} (\rm{UV})}{\nu L_\nu ({\rm WISE}) } \right)~.
\end{equation}

\noindent Again SFR~(Total) refers to the total SFR from the GSWLC catalog. We calculate $C_{\rm IR}$ for each of: FUV+WISE4, FUV+WISE3, NUV+WISE4, NUV+WISE3, WISE4 only (i.e., no UV), and WISE3 only.

Note that our formulation for the hybrid tracers differs slightly from \citet{KENNICUTT12}. They first combine UV and IR emission to estimate the total, extinction-correction UV luminosity. Then they apply $C_{\rm UV}$ to translate from extinction-corrected UV luminosity to SFR. They provide an empirical coefficient to be applied to the IR term, which is then added to the UV term to correct for extinction. This coefficient on the IR term is empirical, and modulo a factor of $C_{\rm UV}$ is identical in meaning to our $C_{\rm IR}$. As long as the approach is linear, the difference is only semantic.

Thus for each galaxy we estimate:

\begin{enumerate}
\item The WISE mass to light ratio, \mtolwise , from comparing the WISE1 luminosity to the GSWLC stellar mass.
\item The coefficient to translate FUV to SFR, $C_{\rm FUV}$, from comparing the GSWLC extinction-corrected FUV luminosity to the GSWLC SFR.
\item The coefficients, $C_{\rm WISE3}$ and $C_{\rm WISE4}$, to translate WISE3 and WISE4, respectively, to SFR using only WISE luminosities, from comparing the GSWLC SFR to the WISE luminosity. We refer to these as $C$ ``WISE4 only'' or ``WISE3 only.''
\item The coefficients, $C_{\rm WISE3}$ and $C_{\rm WISE4}$, to translate WISE3 and WISE4 to SFR when using WISE in conjunction with FUV and NUV. We calculate these from the GSWLC SFR, less the unobscured UV contribution. We refer to these as, e.g., $C$ ``WISE3+FUV'' or ``WISE4+NUV,'' indicating the WISE and UV bands used. When written this way, $C$ refers to the coefficient on the WISE term.
\end{enumerate}

\subsection{Trends in SFR/M$_\star$ vs. M$_\star$ Space}

\begin{figure*}
\centering
\plottwo{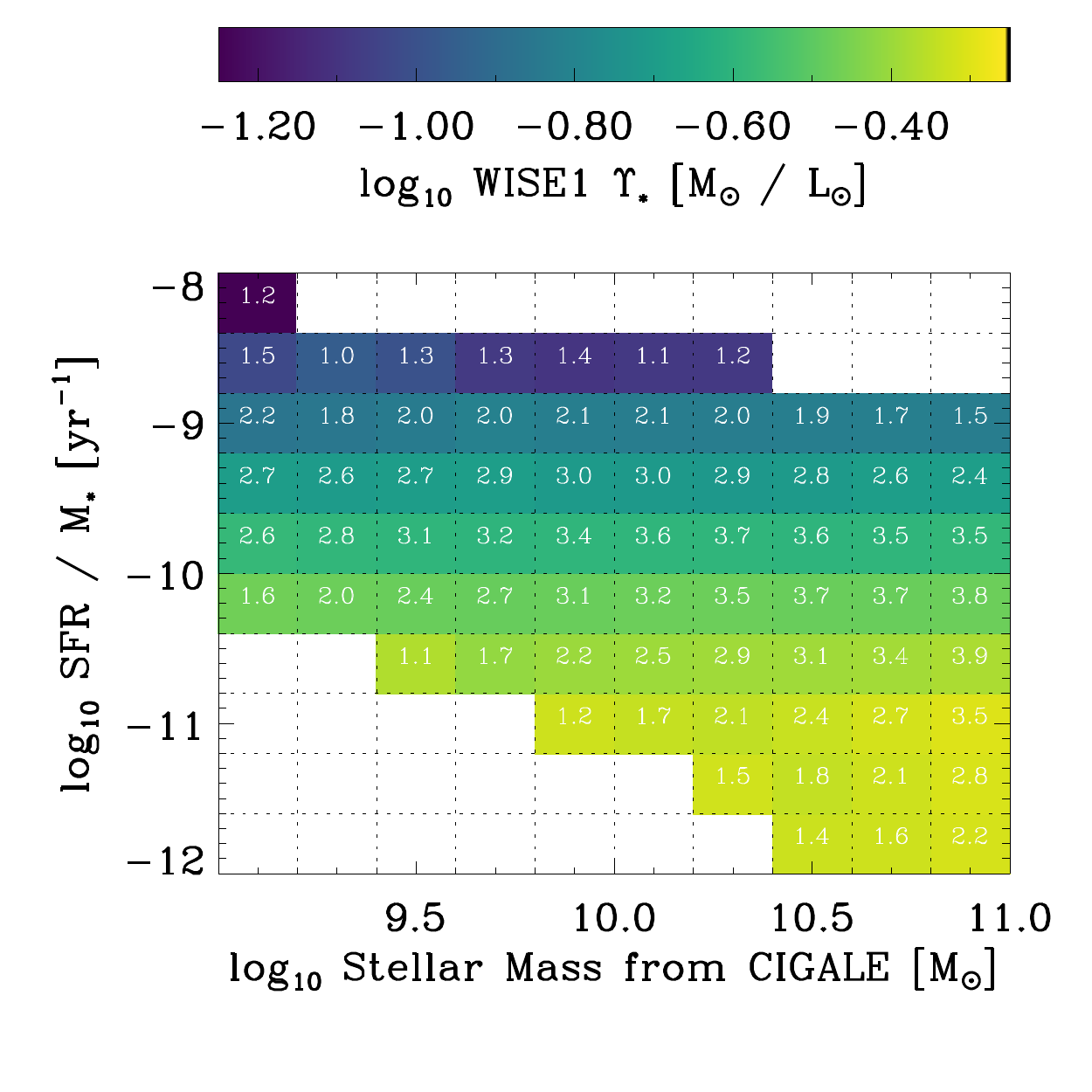}{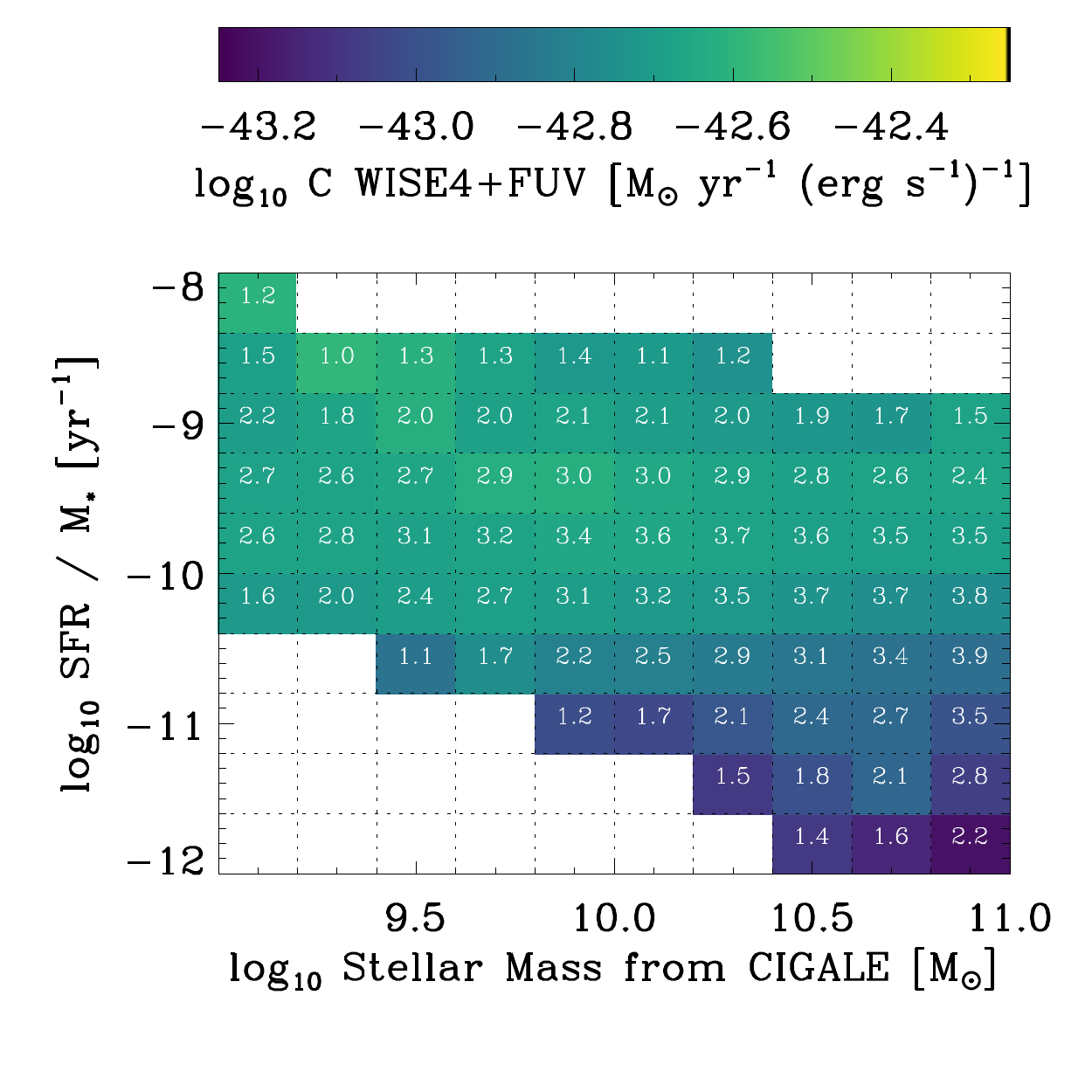}
\plottwo{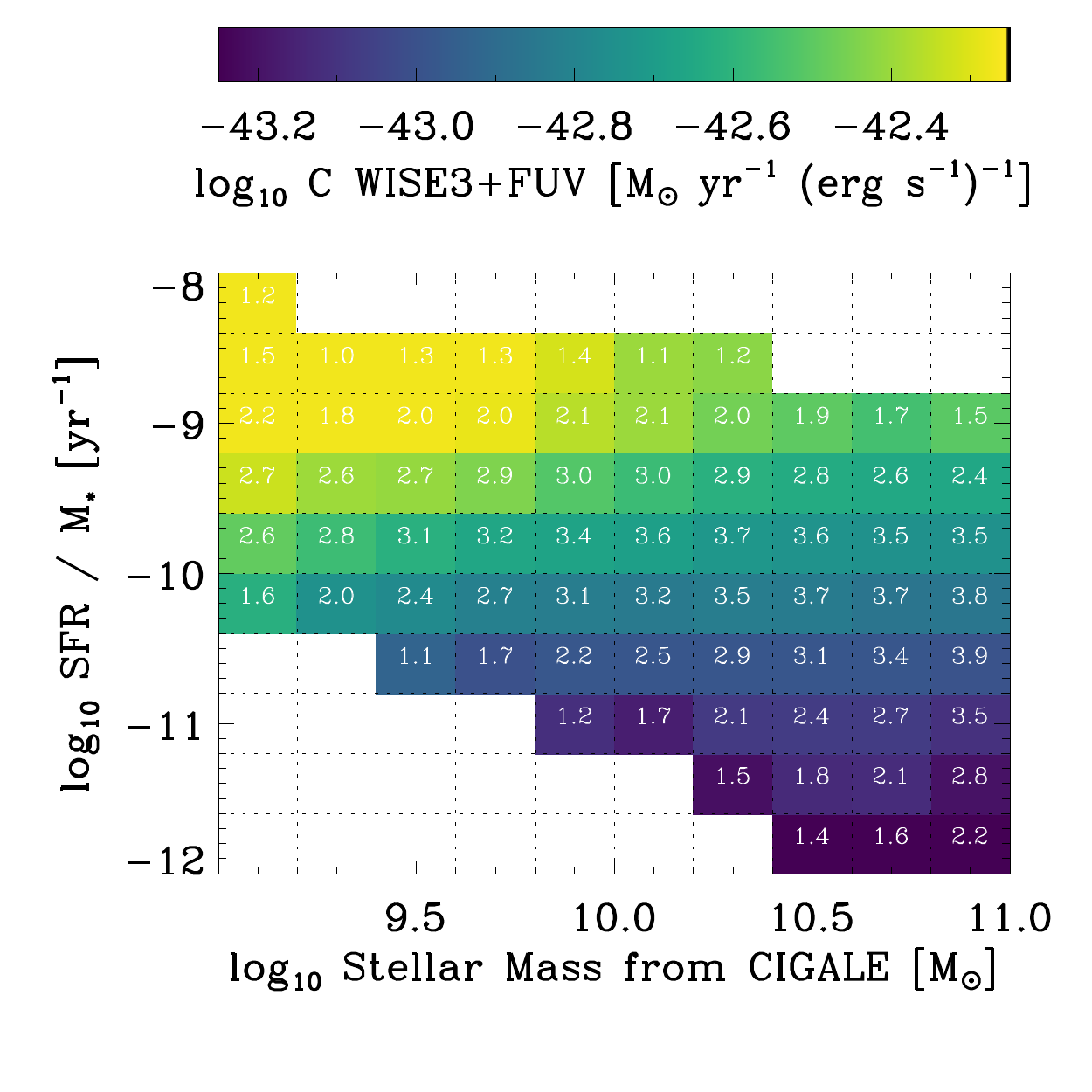}{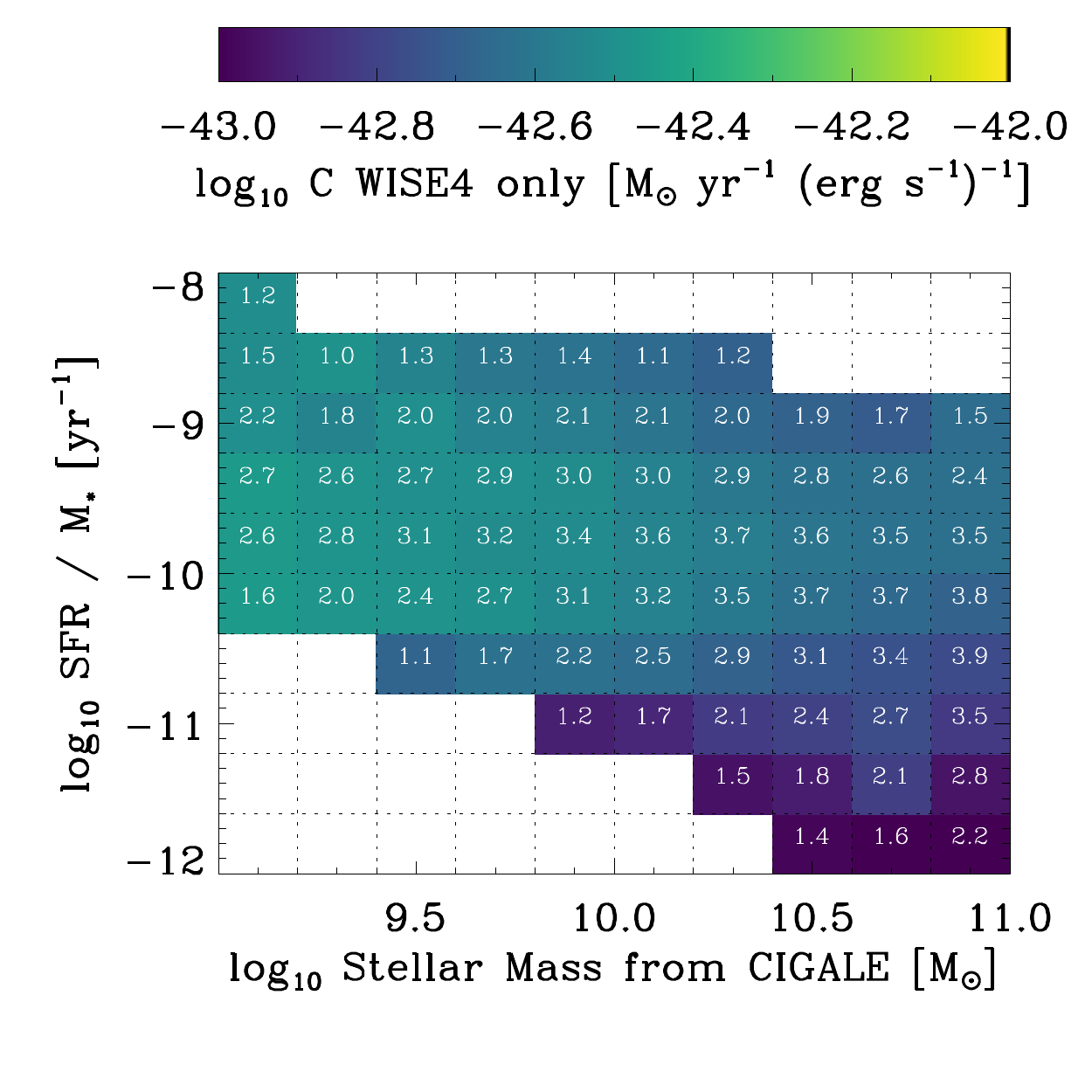}
\caption{{\bf Trends in Mass-to-Light Ratio and Star Formation Rate Calibrations in the GSWLC \citep{SALIM16,SALIM18}.} Calibrations to convert ({\em top left}) WISE1 luminosity to stellar mass, \mtolwise , ({\em top right}) WISE4 to SFR in a hybrid with FUV, ({\em bottom left}) WISE3 to SFR in a hybrid with FUV, and ({\em bottom right}) WISE4 only to SFR, i.e., with no UV term. The images show the average calibrations derived from comparing GSWLC population synthesis-based SFR and M$_\star$ to WISE and GALEX luminosities. The number in each cell reports $\log_{10}$ of the number of galaxies averaged to determine the mean conversion. All plots show the conversion factor on a log stretch with dynamic range of $1$~dex. Key points from these plots include: (1) a WISE1 mass-to-light ratio depends on specific star formation rate; (2) a simple linear hybridization of FUV and WISE4 appears stable for star-forming galaxies but differs slightly in quiescent galaxies; (3) even as part of a hybrid with FUV, the ratio of WISE3 to SFR varies, with less WISE3 emission in low mass galaxies and galaxies with intense radiation fields; and (4) ``WISE4 only'' has a mass-dependent relationship to SFR, in addition to the trends with specific star formation rate noted in point 2. All of these trends qualitatively agree with physical expectations. We provide these grids as a machine readable table and they inform our approach to parameter estimation in our local targets.}
\label{fig:sfr_mtol_grids}
\end{figure*}

\begin{deluxetable}{lccccccccc}
\tabletypesize{\scriptsize}
\tablecaption{WISE Mass-to-Light Ratio and SFR Calibrations in M$_\star$-sSFR Space \label{tab:grid_ssfr_mstar}}
\tablewidth{\textwidth}
\tablehead{
\colhead{$\log_{10} M_\star$} &
\colhead{$\log_{10} {\rm SFR}/M_\star$} & 
\colhead{$\log_{10} n_{\rm gal}$} &
\colhead{$\mtolwise$} & 
\colhead{$C$ WISE4+FUV} &
\colhead{$C$ WISE4+NUV} & 
\colhead{$C$ WISE4 only} & 
\colhead{$C$ WISE3+FUV} &
\colhead{$C$ WISE3+NUV} & 
\colhead{$C$ WISE3 only} \\
}
\startdata
$  9.10$ & $-11.80$ & $\ldots$ & $\ldots$ & $\ldots$ & $\ldots$ & $\ldots$ & $\ldots$ & $\ldots$ & $\ldots$ \\
$  9.10$ & $-11.40$ & $\ldots$ & $\ldots$ & $\ldots$ & $\ldots$ & $\ldots$ & $\ldots$ & $\ldots$ & $\ldots$ \\
$  9.10$ & $-11.00$ & $\ldots$ & $\ldots$ & $\ldots$ & $\ldots$ & $\ldots$ & $\ldots$ & $\ldots$ & $\ldots$ \\
$  9.10$ & $-10.60$ & $\ldots$ & $\ldots$ & $\ldots$ & $\ldots$ & $\ldots$ & $\ldots$ & $\ldots$ & $\ldots$ \\
$  9.10$ & $-10.20$ & $  1.6$ & $  0.35$ & $-42.70$ & $-42.80$ & $-42.45$ & $-42.61$ & $-42.61$ & $-42.43$ \\
$  9.10$ & $ -9.80$ & $  2.6$ & $  0.27$ & $-42.66$ & $-42.75$ & $-42.45$ & $-42.48$ & $-42.48$ & $-42.26$ \\
$  9.10$ & $ -9.40$ & $  2.7$ & $  0.21$ & $-42.67$ & $-42.75$ & $-42.45$ & $-42.31$ & $-42.31$ & $-42.10$ \\
$  9.10$ & $ -9.00$ & $  2.2$ & $  0.14$ & $-42.69$ & $-42.76$ & $-42.50$ & $-42.21$ & $-42.21$ & $-42.03$ \\
$  9.10$ & $ -8.60$ & $  1.5$ & $  0.09$ & $-42.68$ & $-42.71$ & $-42.51$ & $-42.16$ & $-42.16$ & $-41.95$ \\
$  9.10$ & $ -8.20$ & $  1.2$ & $  0.04$ & $-42.59$ & $-42.65$ & $-42.50$ & $-42.03$ & $-42.03$ & $-41.90$ \\
\enddata
\tablecomments{This is a stub. The full version of the table is available in machine readable format with the online version of the paper.}
\end{deluxetable}

Figure \ref{fig:sfr_mtol_grids} shows the behavior of the the mass-to-light-ratio, \mtolwise , and three WISE-based SFR calibrations in SFR/M$_\star$ vs. M$_\star$ space. The numbers in each cell report log$_{10}$ of the number of galaxies use to compute the average. We use the  SFR/M$_\star$ and M$_\star$ from the GSWLC to compute the grid, so we expect that Figure \ref{fig:sfr_mtol_grids} reflect best-estimate distributions of these parameters in this key parameter space. All plots have the same dynamic range, one decade in the relevant conversion factor, displayed on a logarithmic stretch. We found these grids invaluable to interpret WISE and GALEX emission, and Table \ref{tab:grid_ssfr_mstar} gives them in machine readable form.

The figure demonstrates several key conclusions:

\begin{enumerate}
\item \textit{Top left:} The WISE1 mass to light ratio, \mtolwise , varies primarily as a function of the specific star formation rate, SFR/M$_\star$, in the GSWLC. The dynamic range of this variation is from $\mtolwise \sim 0.2${-}$0.5$~\mtolunits\ ($\log_{10} \mtolwise \sim -0.7$ to $-0.3.$ \textit{The GSWLC strongly implies that accurate stellar mass estimation across the whole local galaxy population requires variable \mtolwise .}

\item \textit{Top right:} The SFR calibration for WISE4 in WISE4+FUV remains quite stable across the ``star forming main sequence,'' i.e., the large population of galaxies with $\log_{10} {\rm SFR}/M_\star$ between $\sim -10.25$ and $-9.25$. Galaxies with lower ${\rm SFR}/M_\star$ show evidence for lower $C$, consistent with ``cirrus'' emission, including circumstellar dust, playing an increasing role in these systems \citep[e.g.,][]{DAVIS14,SIMONIAN17}. Galaxies with very high specific star formation rates also show lower $C$, consistent with previous work on the $24\mu$m-to-TIR ratio. \textit{For most galaxies, WISE4+FUV appears to be a remarkably stable SFR indicator.}

\item \textit{Bottom left:} Combining WISE3 with FUV appears far less robust than the combination with WISE4. The plots show strong changes in $C$ for WISE3+FUV as a function of both stellar mass and specific star formation rate. The sense is that quiescent and high mass galaxies show lower $C$, implying more WISE3 emission per unit SFR. High specific star formation rate galaxies and low mass galaxies show low $C$, implying less WISE3 emission. The WISE3 band captures a strong PAH feature and the trends agree with a drop in PAH fraction as a function of both metallicity \citep[e.g.,][]{ENGELBRACHT05,DRAINE07B,REMYRUYER15} and interstellar radiation field \citep{CHASTENET19}. \textit{Using WISE3 to trace SFR, even as part of a hybrid, requires a context-dependent SFR calibration.}

\item \textit{Bottom right:} Using WISE4 only leads to a strong dependence of the SFR prefactor, $C$, on stellar mass compared to hybridizing WISE4 with FUV. This reflects a mass-dependent fraction of ``obscured'' vs. ``unobscured'' star formation, meaning that the WISE term contributes less relative to FUV term in low mass galaxies. This is be expected from the lower dust-to-gas ratio and overall gas column densities in low mass galaxies \citep[e.g.,][]{SANDSTROM13,REMYRUYER14}. \textit{Using WISE without hybridizing it with an ``unobscured'' tracer requires a conversion factor, $C$, that depends on host galaxy mass or some related property.} Though not shown, the situation appears even worse for WISE3.

\end{enumerate}

In the rest of this appendix, we examine these trends in more detail. Then we use them to build prescriptions to translate our GALEX and WISE measurements of local galaxies in SFR and M$_\star$ estimates.

\subsection{Trends in the WISE1 Mass-to-Light Ratio}

\begin{figure*}
\centering
\plottwo{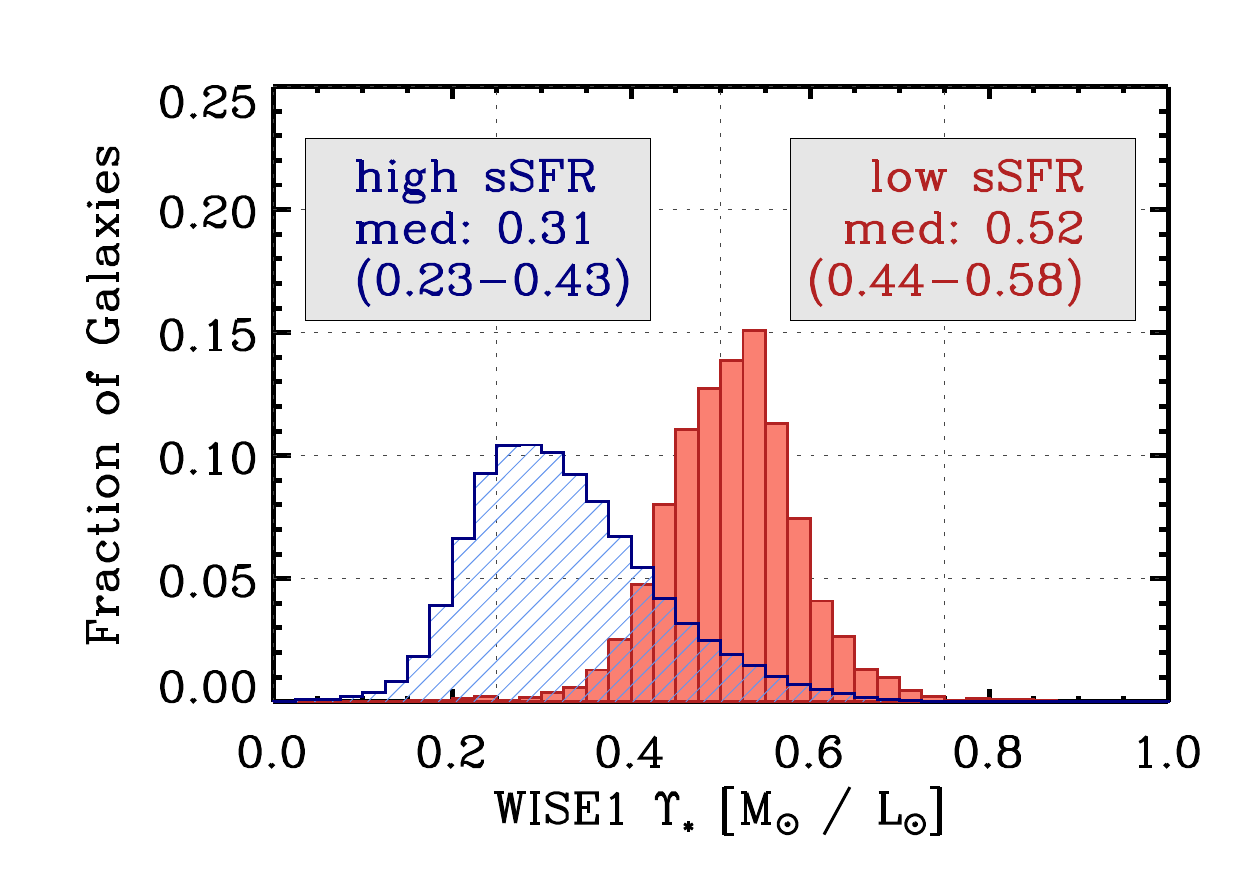}{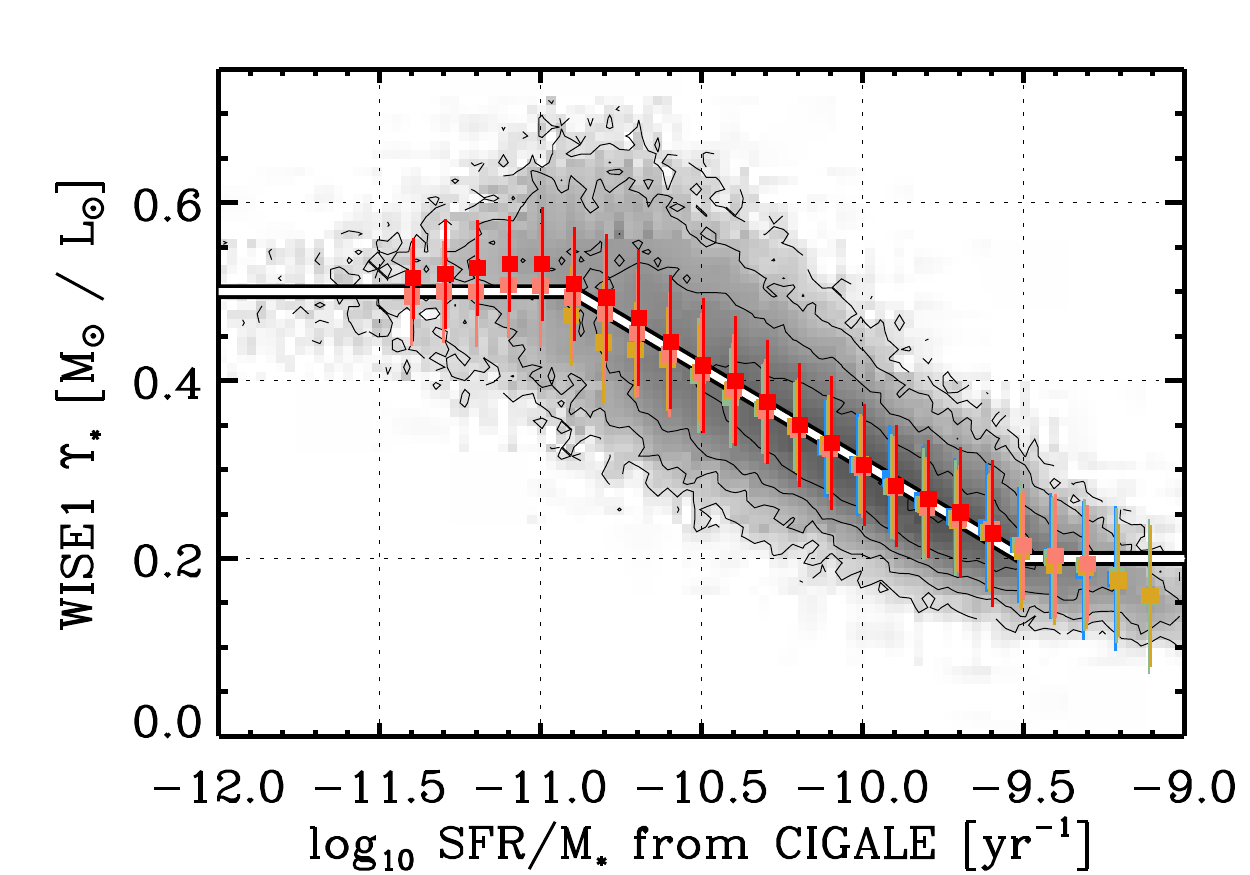}
\caption{{\bf WISE1 Mass-to-Light Ratio in the GSWLC \citep{SALIM16,SALIM18}.} ({\em left}) Histograms of \mtolwise, calculated based on WISE1 photometry and SED fitting by \citet{SALIM16,SALIM18}. We show the distributions for low specific star formation rate (sSFR = SFR/M$_\star$) quiescent galaxies and high sSFR star-forming galaxies separately. The low sSFR galaxies show a narrow range of \mtolwise\ while the high sSFR galaxies show a wider range of values. ({\em right}) \mtolwise\ as a function of sSFR = SFR/M$_\star$. Gray points show data density of galaxies in the GSWLC. Colored points and error bars show binned trends for different stellar mass ranges with error bars indicating the $1\sigma$ scatter. The black and white line shows Equation \ref{eq:pred_mtol}.
}
\label{fig:mtol_hist}
\end{figure*}

\begin{figure*}
\centering
\plottwo{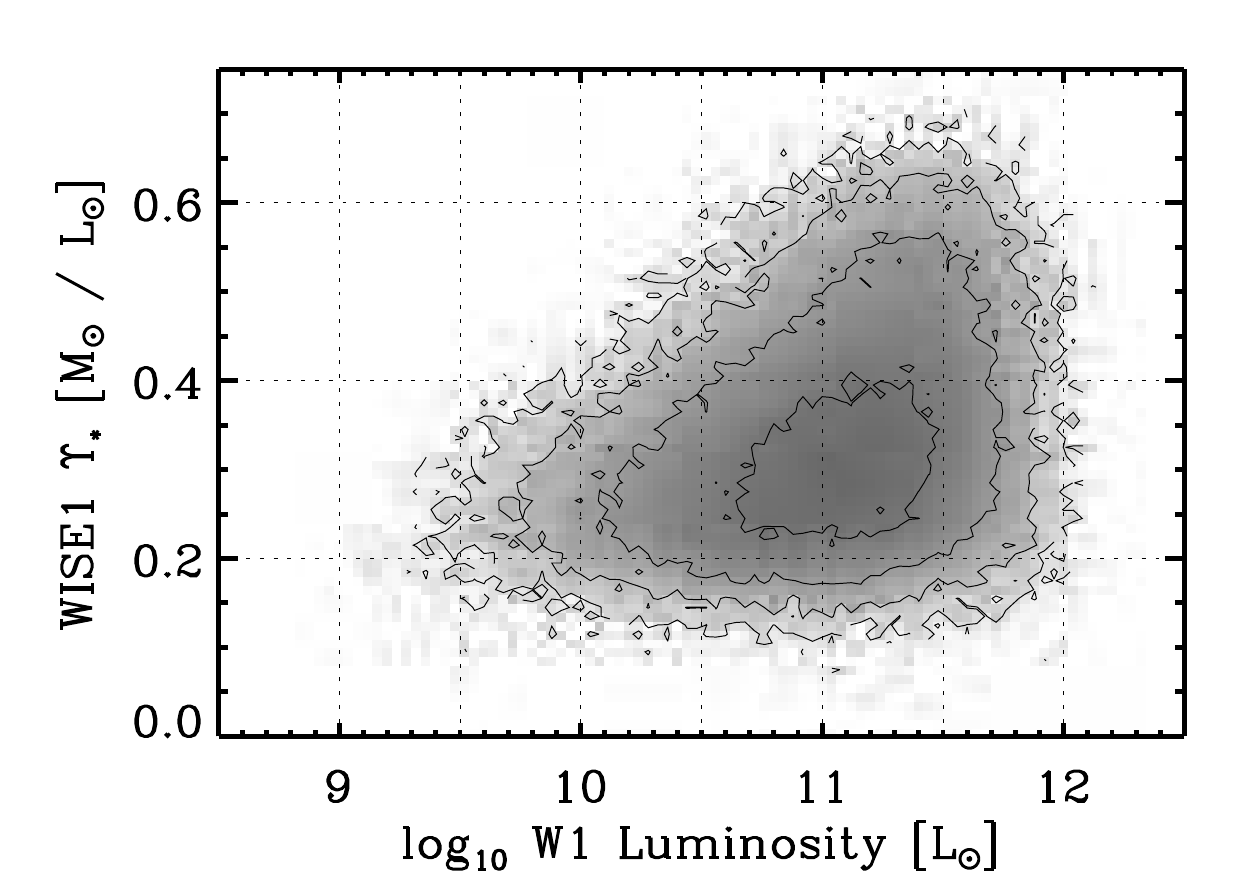}{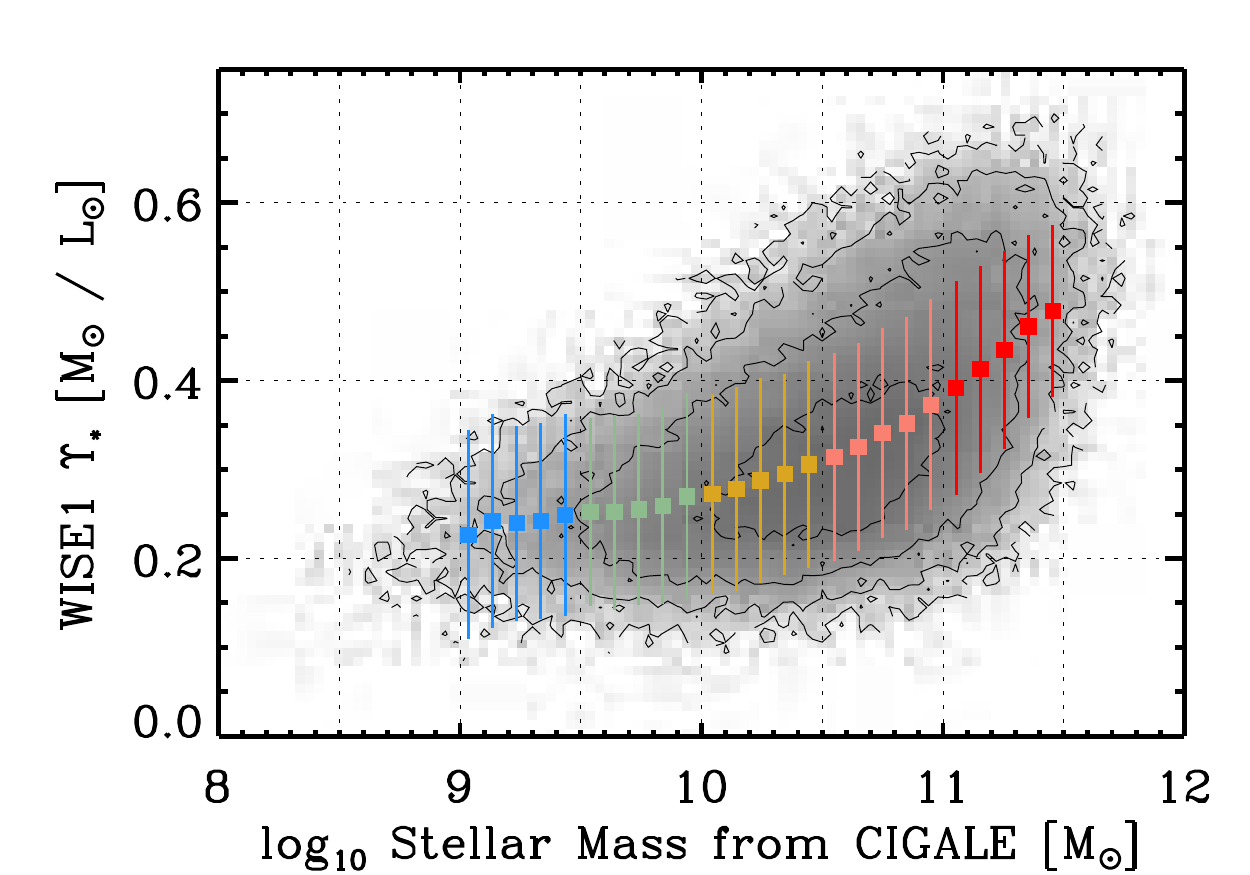}
\plottwo{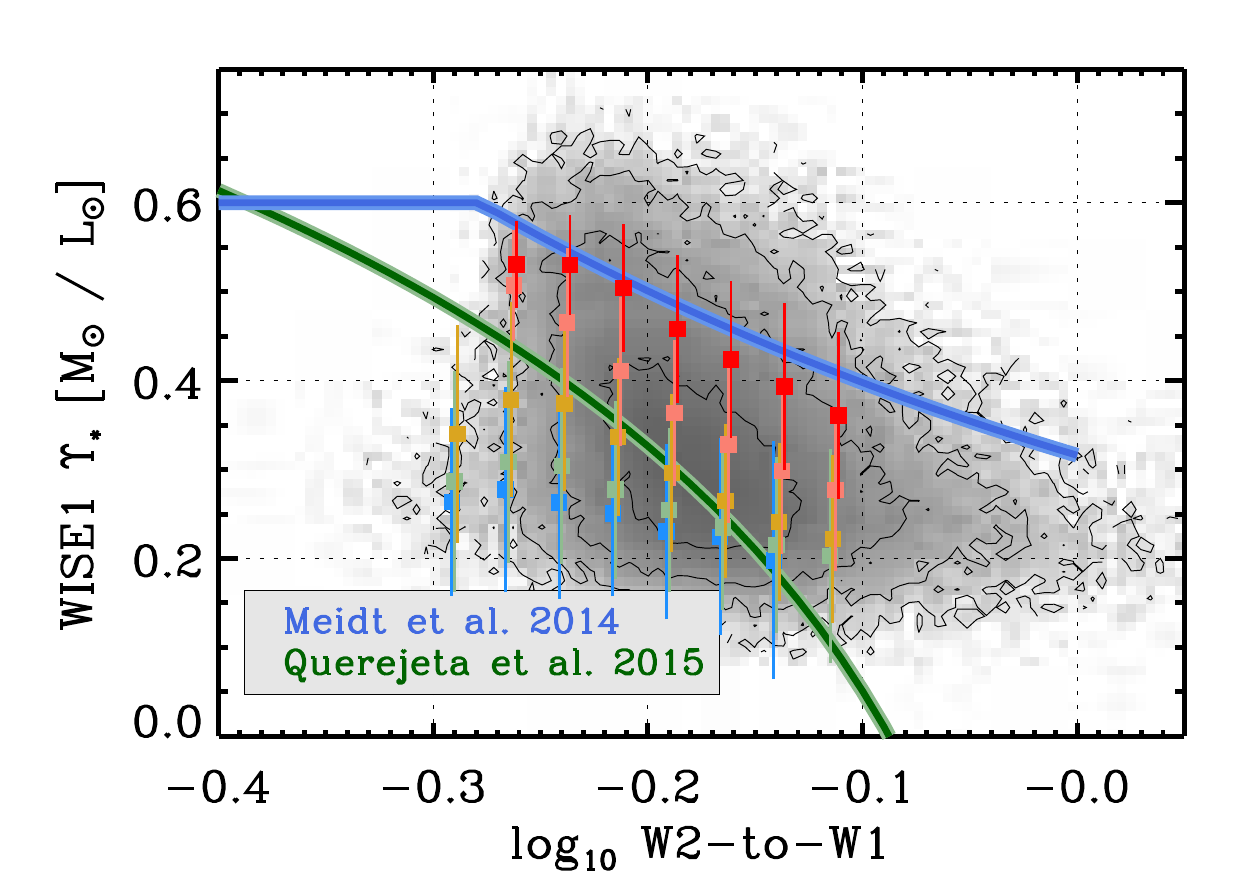}{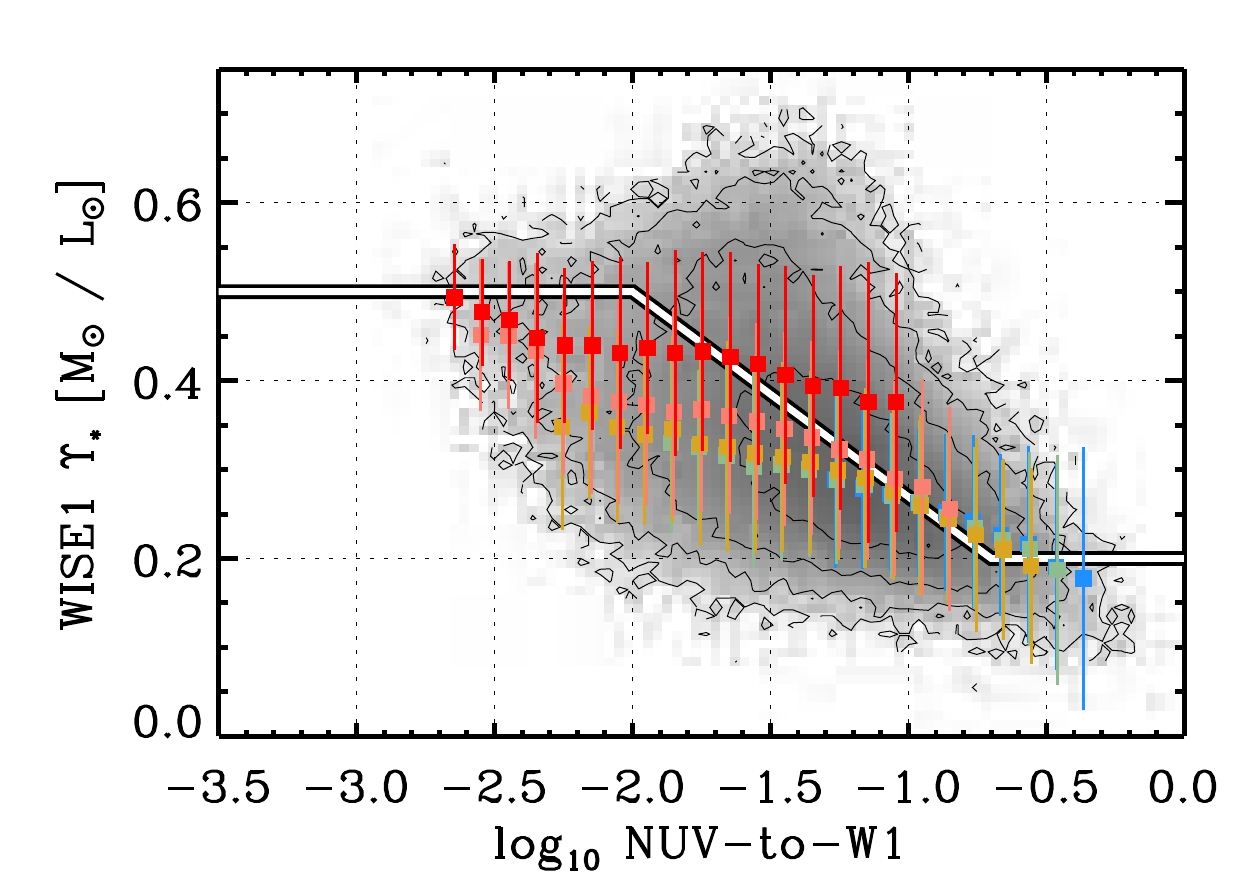}
\plottwo{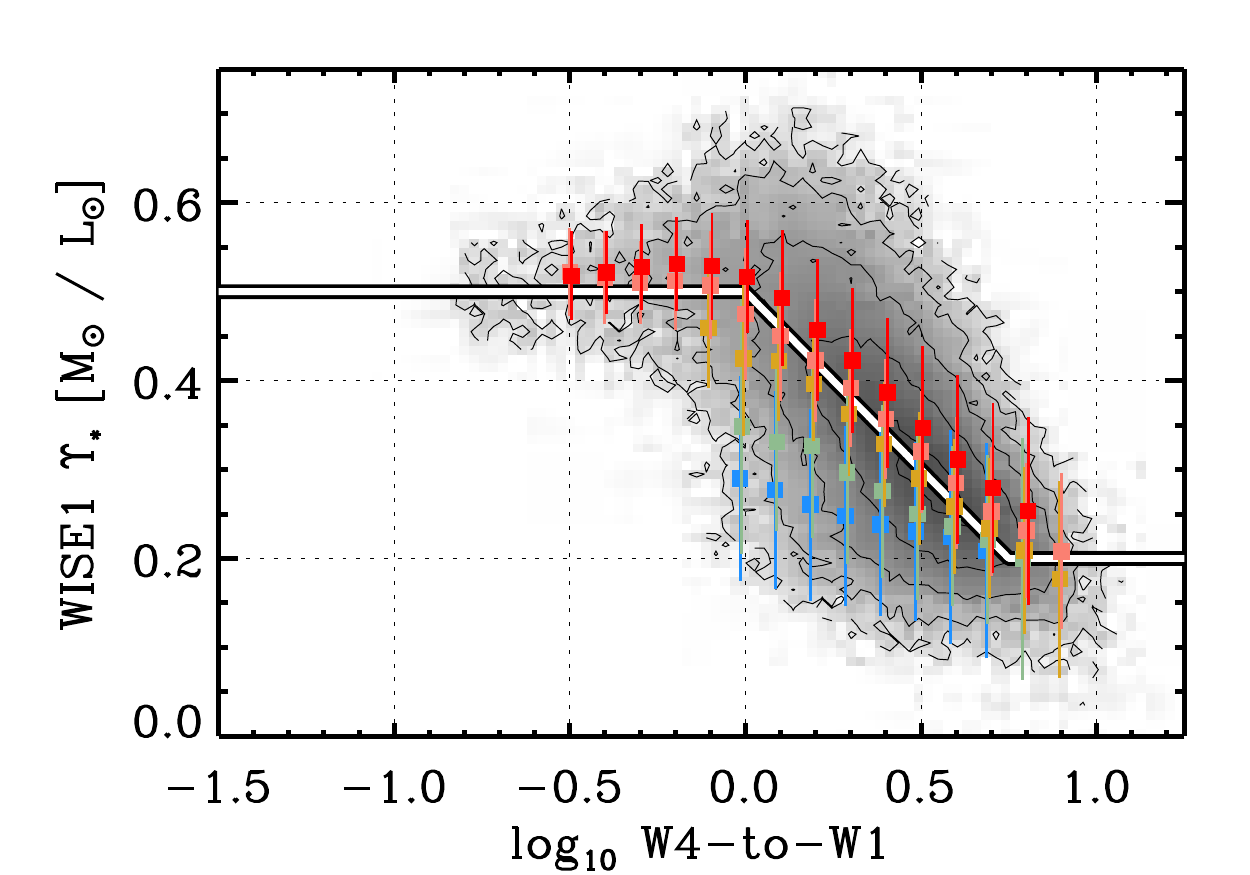}{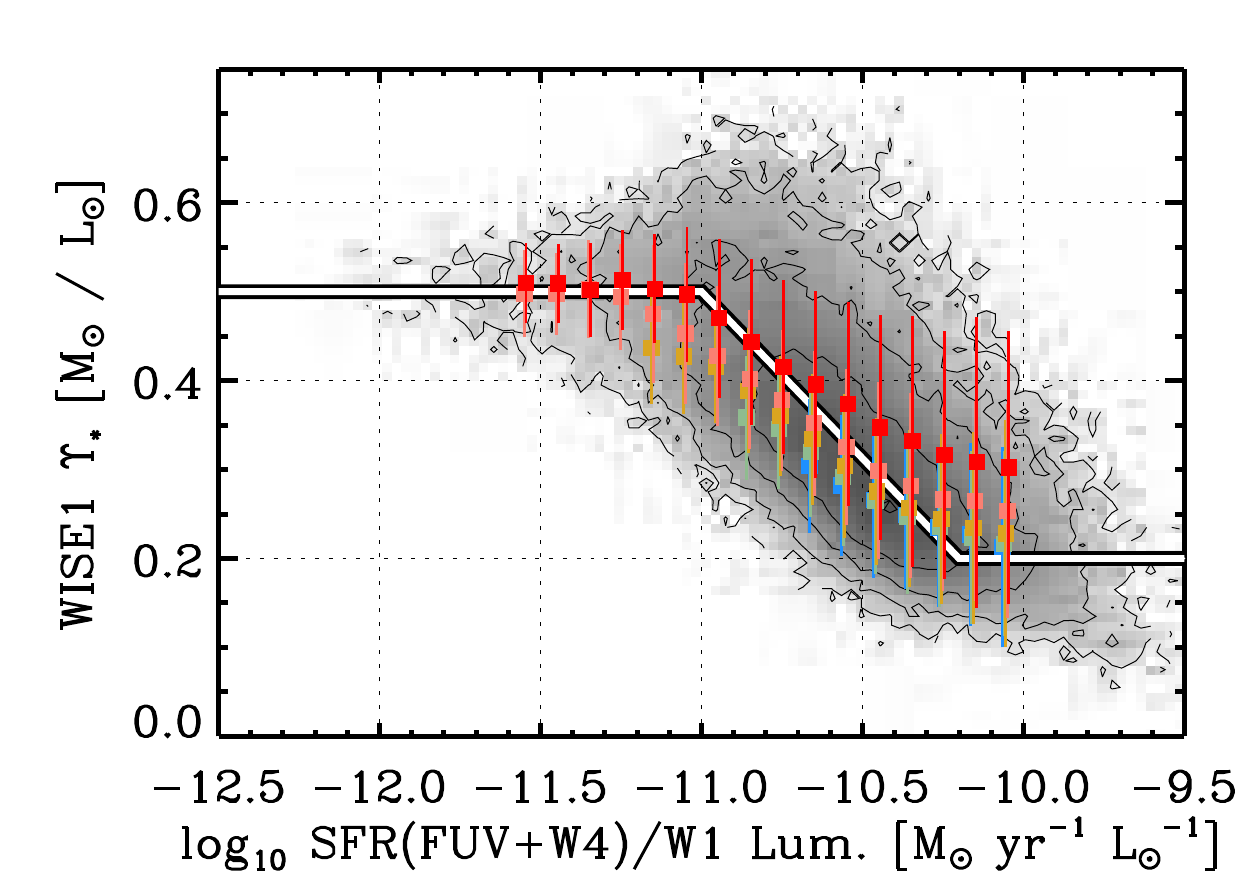}
\caption{{\bf Trends in WISE1 Mass-to-Light Ratio as a Function of GALEX and WISE Observables.} \mtolwise , based on the GSWLC as a function of parameters observed in our nearby galaxy atlas. Gray points show data density of galaxies in the GSWLC. Colored points and error bars show binned trends for different stellar mass ranges with error bars indicating the $1\sigma$ scatter. ({\em top left}) \mtolwise\ as a function of only WISE1 luminosity. The changing \mtolwise\ seen in Figures \ref{fig:sfr_mtol_grids} and \ref{fig:mtol_hist} leads to a multi-valued \mtolwise , and so multivalued mass, at fixed WISE1 luminosity. ({\em top right}) \mtolwise\ as a function of WISE2-to-WISE1 color with prescriptions by \citet{MEIDT14} and \citet{QUEREJETA15}. Though the similar IRAC2-to-IRAC1 color was widely available from warm \textit{Spitzer} observations, WISE2-to-WISE1 has smaller dynamic range and less power to predict \mtolwise\ than the other colors, which better trace SFR/M$_\star$.  \mtolwise\ as a function of the stellar mass and specific star formation rate inferred from SED modeling by \citet{SALIM16}. The figure shows a factor of $\sim 2$ dynamic range in \mtolwise\ \citep[e.g., see][]{BELL01,BELL03}. The figure shows a clear, strong relationship between \mtolwise\ and specific star formation rate. ({\em middle row}) \mtolwise\ as a function of two colors related to specific star formation rate observed in our atlas, NUV-to-WISE1 and WISE3-to-WISE1. Both show the qualitative trends seen in the top right panel, but also second-order dependence on stellar mass. The black and white in each panel offers an approximate prescription for \mtolwise, which we report in the text. }
\label{fig:mtol_obs}
\end{figure*}

\begin{deluxetable}{lcccc}
\tabletypesize{\scriptsize}
\tablecaption{Predictions for $3.4\mu$m Mass-to-Light Ratio \mtolwise\ \label{tab:mtolfits}}
\tablewidth{\textwidth}
\tablehead{
\colhead{Description} &
\colhead{Quantity, $Q$} & 
\colhead{$a$} & 
\colhead{$b$} &
\colhead{$c$} 
}
\startdata
GSWLC specific star formation rate & $\log_{10} {\rm SFR/M}_\star $ & $-10.9$ & $-0.21$ & $-9.5$ \\
``Specific SFR-like'' SFR-to-WISE1 estimate\tablenotemark{x} & $\log_{10} \frac{{\rm SFR(FUV+W4)}}{(\nu L_\nu)_{\rm WISE1}}$ & $-11.0$ & $-0.375$ & $-10.2$ \\
FUV-to-WISE1 color\tablenotemark{y} & $\log_{10} L_{\rm FUV}/L_{\rm WISE1}$ & $-2.5$ & $-0.167$ & $-0.7$ \\
NUV-to-WISE1 color\tablenotemark{y,z} & $\log_{10} L_{\rm NUV}/L_{\rm WISE1}$ & $-2.0$ & $-0.23$ & $-0.7$ \\
WISE3-to-WISE1 color\tablenotemark{y,z} & $\log_{10} L_{\rm WISE3}/L_{\rm WISE1}$ & $0.1$ & $-0.46$ & $0.75$ \\
WISE4-to-WISE1 color\tablenotemark{y} & $\log_{10} L_{\rm WISE4}/L_{\rm WISE1}$ & $0.0$ & $-0.4$ & $0.75$ \\
\enddata
\tablenotetext{x}{This is SFR estimated using FUV+WISE4 or NUV+WISE4 following the prescriptions in Table \ref{tab:sfr_gswlc} then divided by WISE1 luminosity, $\nu L_\nu$. That is, it represents a first order guess at a quantity proportional to SFR/M$_\star$. This is then used to predict \mtolwise .}
\tablenotetext{y}{All colors are logarithmic ratios of specific luminosity in Jy, i.e., $L_{\nu}$ and not $\nu L_\nu$.}
\tablenotetext{z}{Not plotted in Figures \ref{fig:mtol_hist} or \ref{fig:mtol_obs}. Results similar to those shown.}
\tablecomments{Coefficients for approximate predictions of the WISE1 mass-to-light ratio using Equation \ref{eq:pred_mtol}. There, $a$ corresponds to the cutoff in $Q$ below which $\mtolwise = 0.5$~\mtolunits , $c$ corresponds to the cutoff in $Q$ above which $\mtolwise =  0.2$~\mtolunits\ and $b$ indicates the slop in the intermediate regime. These fits appears as black-and-white lines in Figures \ref{fig:mtol_hist} and \ref{fig:mtol_obs}.}
\end{deluxetable}

The grids in Figure \ref{fig:sfr_mtol_grids} indicate that in the GSWLC, specific star formation rate (sSFR $\equiv$ SFR/M$_\star$) alone predicts \mtolwise\ well. Figure \ref{fig:mtol_hist} shows this directly. In the left panel, we plot histograms of \mtolwise\ for high (blue) and low (red) SFR/M$_\star$ galaxies with the cut at $\log_{10}$SFR/M$_\star = -11$~yr$^{-1}$. In low sSFR galaxies, \mtolwise\ appears high $\sim 0.5~\mtolunits$. The distribution for these targets also appears narrow, with only about $0.05$~dex scatter and a normal distribution.

Star-forming galaxies show a wider range of \mtolwise , mostly $0.2{-}0.5~\mtolunits$. The single best value would be $\sim 0.3$~\mtolunits , but with large scatter and dependent on sample selection. The black and white line in the right panel of Figure~\ref{fig:mtol_hist} shows that a bounded power law model offers a good approximation to \mtolwise. In Figures \ref{fig:mtol_hist} and \ref{fig:mtol_obs}, these lines have the form:

\begin{equation}
\label{eq:pred_mtol}
\mtolwise~\left[ \mtolunits \right] =
\begin{cases}
    0.5, & \text{if}~ Q < a \\
    0.5 + b~\left( Q - a\right) , & \text{if}~a < Q <c \\
    0.2 , & \text{if}~Q>c
\end{cases}
\end{equation}

\noindent with $Q$ the quantity of interest. $Q$ is always a logarithmic quantity in our fits, either $\log_{10}$ of the specific star formation rate or an observable color, so that the intermediate regime represents a power law. We cap \mtolwise\ at a high value, $0.5$~\mtolunits , for the most quiescent galaxies and a low value, $0.2$~\mtolunits , for the most active star-forming galaxies. Though higher or lower values may exist, they do not appear common in the GSWLC and adopting this form avoids extrapolating to unphysical values.

Table \ref{tab:mtolfits} gives approximate fits of this form for several quantities, $Q$, including $\log_{10}$ sSFR. The right panel in Figure \ref{fig:mtol_hist} shows that this is a good representation of the data. Even a simple prescription using SFR/$M_\star$ captures the main variations in \mtolwise\ in the GSWLC.

This agrees with expectations and many previous studies. Younger stellar populations produce more near infrared light per unit mass, on average \citep[e.g.,][among many others]{BELL01,BELL03,KANNAPPAN04,COURTEAU14,SIMONIAN17,STANWAY18} because SFR/M$_\star$ relates to the age of the stellar population \citep[e.g.,][]{KANNAPPAN13}. Again consistent with previous work, including \citet{BELL01} and \citet{BELL03}, the near-infrared $\Upsilon_\star$ has a dynamic range of about a factor of $\sim 2$ over the massive galaxy population. But the correlation of this quantity with SFR/$M_\star$ means that it must be accounted for to reproduce galaxy scaling relations using stellar masses estimated based on WISE1.

\subsection{Calculating \mtolwise\ From WISE and GALEX Data} 

If we knew SFR/M$_\star$ \textit{a priori} then we would use this to predict \mtolwise\ and be done. Unfortunately, we have only GALEX and WISE fluxes for our sample. We need to predict \mtolwise\ based on these measurements. This is further complicated by the uncertainties in distances to nearby galaxies and lack of completeness in the SDSS comparison sample. This leads us to prefer predictions based on colors, i.e., band ratios, which do not depend on the distance. Further complicating matters, the coefficients to translate WISE to star formation rate can also vary across the galaxy population.

With these complications in mind, Figure \ref{fig:mtol_obs} shows \mtolwise\ as a function of a GALEX- and WISE-based observables, again using GSWLC data. Grayscale and contours again show logarithmic data density of GSWLC galaxies. Colored points show the median trend and $1\sigma$ scatter for galaxies in narrow bins of stellar mass ($\log_{10} M_\star = 9.0$ to $11.5$~dex in bins of $0.5$~dex). Black-and-white lines show the prescriptions in Table \ref{tab:mtolfits}. The table also reports fits to several additional colors that we do not plot.

The top right panel shows the problem. Plotting \mtolwise\ as a function of WISE1 luminosity yields a large spread in \mtolwise\ at fixed luminosity. This appears particularly striking at the high luminosity end, where selection effects due to requiring GALEX and WISE detections will be smaller. Only knowing that a galaxy has WISE1 luminosity $\sim 10^{11}$~L$_\odot$ leaves the mass uncertain by a factor of $\sim 2$.

This issue has been known for decades and numerous prescriptions suggested to deal with it. The middle left panel shows WISE2-to-WISE1 color. The closely related IRAC2-to-IRAC1 color could be measured by \textit{Spitzer} during its warm mission. Because of this \citet{MEIDT14} and \citet{QUEREJETA15}, among others, use this color to infer $\Upsilon_\star^{3.6}$ for S4G targets. Their prescriptions, shown in the figure, bracket the behavior in the GSWLC. In the GSWLC, galaxies do show the expected anti-correlation between \mtolwise\ and WISE2-to-WISE1. However, galaxies exhibit only a small range of WISE2-to-WISE1 ratios, with \mtolwise\ heavily multi-valued around $\log_{10}$WISE2/WISE $\sim -0.1$. Galaxies with different masses also separate in the plot, indicating that more information than only this color is needed. In short, \mtolwise\ vs. WISE2-to-WISE1 trend in the GSWLC appears steep, multi-valued, and dependent on stellar mass. 

The other panels plot colors more directly related to SFR/M$_\star$. These do a better job, offering more dynamic range and showing a more direct relationship to \mtolwise\ than WISE2-to-WISE1. The middle right panel shows NUV-to-WISE1 and the bottom left panel shows WISE4-to-WISE1. Both ratios divide a star formation-tracing band by a stellar mass-tracing band and so trace sSFR, though with caveats. In both cases, we see a clear relation between \mtolwise\ and the observed color, though with considerable scatter. Galaxies with different masses show somewhat different trends, that is, the colored bins separate. Still, we report the fits using these colors in Table \ref{tab:mtolfits}. These may be useful, e.g., when only GALEX or WISE are available. We suggest that these represent a better option than WISE2-to-WISE1 or IRAC2-to-IRAC1 moving forward. WISE3 and WISE4 measurements, and often NUV and FUV, should now be available for almost all galaxies, including the S4G sample \citep[but were not available to][and earlier work]{MEIDT14,QUEREJETA15}.

\textbf{A specific star formation-like quantity:} Rather than a single observed color, one can estimate SFR, divide by the WISE1 luminosity, and use this to predict \mtolwise . The SFR divided by the WISE1 luminosity yields an ``sSFR-like'' quantity that predicts \mtolwise\ well. The bottom right panel in Figure \ref{fig:mtol_obs} shows only weak systematics as a function of stellar mass.

This appears to be the best option among the ratios that we consider. We find similar results using FUV+WISE4 and NUV+WISE4. In both cases we use the prescriptions calculated below, but would achieve qualitatively similar results for prescriptions from the literature, e.g., those in \citet{KENNICUTT12}.

\textbf{Color-color grids:} We also experimented with using grids in FUV-to-WISE1 vs. WISE4-to-WISE1 space or NUV-to-WISE1 vs. WISE3-to-WISE1 space. These offer an appealing way to take full advantage of the GALEX and WISE data. Unfortunately we found that a small fraction of our local targets showed colors not well-represented in the GSWLC. Applying this method to the GSWLC also yielded more catastrophic outliers than the other approaches. A grid-based approach still offers a sensible way forward, but will require properly treated likelihoods and sensible behavior outside the parameter space well-covered by the GSWLC. This approach would benefit from the inclusion of additional bands and direct interface with CIGALE. This remains beyond the scope of this paper.

\textbf{Adopted approach and results:} For our purpose, predicting \mtolwise\ using the specific star formation-like quantity, SFR-to-WISE1 luminosity, offers the best combination of robustness and accuracy. We adopt this as our preferred approach whenever GALEX and WISE data are both available. We always prefer WISE4 to WISE3 in the SFR estimate. Despite the better quality of the WISE3 data, the strong systematic trends visible in Figure \ref{fig:sfr_mtol_grids} render WISE3 a less robust indicator of the SFR than WISE4. Unfortunately, this approach requires both GALEX and WISE data, and so will not always be available. Approximately $1/3$ of our targets have only WISE data. In these cases, the WISE4-to-WISE1 ratio appears to offer the best option.

In detail our approach is:

\begin{enumerate}
\item If FUV and WISE4 are both available, estimate SFR based on the prescription combining FUV and WISE4 in Table \ref{tab:sfr_gswlc}, divide by WISE1 luminosity, and use Table \ref{tab:mtolfits} to estimate \mtolwise . WISE1 is always available.
\item If NUV is available and FUV is not, estimate SFR based on the prescription combining NUV and WISE4 in Table \ref{tab:sfr_gswlc}, divide by WISE1 luminosity, and use Table \ref{tab:mtolfits} to estimate \mtolwise .
\item If only WISE4 is available, estimate \mtolwise\ based on the WISE4-to-WISE1 color in Table \ref{tab:mtolfits}.
\item Lacking any other indicator, we adopt $\mtolwise = 0.35$~\mtolunits\ for star-forming or late type galaxies and $\mtolwise = 0.5$~\mtolunits\ for quiescent early type galaxies. In practice, this case almost never arises. Given no information, we assume that we consider a star-forming galaxy and adopt $\mtolwise = 0.35$~\mtolunits .
\end{enumerate}

\begin{figure*}
\centering
\plottwo{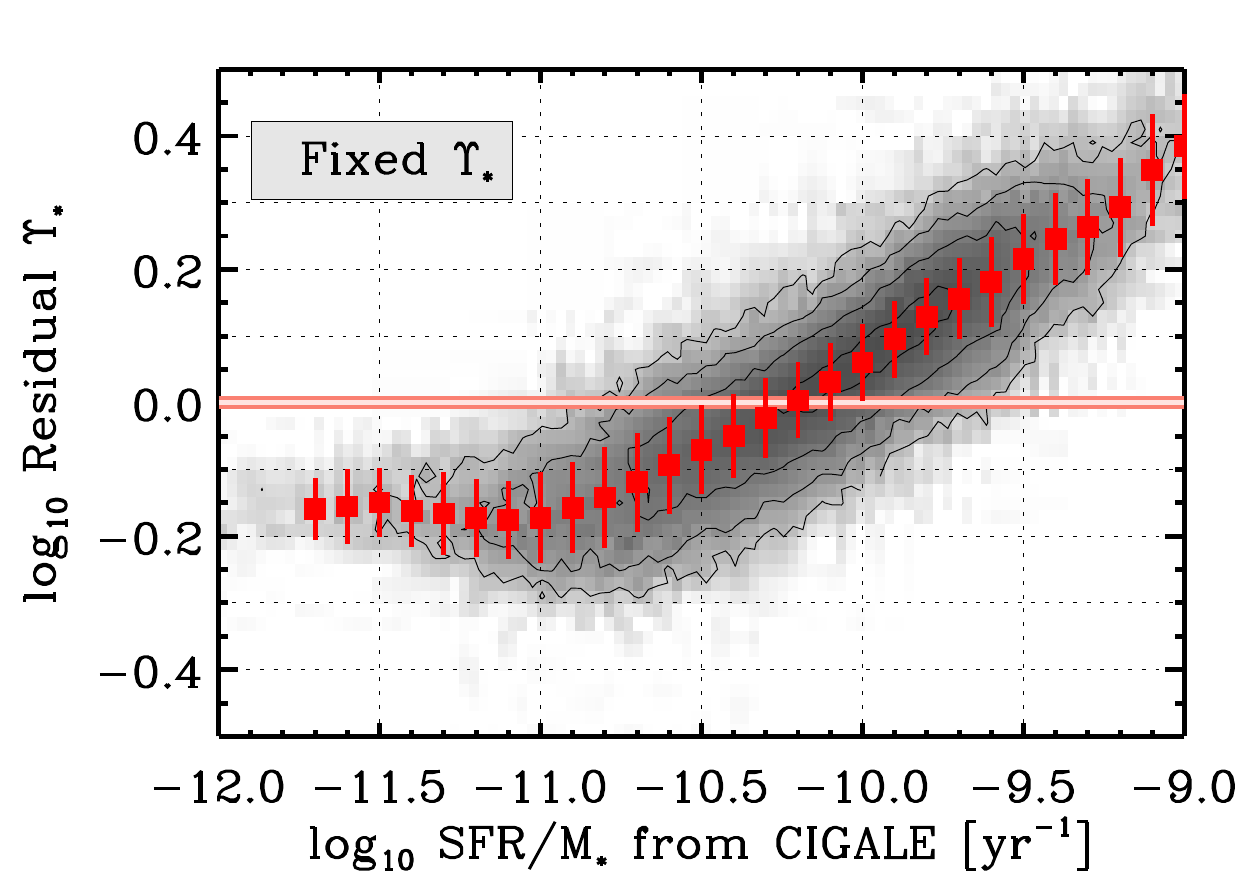}{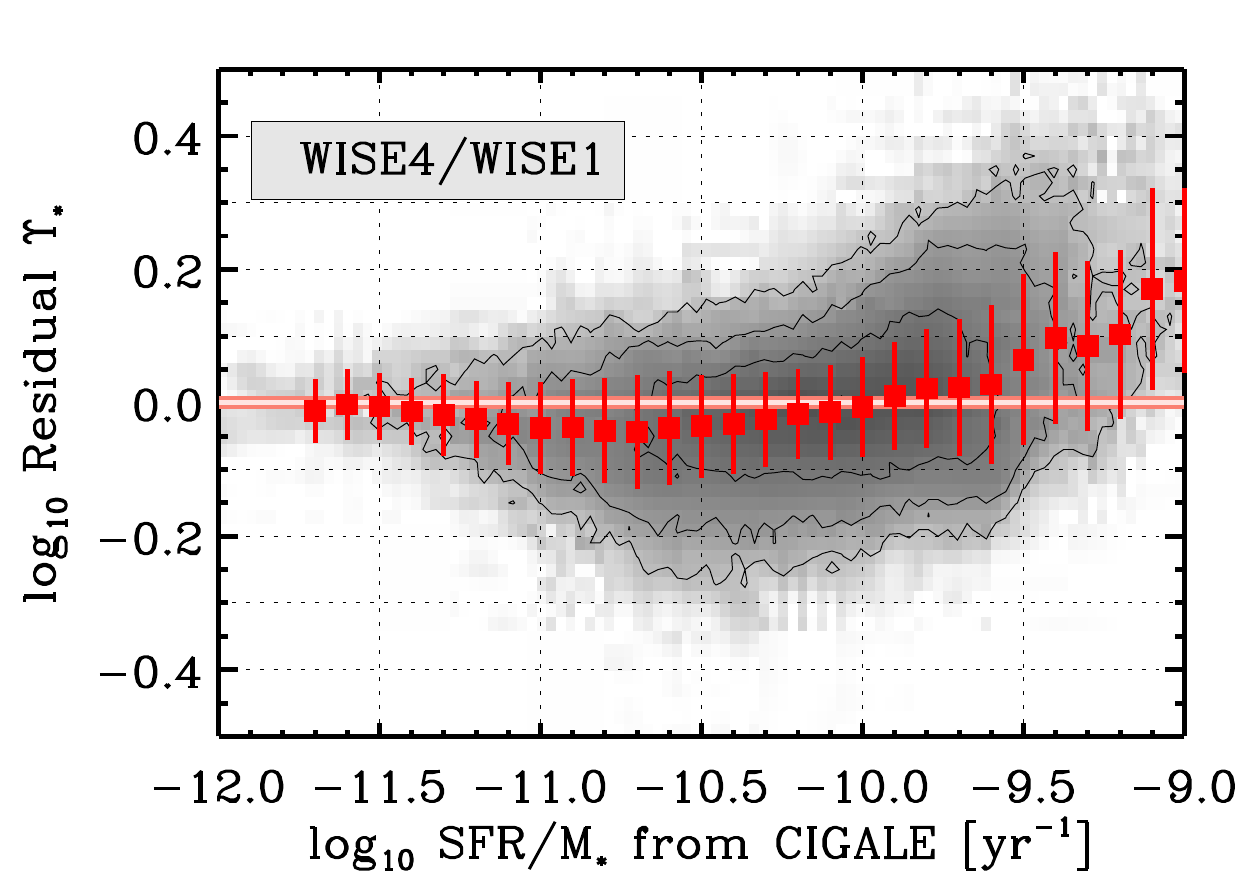}
\plottwo{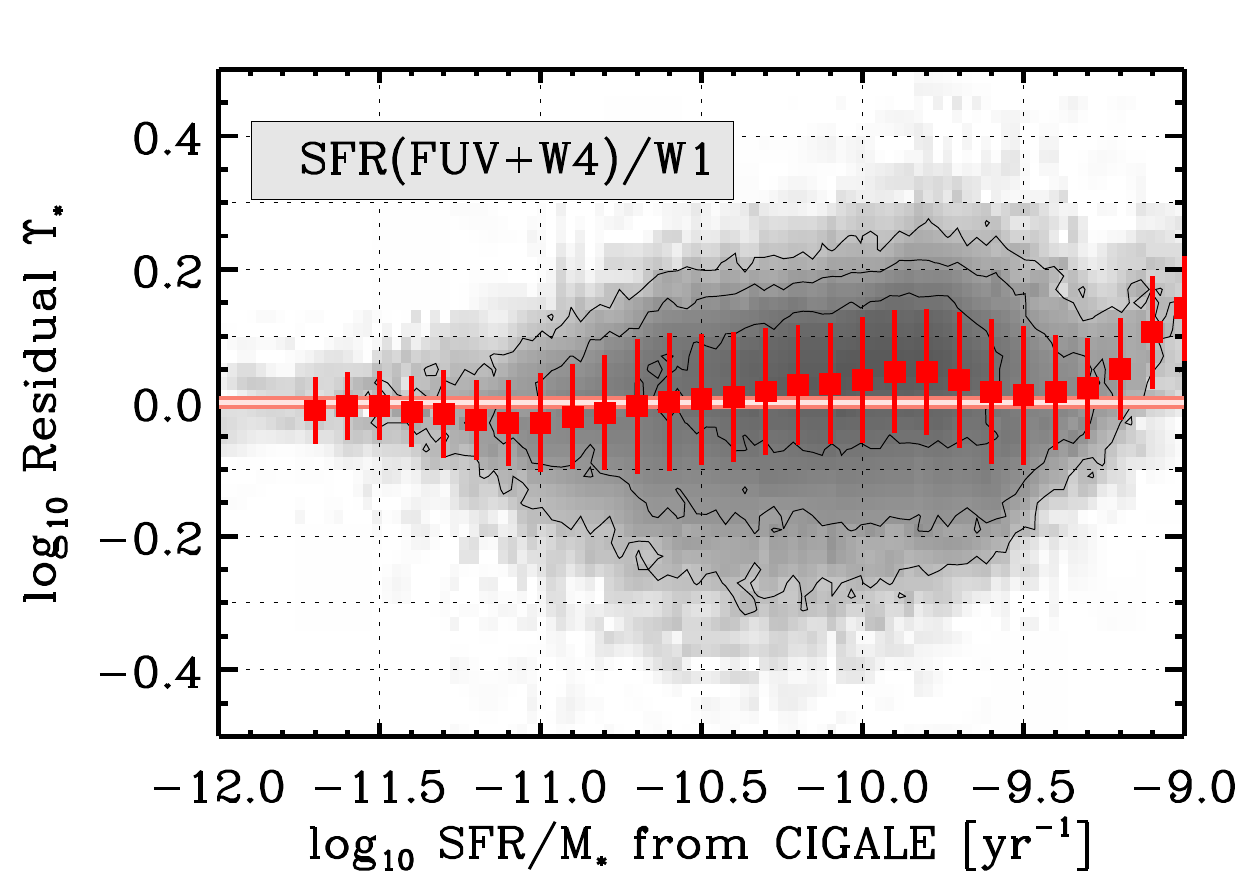}{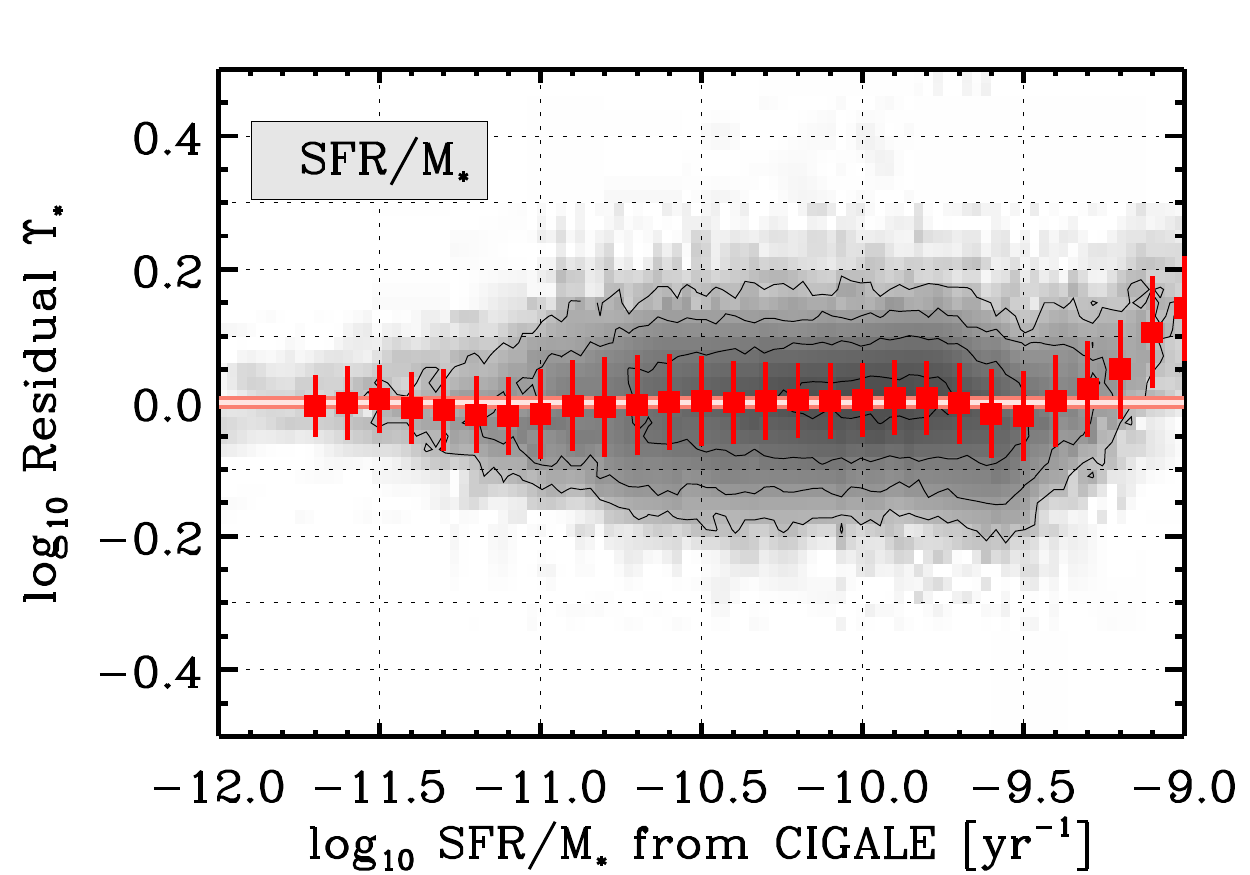}
\caption{{\bf Residuals About Mass-to-Light Ratio Prescriptions.} Logarithmic residuals ($\log_{10}$ predicted-to-actual \mtolwise ) comparing prescriptions to predict \mtolwise\ (Table \ref{tab:mtolfits}) to the best-fit \mtolwise\ from \citet{SALIM16,SALIM18}. In all panels the gray image show logarithmic data density for individual GSWLC galaxies. The red points show the median residual in bins of fixed SFR/M$_\star$ (with SFR and M$_\star$ the best fit GSWLC values). Error bars on the bins show the scatter in the residuals. The {\em top left} panel shows results adopting a fixed \mtolwise. The {\em top right} panel shows results using only a WISE color to predict \mtolwise . This performs well but fails in the highest SFR/M$_\star$ cases, which also tend to be lower mass, dust-poor galaxies in our GSWLC selection. The {\em bottom left} panel shows residuals using the ratio of SFR-to-WISE1 luminosity with SFR computed from GALEX and WISE photometry. The {\em bottom right} panel shows the best case, when SFR/M$_\star$ is known. For our atlas, we use the approach shown in the bottom left panel or a closely related approach (substituting NUV for FUV) when available. When only WISE data are available, we use the approach shown in the top right panel.}
\label{fig:mtol_resid}
\end{figure*}

To test this approach, we predict \mtolwise\ for the GSWLC galaxies using our method and compare this to the values based on SED modeling. Figure \ref{fig:mtol_resid} plots the residuals in \mtolwise , defined as $\log_{10}$ predicted value divided by the GSWLC value. We show the performance of a fixed mass to light ratio, only WISE4-to-WISE1 color, and our specific star formation-like ratio. We also illustrate the case where we knew the true GSWLC SFR/M$_\star$. 

The figure shows that not accounting for a variable \mtolwise\ leads to $\pm 0.4$~dex of systematic uncertainty. On the other hand, if we knew SFR/M$_\star$, our prescription would match the GSWLC with better than $0.1$~dex accuracy. Using WISE4-to-WISE1 works well for more massive and lower SFR/M$_\star$ galaxies but overpredicts \mtolwise\ in high SFR/M$_\star$ GSWLC galaxies. Note that these tend to be lower mass galaxies (see Figure \ref{fig:sfr_mtol_grids}) in the GSWLC. Our sSFR-like quantity works well, though with more uncertainty than the true SFR/M$_\star$. On average, this approach recovers \mtolwise\ within $\sim 0.1$~dex with only weak systematic trends.

\subsection{Coefficient on the UV Term in SFR Prescriptions}

\begin{figure*}
\centering
\plottwo{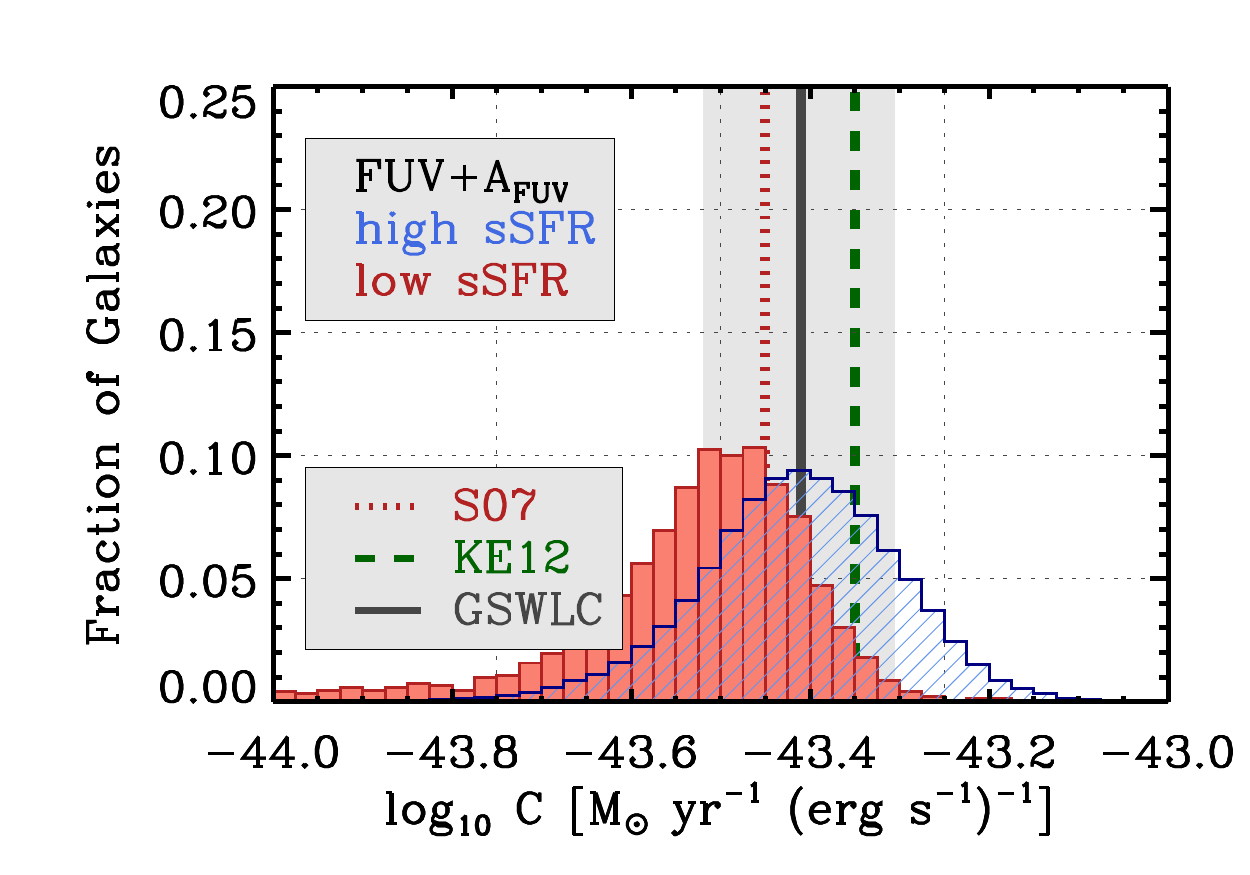}{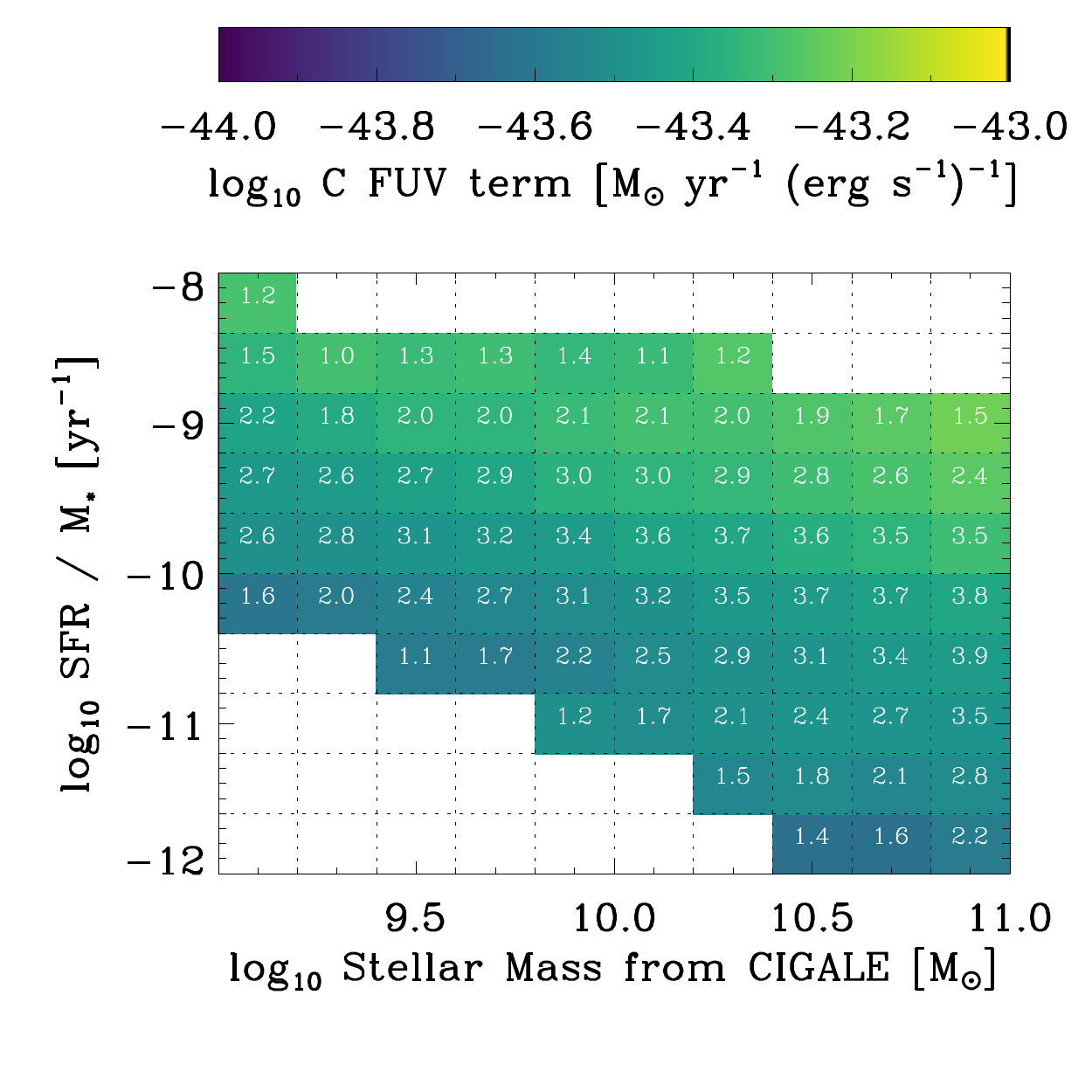}
\caption{{\bf Coefficient to Convert FUV to SFR.} Conversion factor, $C$ (FUV), between extinction-corrected FUV luminosity, $\nu L_\nu$, and SFR in the GSWLC \citep{SALIM18}. ({\em left}) Histogram of $C$ for all high specific star formation rate targets (blue) and the smaller set of low specific star formation rate targets with a WISE detection (red). For reference, we plot $C$ (FUV) from \citet[][$\log_{10} C = -43.45$]{SALIM07} and \citet[][$\log_{10} C = -43.35$]{KENNICUTT12}. ({\em right}) $C$ (FUV) in the stellar mass-sSFR space. As in Figure \ref{fig:sfr_mtol_grids} with use a log stretch with dynamic range is $1$~dex, numbers indicate the $\log_{10}$ of the galaxies used in the average. The GSWLC implies $\log_{10} C = -43.42$ with $\pm 0.1$~dex scatter (gray region), intermediate between the two. We use this for the unobscured term in FUV-based hybrids and adopt a simple extrapolation for NUV-based hybrids.}
\label{fig:sfr_fuv}
\end{figure*}

We use the GSWLC to check the prefactor on the UV term in SFR prescriptions. In Figure \ref{fig:sfr_fuv} shows we plot the $C$ to convert FUV luminosity, $\nu L_\nu$, to SFR for galaxies in \citet{SALIM18}. Here we calculate $C$ by dividing the SFR fit by \citet{SALIM18} by the extinction-corrected FUV luminosity. We correct for extinction using the best-fit GSWLC FUV extinction, $A_{\rm FUV}$. Vertical lines show the conversion factors from \citet{KENNICUTT12} and \citet{SALIM07}. In the histograms, blue shows results for high sSFR galaxies, defined as $\log_{10}$~SFR/$M_\star < -11$~yr$^{-1}$. The red histogram, which appears shifted to slightly lower values than the blue one, shows results for low sSFR galaxies that have GALEX and WISE detections. Note that due to the signal to noise requirement and intrinsically low SFR for quiescent galaxies, we expect bias in the measurement for these low sSFR galaxies. We will include those with unusually bright FUV, leading $C$ to be biased low. Our focus here is on the star forming galaxies, the blue histogram.

In Figure \ref{fig:sfr_fuv} solid gray line shows the median $C$ implied by the GSWLC and the gray band shows the 16-84$^{\rm th}$ percentile range. This is 

\begin{equation}
\label{eq:sfr_fuv}
\log_{10} C ({\rm FUV}) = -43.42~{\rm M_\odot~yr^{-1}~(erg~s^{-1})^{-1}}
\end{equation}

\noindent with $\sim \pm 0.1$~dex scatter. This conversion lies between the $\log_{10} C_{\rm FUV} = -43.45$ of \citet{SALIM07} and $\log_{10} C_{\rm FUV} = -43.35$ of \citet{KENNICUTT12}. It is close to both, implying no major imprecision from using these existing calibrations. It agrees better with \citet{SALIM07} than \citet{KENNICUTT12}. This might be expected given that \citet{SALIM07} used a similar approach to \citet{SALIM18} to derive their values. We proceed using Equation \ref{eq:sfr_fuv} but would reach similar results using the calibration of \citet{SALIM07} or \citet{KENNICUTT12}.

The right panel in Figure \ref{fig:sfr_fuv} shows only weak correlation of $C$ (FUV) with stellar mass and sSFR. A single $C$~(FUV) represents a reasonable first-order assumption and the $0.1$~dex scatter appears consistent with the $\approx 0.3$~mag typical uncertainty on the fit $A_{\rm FUV}$ in the GSWLC. Weak systematic variations do exist, however. The appropriate $C$ resembles the lower \citet{SALIM07} value in more quiescent galaxies and the \citet{KENNICUTT12} value in the most massive, actively star-forming galaxies.

We extrapolate from Equation \ref{eq:sfr_fuv} to $C$ (NUV) adopting the standard assumption that the UV spectral shape of a star-forming population is approximately flat \citep[see][]{KENNICUTT12}. Ideally, we would check the NUV-to-SFR directly from the results of \citet{SALIM18}, but the presence of the UV bump in the bandpass requires uncertain assumptions about dust properties \citep[see extensive discussion in][]{SALIM18}. This yields:

\begin{equation}
\label{eq:sfr_nuv}
\log_{10} C ({\rm NUV}) = -43.24~{\rm M_\odot~yr^{-1}~(erg~s^{-1})^{-1}}
\end{equation}

\noindent Again, this closely resembles the $\log_{10} C_{\rm NUV} = -43.17$ from \citet{KENNICUTT12} and $\log_{10} C_{\rm NUV} \approx -43.28$  implied by the FUV data from \citet{SALIM07}.

We report both $C ({\rm FUV})$ and $C ({\rm NUV})$ in Table \ref{tab:sfr_gswlc}. There, we also report the galaxy-to-galaxy scatter in the coefficient across our sample of GSWLC galaxies (in dex). We also report the scatter in $C$ from cell-to-cell across the SFR-M$_\star$ plane. This scatter, which weights each cell equally regardless of the number of galaxies, captures the degree of systematic variation in the quantity.

\subsection{Coefficient on the mid-IR Term}

\begin{figure*}
\centering
\plotone{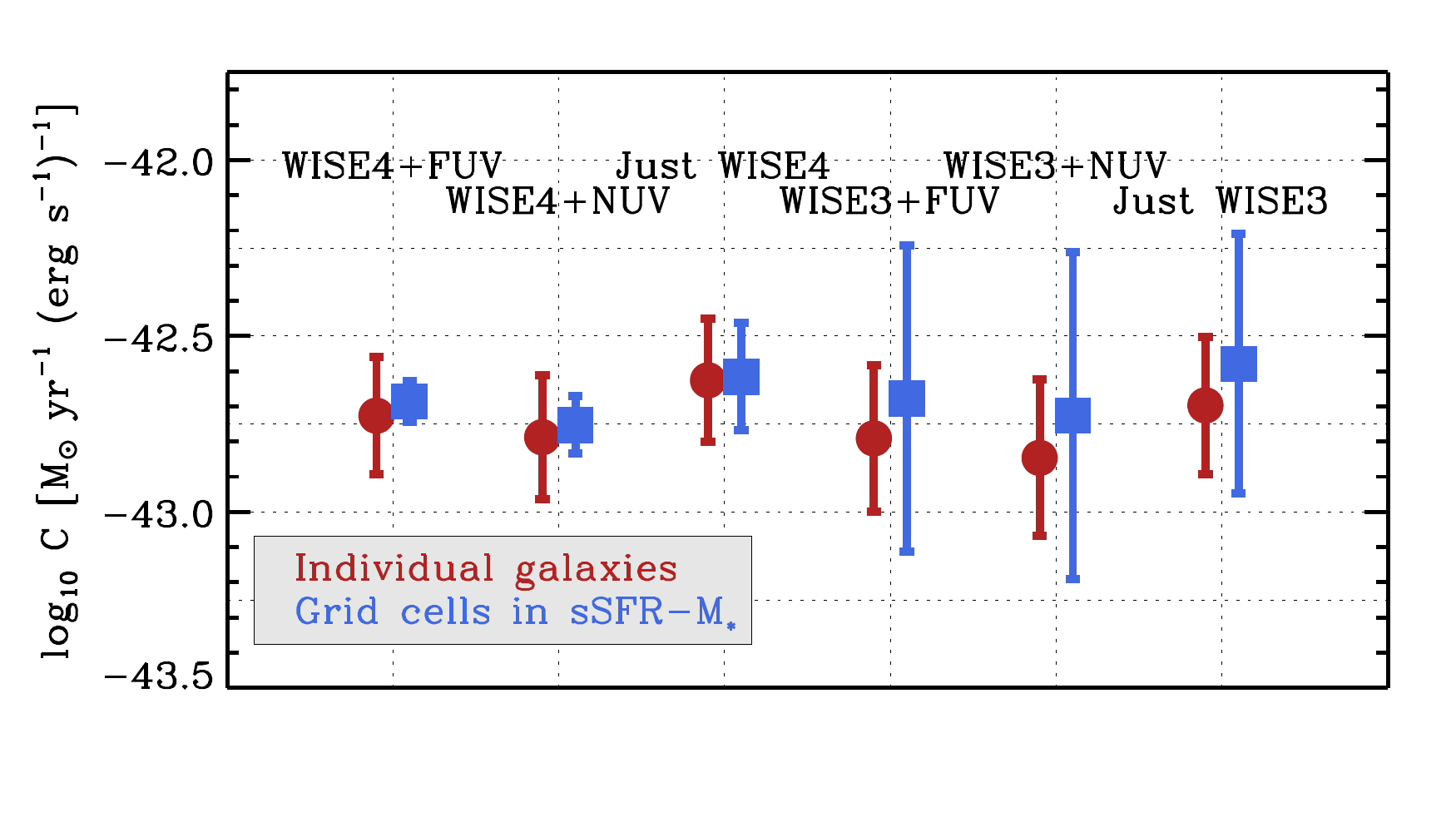}
\caption{{\bf Coefficients, $C$, on the WISE term in hybrid SFR indicators implied by the GSWLC.} The coefficient, $C$, to translate WISE luminosity, $\nu L_\nu$, to a star formation rate estimate. We calculate $C$ by comparing the SFR from \citet{SALIM18} to the WISE flux, after subtracting the SFR implied by the UV alone in the cases of hybrids. We show the median and $1\sigma$ scatter in $C$ treating all GSWLC galaxies equally in red. In blue, we show the median $C$ and the scatter treating each cell in the SFR/M$_\star$-M$_\star$ plane equally (Figure \ref{fig:sfr_mtol_grids}). Thus, the blue error bars capture the level of systematic seen across GSWLC. Table \ref{tab:sfr_gswlc} reports these values.}
\label{fig:cwise}
\end{figure*}

\begin{deluxetable}{lcccc}
\tabletypesize{\scriptsize}
\tablecaption{Linear ``Hybrid'' Star Formation Rate Calibrations Implied by the GSWLC \label{tab:sfr_gswlc}}
\tablewidth{0pt}
\tablehead{
\colhead{Band} &
\colhead{Combined with} & 
\colhead{$\log_{10} C_X$\tablenotemark{a}} &
\colhead{Scatter\tablenotemark{b}} &
\colhead{Scatter in in grid\tablenotemark{c}} \\
\colhead{} &
\colhead{} & 
\colhead{($\log_{10}$~M$_\odot$~yr$^{-1}$~(erg~s$^{-1}$)$^{-1}$)} &
\colhead{(dex)} &
\colhead{(dex)}
}
\startdata
\multicolumn{5}{c}{``Unobscured'' UV term} \\
FUV\tablenotemark{d} & GSWLC $A_{\rm FUV}$ & $-43.42$ & $0.1$ & $\ldots$ \\
NUV\tablenotemark{d} & Extrapolated from FUV & $-43.24$ & $0.1$ & $\ldots$ \\
\hline
\multicolumn{5}{c}{WISE Term} \\
WISE4 & alone & $-42.63$ & $0.17$ & $0.15$ \\
WISE4 & with FUV & $-42.73$ & $0.17$ & $0.06$ \\
WISE4 & with NUV & $-42.79$ & $0.18$ & $0.08$ \\
\hline
WISE3 & alone & $-42.70$ & $0.20$ & $0.37$ \\
WISE3 & with FUV & $-42.79$ & $0.21$ & $0.44$ \\
WISE3 & with NUV & $-42.86$ & $0.22$ & $0.47$ \\
\enddata
\tablenotetext{a}{$C_X$ is the factor to convert $\nu L_\nu$ to SFR \citep[notation follows][]{KENNICUTT12}. For the WISE term in linear hybrids, this is the factor to apply to WISE and then add to the ``unobscured term.''}
\tablenotetext{b}{Galaxy-to-galaxy scatter, in dex, in $C_X$ treating all GSWLC galaxies equally.}
\tablenotemark{c}{Scatter, in dex, across the specific star formation rate-stellar mass plane (Figure \ref{fig:sfr_mtol_grids}) treating each cell equally. This indicates the amount of systematic uncertainty in the tracer.}
\tablecomments{Conversions from luminosity to star formation taking the \citet{SALIM18} fits to the GSWLC as a reference. We compare GALEX and WISE photometry to SFR to calculate $C_X$ for each galaxy and band combination. This table reports the median values and scatter.}
\end{deluxetable}

We adopt an empirical approach to calibrate the mid-IR part of the SFR estimate. This approach has been discussed extensively in reviews by \citet{KENNICUTT12} and \citet{CALZETTI13}. We will focus on the linear ``hybrid'' tracer formalism, which combines an unobscured and infrared term to estimate the star formation rate. Following groundbreaking early work by \citet{CALZETTI97} and \citet{MEURER99}, \citet{CALZETTI07} and \citet{KENNICUTT07} pioneered an extension of this approach combining H$\alpha$ and mid-IR emission. \citet{LEROY08} and \citet{THILKER07} carried out early combinations of GALEX and mid-IR data, the bands that we use here. Also see \citet{LEROY12} for an extensive discussion.

For each selected \citet{SALIM18} galaxy and each combination of NUV and FUV with WISE3 and WISE4, we calculate the scaling factor that must be applied to WISE in order to match the best GSWLC SFR estimate. In the cases where we hybridize a UV and a mid-IR band, we first subtract the contribution of the UV band in question, estimated using $C_{\rm UV}$, from the SFR and then compare to WISE. Thus,

\begin{equation}
\label{eq:sfr_wise}
C_{\rm WISE} \equiv \log_{10} \left( \frac{{\rm SFR} (\rm{GSWLC}) - {\rm SFR} (\rm{UV})}{\nu L_\nu ({\rm WISE}) } \right)
\end{equation}

\noindent Here again $C_{\rm WISE}$ represents the conversion factor to translate a WISE luminosity, $\nu L_\nu$, into an estimate of the SFR. 

For the case of a hybrid tracer, one would calculate SFR from:

\begin{equation}
{\rm SFR} = C_{\rm UV} \nu_{\rm UV} L_{\nu, {\rm UV}} + C_{\rm WISE} \nu_{\rm WISE} L_{\nu, {\rm WISE}}~.
\end{equation}

\noindent For a ``WISE only'' tracer, one would use only the second term and use the ``WISE3 alone'' or ``WISE4 alone'' $C_{\rm WISE}$ values.

Figure \ref{fig:cwise} shows the median and scatter in $C$ for WISE3 and WISE4 alone and combination with FUV and NUV. We plot two results for each coefficient, one calculated treating all GSWLC galaxies equally (in red) and one treating each cell in SFR/M$_\star$-M$_\star$ space equally (in blue). A large scatter in the second estimate, treating all grid cells equally, indicates significant systematic variations in that $C$ across the galaxy population.

As expected, we find the highest $C$ when using WISE alone, the lowest $C$ when hybridizing with NUV, and intermediate values combining WISE and FUV. The higher value for FUV than NUV reflects that FUV emission tends to be more extinguished than NUV emission. The even higher values for WISE alone reflects that the ``unobscured'' UV term contributes significantly to the SFR. Lacking an unobscured component entirely requires the mid-IR component be weighted more heavily to account for this. All of these results also underscore that this application of mid-IR to trace the SFR is fundamentally empirical.

\begin{figure*}
\centering
\plottwo{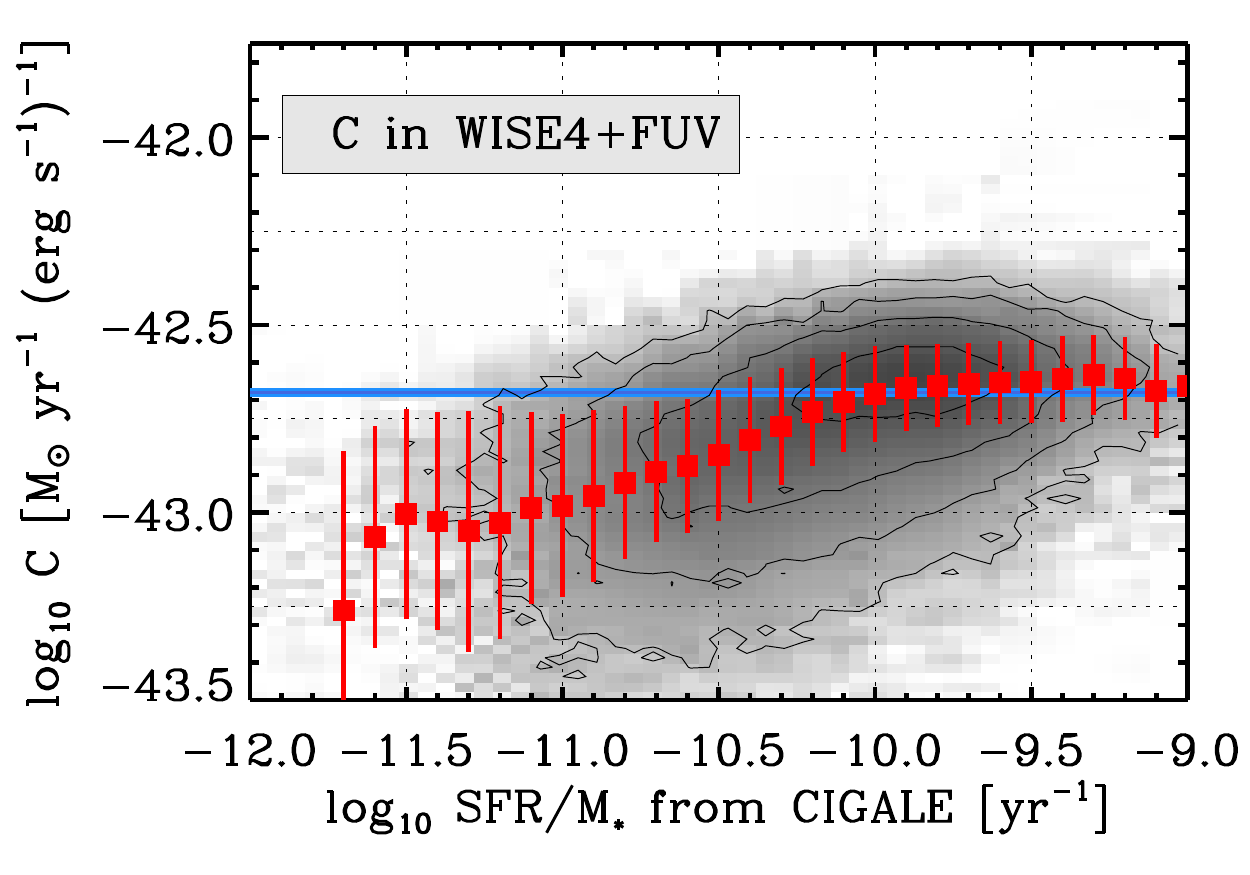}{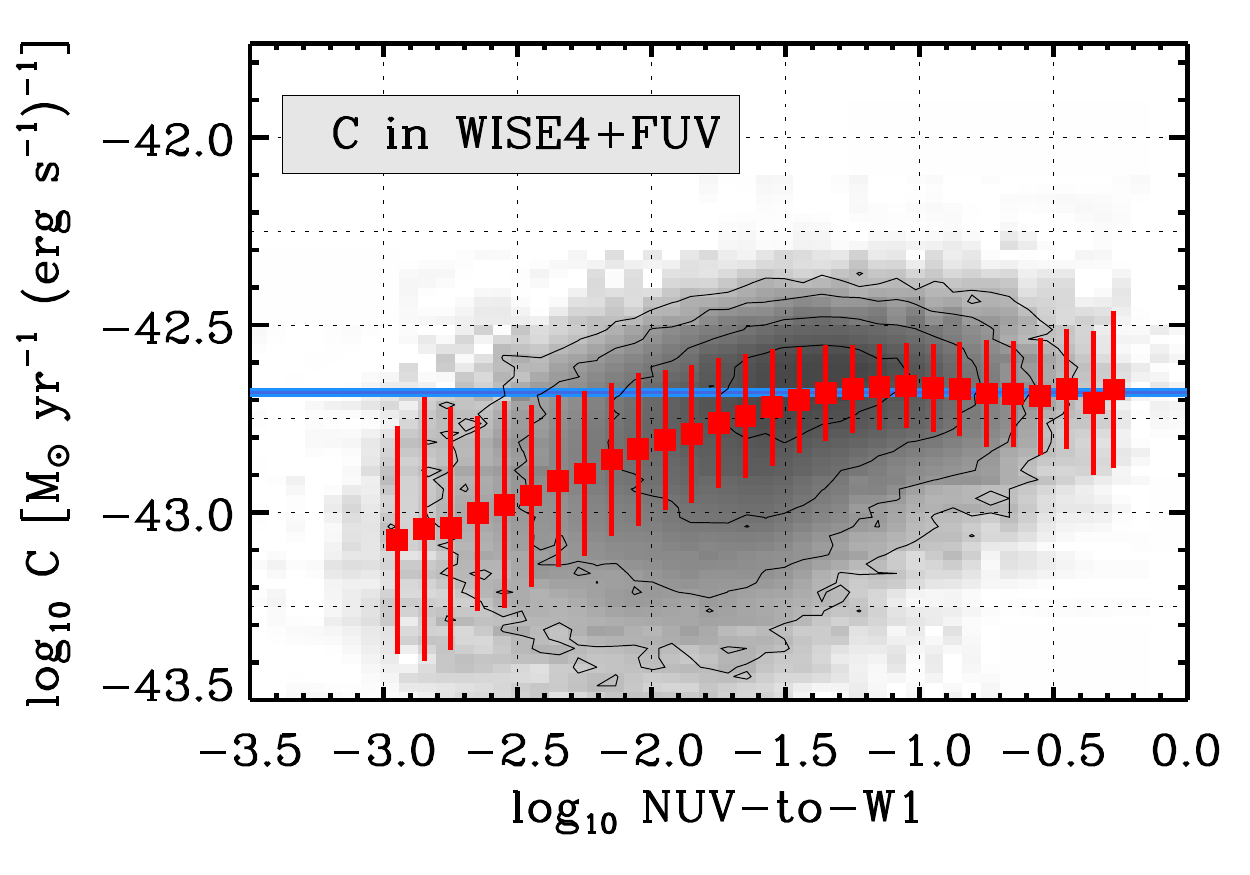}
\caption{{\bf Coefficient to translate WISE4 to SFR in WISE4+FUV hybrid.} The coefficient, $C$, to convert from WISE4 to SFR when using WISE4 as a hybrid tracer with FUV. Grayscale shows data density from the GSWLC \citep{SALIM18} on a logarithmic stretch. Red points show the median trend as a function SFR/M$_\star$ and error bars indicate the scatter. The blue line shows the median over galaxies. Quiescent galaxies, which can be picked out by their low NUV-to-WISE1 ratios, show more WISE4 per unit SFR. This agrees with a scenario where contaminants including optically heated ISM dust and circumstellar dust contribute heavily to WISE4 in these systems \citep[e.g., see][]{DAVIS14,SIMONIAN17}.}
\label{fig:sfr_trends}
\end{figure*}

\textbf{WISE4:} Overall, Figures \ref{fig:sfr_mtol_grids} and \ref{fig:cwise} and Table \ref{tab:sfr_gswlc} show that a single coefficient, $C$, combined with either FUV or NUV does a reasonable or reproducing the GSWLC SFR across a wide range of stellar masses. The cell-by-cell scatter in Figure \ref{fig:cwise} and Table \ref{tab:sfr_gswlc} implies low systematic uncertainty, $< 0.1$~dex, for FUV+WISE4 and NUV+WISE4. Similar results, albeit derived from $\sim 100$ times smaller samples, have motivated the widespread use of these tracers for the last decade. We find moderately higher mid-IR conversion factors than \citet{HAO11}, who derive the value summarized in \citet{KENNICUTT12}. 

Using WISE4 alone implies larger systematic uncertainty. This reflects the trend with stellar mass seen in Figure \ref{fig:sfr_mtol_grids}; low mass galaxies, with lower dust content, require a larger $C$ applied to only WISE4 in order to make up for the large missing unobscured term.

The top right panel in Figure \ref{fig:sfr_mtol_grids} shows that though these variations appear comparatively weak, $C$ to combine WISE4 with UV does vary as a function of SFR/M$_\star$ and M$_\star$. The sense is that in low SFR-to-M$_\star$ galaxies and massive galaxies $C$ drops, reflecting more WISE4 emission relative to star formation. As discussed in \citet{DAVIS14} and \citet{SIMONIAN17}, this should be expected in environments with little star formation. There will be more optical heating of interstellar dust grains, producing interstellar cirrus \citet{DAVIS14} and a floor provided by emission from circumstellar dust \citep{SIMONIAN17}.

Figure \ref{fig:sfr_trends} shows the variations in $C$ for WISE4 in WISE4+FUV. We plot $C$ as a function of SFR/M$_\star$, capturing the vertical variations in Figure \ref{fig:sfr_mtol_grids} from that figure. The figures show a gradual decline in $C$ below $\log_{10}$ SFR/M$_\star \sim -10.25$~yr$^{-1}$ The median $C$ drops by $\sim 0.3$~dex (a factor of 2) from the most active to most quiescent galaxies. At fixed SFR/M$_\star$ the plot still show significant scatter in $C$ for quiescent galaxies.

For our main SFR estimates, we do not suggest a functional correction to $C$ based on Figure \ref{fig:sfr_trends}. In our view, the requirement of WISE4 detections for quiescent galaxies introduces a nontrivial bias that will exacerbate the trends in the figure. In future work, using forced photometry and stacking could allow a more careful, general estimate of $C$ (and so ``IR cirrus'' effects). Our eventual goal is a color or SED-based IR cirrus estimate \citep[e.g., improving on][]{LEROY12}.

Alternatively, after estimating SFR using only the coefficients in Table \ref{tab:sfr_gswlc} and estimating \mtolwise , one can use the grid in Table \ref{tab:grid_ssfr_mstar} to iterate towards an appropriate $C$ (WISE4+FUV) given SFR/M$_\star$ and M$_\star$. This approach requires knowing the distance to the galaxy, but represent our best second-order prediction for SFR.

\textbf{WISE3:} The coefficient to convert WISE3 to SFR shows large cell-to-cell scatter in all combinations, rms $\sim 0.3{-}0.4$~dex compared to $\lesssim 0.1$~dex for WISE4. As discussed above, this reflects both a trend with stellar mass and one with specific star formation rate. These trends appear stronger than what we see for WISE4. They have the sense that WISE3 becomes weaker in low mass galaxies and high SFR/M$_\star$ galaxies As a result, $C_{\rm WISE3}$ increases to compensate, in low mass galaxies. The trends could be expected based on the known metallicity-dependence of PAH emission \citep[e.g.,][]{ENGELBRACHT05} and the strong contribution of the 11.3$\mu$m PAH feature to the WISE3 band. 

Based on these variations, we do not treat WISE3 as a primary star formation tracer. The higher resolution and sensitivity of the WISE3 data compared to WISE4 do make them useful, however. A full calibration remains the topic of future work, but in the meantime we use WISE3 to estimate the SFR in two ways:

\begin{enumerate}
\item Take the local ratio of WISE3-to-WISE4, measured, e.g., for the whole galaxy or similar types of galaxies. Then use the WISE4 calibration, modified by this ratio. That is take $C_{\rm WISE3} = W3/W4 \times C_{\rm WISE4}$ with $W3/W4$ the measured ratio. This approach assumes a locally fixed or otherwise known color. It could be useful, e.g., in the case where the mean W3/W4 color is known for a galaxy and we wish to use the higher quality WISE3 data to explore the resolved distribution SFR.

\item Make iterative estimates of M$_\star$ and SFR/M$_\star$ and use Table \ref{tab:grid_ssfr_mstar} to estimate $C_{\rm WISE3}$. This requires a first guess at SFR, because both $C_{\rm WISE3}$ and \mtolwise\ depend on SFR. One could make this first guess, using WISE4 or an average WISE3. Then one would refine $C_{\rm WISE3}$ iteratively, looking up the appropriate value given the current estimates in Table \ref{tab:grid_ssfr_mstar}.
\end{enumerate}

As with WISE4, we aim eventually for a calibration that uses only colors and could be applied to large regions of galaxies with confidence. This remains a topic for future work. Here we suggest to bootstrap via WISE4 to take advantage of the better WISE3 data quality \citep[in good agreement with the views of][]{JARRETT13,CLUVER17}.

\end{appendix}

\end{document}